\documentclass{aa}  

\usepackage{graphicx}
\usepackage{txfonts}
\usepackage[utf8]{inputenc}
\usepackage{dcolumn}
\usepackage{supertabular}
\usepackage{rotating}
\usepackage{longtable}

\usepackage{lscape}
\usepackage{float}
\usepackage{subfig}
\usepackage{url}
\usepackage{color}
\usepackage{natbib}
\usepackage[dvipsnames]{xcolor}
\usepackage{ulem}
\usepackage{soul}
\usepackage[colorinlistoftodos]{todonotes}
\usepackage[version=4]{mhchem} 
\usepackage{supertabular} 
\usepackage{amssymb}
\usepackage[colorlinks=true,
            linkcolor=blue,
            urlcolor=blue,
            citecolor=blue]{hyperref}

\newcommand{\kms}{\mbox{km s$^{-1}$}}
\newcolumntype{d}[1]{D{.}{\cdot}{#1}}
\newcolumntype{.}{D{.}{.}{-1}}

\newcommand{\hii}{H{\sc ii}}
\newcommand{\uchii}{UCH{\sc ii}}
\newcommand{\uchiis}{UCH{\sc ii}s}
\newcommand{\Msol}{\textrm \,M$_\odot$}
\begin{document}

   \title{A survey of SiO $J=$ 1 -- 0 emission toward massive star-forming regions \thanks{Full tables \ref{tab:soutb}, \ref{tab:sio_gaussianfit}, and \ref{tab:phy_para} are only available in electronic form at the CDS via anonymous ftp to cdsarc.cds.unistra.fr (130.79.128.5)
or via https://cdsarc.cds.unistra.fr/cgi-bin/qcat?J/A+A/.}}

   \author{W.-J.\,Kim\inst{1, 5} 
    \and J.\,S.\,Urquhart\inst{2}
    \and  V.\,S.\,Veena\inst{1, 3} 
    \and  G.\,A.\,Fuller\inst{1, 4}
    \and P.\,Schilke\inst{1}
    \and K-T\,Kim\inst{5, 6}}

   \institute{I. Physikalisches Institut, Universit\"at zu Köln, Z\"ulpicher Str. 77, 50937 K\"oln, Germany\\
              \email{wonjukim@ph1.uni-koeln.de}
        \and Centre for Astrophysics and Planetary Science, University of Kent, Ingram Building, Canterbury, Kent CT2\,7NH, UK       
        \and Max-Planck-Institut f\"{u}r Radioastronomie, Auf dem H\"{u}gel 69, 53121 Bonn, Germany
        \and Jodrell Bank Centre for Astrophysics, Department of Physics and Astronomy, the University of Manchester, Oxford Road, Manchester M13 9PL, UK 
        \and Korea Astronomy and Space Science Institute, 776 Daedeokdae-ro, Yuseong-gu, Daejeon 34055, Republic of Korea 
        \and University of Science and Technology, Korea (UST), 217 Gajeong-ro, Yuseong-gu, Daejeon 34113, Republic of Korea}

   \date{Received August 17, 2023; accepted September 25, 2023 }

 
  \abstract
   {}
   {The application of silicon monoxide (SiO) as a shock tracer arises from its propensity to occur in the gas phase as a result of shock-induced phenomena, including outflow activity and interactions between molecular clouds and expanding \hii\ regions or supernova remnants. For this work, we searched for indications of shocks toward 366 massive star-forming regions by observing the ground rotational transition of SiO ($v=0$, $J=1-0$) at 43\,GHz with the Korean VLBI Network (KVN) 21\,m telescopes to extend our understanding on the origins of SiO in star-forming regions.}
   {We analyzed the thermal SiO 1 -- 0 emission and compared the properties of SiO emission with the physical parameters of associated massive dense clumps as well as 22\,GHz \ce{H2O} and Class I 44\,GHz \ce{CH3OH} maser emission.}
   {We detected SiO emission toward 104 regions that consist of 57 IRDCs, 21 HMPOs, and 26 \uchiis. Out of 104 sources, 71 and 80 sources have 22\,GHz \ce{H2O} and 44\,GHz Class I \ce{CH3OH} maser counterparts, respectively. The determined median SiO column density, $N$(SiO), and abundance, $X$(SiO), relative to $N$(\ce{H2}) are $8.12\times10^{12}$\,cm$^{-2}$ and $1.28\times10^{-10}$, respectively. These values are similar to those obtained toward other star-forming regions and also consistent with predicted values from shock models with low-velocity shocks ($\lesssim$10 -- 15\,\kms). For sources with dust temperatures of ($T_{\rm dust}$) $\lesssim20$\,K, we find that $N$(SiO) and $X$(SiO) derived with the $J=1-0$ transition are a factor $\sim3$ larger than those from the previous studies obtained with SiO 2 -- 1. While the $X$(SiO) does not exhibit any strong correlation with the evolutionary stages of their host clumps, $L_{\rm SiO}$ is highly correlated with dust clump mass, and $L_{\rm SiO}/L_{\rm bol}$ also has a strong negative correlation with $T_{\rm dust}$. This shows that colder and younger clumps have high $L_{\rm SiO}/L_{\rm bol}$ suggestive of an evolutionary trend. This trend is not due to excess emission at higher velocities, such as SiO wing features, as the colder sources with high $L_{\rm SiO}/L_{\rm bol}$ ratios lack wing features. Comparing SiO emission with \ce{H2O} and Class I \ce{CH3OH} masers, we find a significant correlation between $L_{\rm SiO}$/$L_{\rm bol}$ and $L_{\rm \ce{CH3OH}}/L_{\rm bol}$ ratios, whereas  no similar correlation is seen for the \ce{H2O} maser emission. This suggests a similar origin for the SiO and Class I \ce{CH3OH} emission in these sources.}
    {We demonstrate that in cold regions SiO $J=1-0$ may be a better tracer of shocks than a higher $J$ transition of SiO. Lower $T_{\rm dust}$ (and so probably less globally evolved) sources appear to have higher $L_{\rm SiO}$ relative to their $L_{\rm bol}$. The SiO 1 -- 0 emission toward infrared dark sources ($T_{\rm dust}\lesssim$ 20\,K), which do not contain identified outflow sources, may be related to other mechanisms producing low-velocity shocks (5 -- 15\,\kms) for example, arising from cloud-cloud collisions, shocks triggered by expanding \hii\ regions, global infall, or converging flows. }


   \maketitle
%

\section{Introduction}
In the quiescent cold interstellar medium (ISM), silicon monoxide (SiO) is depleted in the gas phase to the point of being undetectable as $X$(SiO) $< 10^{-12}$ \citep{Ziurys1989_sio,Martin-Pintado1992_sio} as silicon (Si) is mostly locked into the dust grain cores, although SiO can be present in the icy mantles of dust grains. SiO is only liberated into the gas phase,  or formed from released Si, and detectable at millimeter and submillimeter wavelengths if the mantle or grain core is disrupted \citep{Schilke1997, Gusdorf_SiO_models_2008}.

Most commonly, this disruption takes place in the high-velocity shocks ($\varv_{\rm s} > $ 20 -- 25\,\kms\ and up to 100\,\kms) associated with outflows from young stars, with either the jets from the protostars or the interaction of the molecular outflow with the surrounding ambient gas (e.g., \citealt{Gusdorf_SiO_models_2008,Guillet2011,Kelly2017}). In addition to outflow regions with strong radiation regions such as photodissociation regions (PDRs), photo-desorption can also lead to an enhancement of SiO in the gas phase \citep{Walmsley1999_pdr, Schilke_PDR_2001}. Recently, SiO emission of a different nature has been detected in several regions, including massive star-forming regions (e.g., \citealt{Jimenez-Serra2010_extend_sio,Nguyen-Luong_W43_2013,Louvet_2016,Cosentino_SiO_IDRC_2020}) and even diffuse molecular clouds at high latitudes (e.g., \citealt{Rybarczyk2023_sio_diffuseISM}). This emission found toward active star-forming regions is spatially extended and does not always coincide with outflow sources or CO outflows (e.g., \citealt{Widmann2016}), which are concentrated on smaller scales ($<$\,0.05\,pc). 

Spatially extended SiO emission with a typical fractional SiO abundance, $X$(SiO), of $10^{-11} - 10^{-10}$ is not only weaker than expected from energetic outflows associated with massive stars, which produce bright and broad lines with $X$(SiO) from $10^{-9}$ to $\sim 10^{-7}$ (e.g., \citealt{Jimenez-Serra2010_extend_sio,Lopez-Sepulcre_2016}), but often also has a narrower line width range, typically $\sim$ 2 -- 4\,\kms\ and even smaller than 1\,\kms\ in some cases (e.g., \citealt{Jimenez-Serra2010_extend_sio, Cosentino_SiO_IDRC_2020}). Such narrow SiO emission lines imply the existence of low-velocity shocks. According to several observational studies with shock models (e.g., \citealt{Nguyen-Luong_W43_2013, Jimenez-Serra2008_shock_model_low_vel, Duarte-Cabral2014,Louvet_2016, Csengeri2011_a,Csengeri2011_b}), shock velocities of 5 -- 15\,\kms\ can produce abundances of SiO about $10^{-11} - 10^{-10}$. Three possible explanations have been proposed for the origin of narrow and spatially extended SiO emission: (1) an association with the outflows from an undetected dispersed population of low-mass protostars \citep{Jimenez-Serra2010_extend_sio}; (2) produced in low velocity, large-scale shocks associated with the formation of the molecular cloud themselves, perhaps as a result of colliding or converging flows (e.g., \citealt{Csengeri2011_a,Csengeri2011_b,Csengeri2016}), cloud-cloud collisions (e.g., \citealt{Cosentino_SiO_IDRC_2020,Armijos-Abendano_2020_SrgB2, Zhu2023_BCGS}), or accretion flows associated with global collapse (e.g., \citealt{Duarte-Cabral2014}); or (3) low-velocity shocks ($\sim$ 20\,\kms) driven by supernova remnants, interacting with adjacent molecular clouds (e.g., \citealt{Vaupre2014_SNRs_shock, Dumas2014_SNR_sio_shocks, Cosentino2019_SNR_shocks_sio,Cosentino2022_SNR_shock_sio}).

\begin{table*}[h!]
\centering
\tiny
\caption{Summary of sources.}
\begin{tabular}{l c c c c c c c c c}
\hline 
\hline
Source name & R.A. & Dec. & $\varv_{\rm sys}$ & SiO & \ce{H2O} & \multicolumn{1}{c}{\ce{CH3OH}} & Selected & ATLASGAL & Classification \\
 & $\alpha$ (J2000) & $\delta$ (J2000) & (\kms) & (1 -- 0) & Maser & \multicolumn{1}{c}{Maser} & Category & Name &  \\
\hline
IRAS18032$-$2032&18:06:14&$-$20:31:47&5.0&Y$^b$&Y&Y&\uchii&009.621$+$00.194&MSF (radio-\hii)$^*$\\
G8.67$-$0.36&18:06:19&$-$21:37:32&38.0&Y$^b$&Y&Y&\uchii&008.671$-$00.356&MSF (radio-\hii)$^*$\\
G10.47$+$0.03&18:08:38&$-$19:51:52&66.2&Y$^b$&Y&Y&\uchii&010.472$+$00.027&MSF (radio-\hii)$^*$\\
G10.30$-$0.15&18:08:56&$-$20:05:54&11.0&Y$^b$&Y&Y&\uchii&010.299$-$00.147&MSF (radio-\hii)$^*$\\
G10.15$-$0.34&18:09:25&$-$20:19:15&5.0&N&N&N&\uchii&--&MSF (radio-\hii)$^*$\\
IRAS18073$-$2046&18:10:17&$-$20:45:43&28.0&N&Y&Y&\uchii&009.879$-$00.751&MSF (radio-\hii)$^*$\\
G10.62$-$0.38&18:10:24&$-$19:56:15&$-$3.0&N&Y&Y&\uchii&--&MSF (radio-\hii)$^{*}$\\
IRAS18085$-$1931&18:11:33&$-$19:30:39&$-$1.1&Y$^e$&N&N&\uchii&011.109$-$00.397&MSF (radio-\hii)$^*$\\
HM18089$-$1732&18:11:51&$-$17:31:28&33.0&Y$^c$&Y&Y&HMPO&012.888$+$00.489&MSF (radio-\hii)\\
HM18090$-$1832&18:12:02&$-$18:31:56&110.0&N&N&N&HMPO&012.024$-$00.031&YSO\\
G12.21$-$0.10&18:12:40&$-$18:24:21&20.0&Y$^{b,c}$&Y&Y&\uchii&012.208$-$00.102&MSF (radio-\hii)$^*$\\
G12.43$-$0.05&18:12:55&$-$18:11:08&40.0&N&N&N&\uchii&012.431$-$00.049&MSF (radio-\hii)$^{*}$\\
HM18102$-$1800&18:13:12&$-$17:59:35&21.0&Y$^b$&N&Y&HMPO&--&unclassified\\
G11.94$-$0.62&18:14:01&$-$18:53:24&38.9&Y$^{b,c}$&Y&Y&\uchii&011.936$-$00.616&MSF (radio-\hii)$^*$\\
HM18144$-$1723&18:17:24&$-$17:22:13&47.0&Y$^{a,c}$&Y&Y&HMPO&013.658$-$00.599&YSO\\
G015.05$+$00.07 MM2&18:17:40&$-$15:48:55&24.7&N&N&N&IRDC&015.053$+$00.089&YSO\\
G015.05$+$00.07 MM1&18:17:50&$-$15:53:38&24.7&Y$^d$&N&N&IRDC&015.006$+$00.009&Quiescent\\
HM18151$-$1208&18:17:57&$-$12:07:22&33.0&Y$^e$&Y&Y&HMPO&--&unclassified\\
HM18159$-$1550&18:18:48&$-$15:48:54&60.0&N&N&N&HMPO&015.184$-$00.159&MSF (\hii)\\
G015.31$-$00.16 MM2&18:18:50&$-$15:43:19&31.1&N&N&N&IRDC&015.271$-$00.122&Protostellar\\
G015.31$-$00.16 MM1&18:18:56&$-$15:45:00&31.1&N&N&N&IRDC&--&unclassified\\
\hline
\end{tabular}
\tablefoot{This table only shows a fraction of the whole table; the full table are available at CDS. $\varv_{sys}$ is systemic velocity, and the velocity information is obtained from several molecular line observations (\ce{NH3}; \citep{Churchwell1990_NH3,Molinari1996_hmpo_cata,Beuther2002, Chira2013_IRDC_NH3}, CS; \citep{Bronfman1996_CS_uchii}). Asterisk (*) in the classification column indicates \uchii\ sources without ATLASGAL classification. The presence of SiO wings in rotational transitions, 1 -- 0 from this survey and 2 -- 1 from the survey presented in \citep{Csengeri2016} are indicated with four different classifications: $^a$ indicates wing features in SiO 1 -- 0 emission, and $^b$ means SiO wing candidates for cases with a SiO FWHM $\geq$\,8\,kms. $^c$ indicates SiO wings in the 2 -- 1 transition, and $^d$ is for the cases that there are no wings in 2 -- 1 transition but a broad Gaussian component has a FWHM line width $\geq$\,8\,kms. $^e$ marks for sources without any sign of wing features in either 1 -- 0 or 2 -- 1 transitions.}
\label{tab:soutb}
\end{table*}

Massive stars ($>$ 8 -- 10\,\Msol) and their clusters of lower mass companions form in massive (up to several $10^3$\Msol), pc-sized clumps within molecular clouds \citep{Urquhart2022,Elia2021}. Understanding how stars form in such environments requires tracing the gas flow from the clouds through the clumps and the progenitors of individual stars, $\sim0.1$\,pc-sized cores, and ultimately on to the protostars. Observational evidence increasingly shows that the formation of most massive clumps is driven by the global collapse of regions several parsec in size (e.g., \citealt{Csengeri2017,Jackson2019,Pillai2023_infall}). \citet{Peretto2013} and \cite{Wyrowski2016_infall} report an infall rate range between (0.3$-$$16)\times10^{-3}$\,\Msol/yr for massive clumps that are in a wide range of evolutionary stages. Blue asymmetric line profiles of \ce{HCO+} ($J=1-0$), which are diagnostic of infall motions, have also been observed on pc-scales in massive but less evolved regions \citep{Fuller2005_infall,Traficante2018, Pillai2023_infall}. However, the exact nature of these gas infall motions, such as how they are initiated and their role in assembling massive clumps, is still unclear.

In this paper, we use SiO ($J=1 - 0$) observations toward 366 massive star-forming regions to explore the connection between extended shocks and star formation activity. In Sect.\,\ref{sec:obs}, we describe the source selection criteria, observational setup, and data reduction. We complement these observations with \ce{H2O} and Class I \ce{CH3OH} masers archival data as well as properties derived from dust surveys (i.e., molecular hydrogen column density and dust temperature, \citealt{Urquhart2018,Urquhart2022,Marsh2017}); these are introduced in Sect.\,\ref{sec:obs}. In Sect.\,\ref{sec:result_analysis}, we present the SiO (1 -- 0) detection statistics, line profiles, and determined column density and abundances. In Sect.\,\ref{sec:discussion}, we compare the SiO column densities and abundances with the archival datasets (dust properties and maser associations) and discuss the origin of SiO (1 -- 0) emission. We summarize our results and highlight our main findings in Sect.\,\ref{sec:summary_conclusion}. 

\section{Observation}\label{sec:obs}

\subsection{Description of target sources}\label{sec:source}
\begin{table}[h!]
    \centering
    \caption{Beam size and efficiency.}
    \begin{tabular}{lcc}
\hline \hline
        Telescope & Beam size ($''$) & Beam efficiency \\
        \hline
        KVN$-$Yonsei & 61.5 & 0.45 \\
        KVN$-$Ulsan & 61.6 & 0.42 \\
        KVN$-$Tamna & 62.1 & 0.48 \\
         \hline
    \end{tabular}
    \tablefoot{ These values are updated in 2013\footnote{\url{{https://radio.kasi.re.kr/kvn/status_report_2013/aperture_efficiency.html}}}.}
    \label{tab:beam_eff}
\end{table}

\begin{figure*}[h!]
    \centering
    \includegraphics[width=0.45\textwidth]{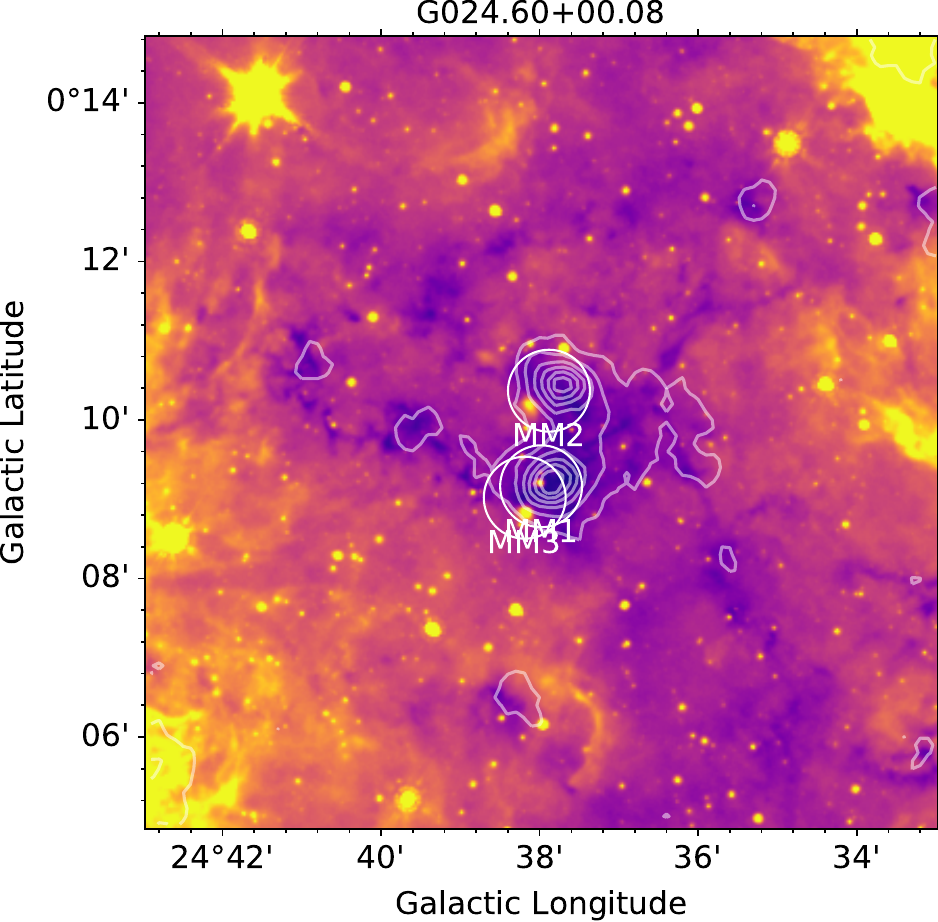}
    \hskip 0.5 cm
    \includegraphics[width=0.45\textwidth]{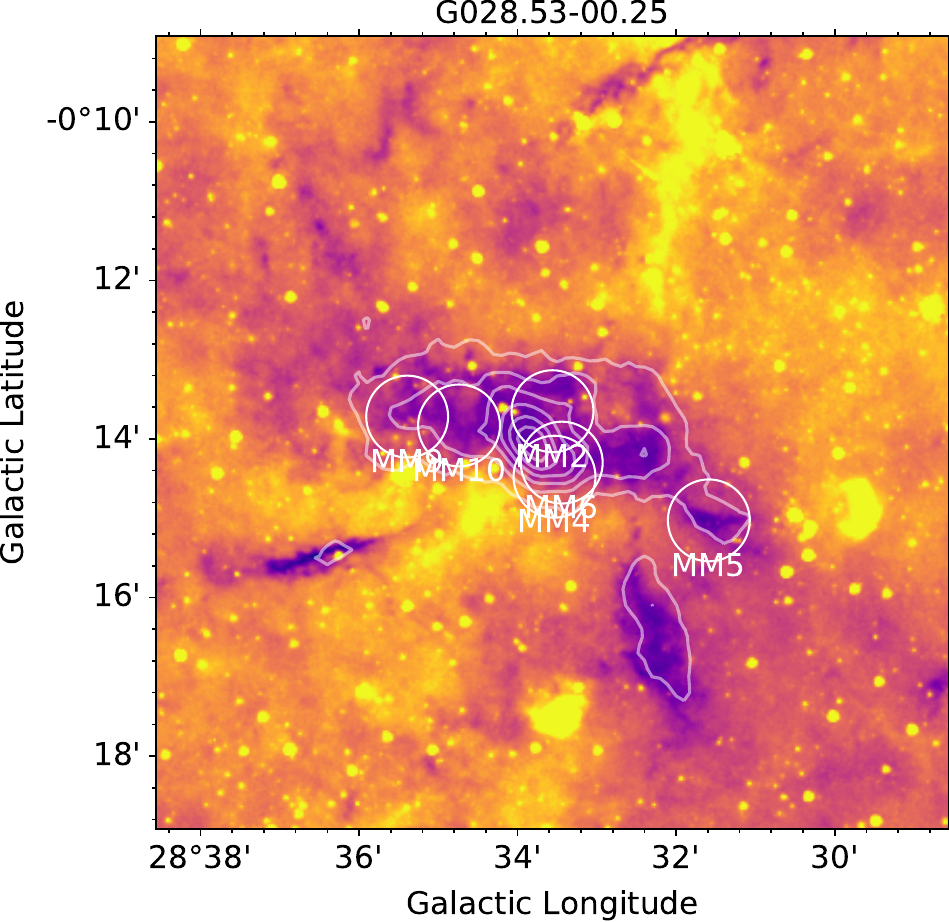}
    \caption{GLIMPSE IRAC 8\,$\mu$m image in color toward G024.50+00.08 and G028.53$-$00.25 from left to right, with the dust emission at 870\,$\mu$m, in gray contours, made from the combined Planck/HFI and APEX/LABOCA maps (\citealt{csengeri2016_planck}) from the ATLASGAL survey (\citealt{schuller2009}). The white circles indicate the beam of KVN observations toward target sources. }
    \label{fig:8micro}
\end{figure*}


\begin{figure*}[h!]
    \centering
    \includegraphics[width=0.33\textwidth]{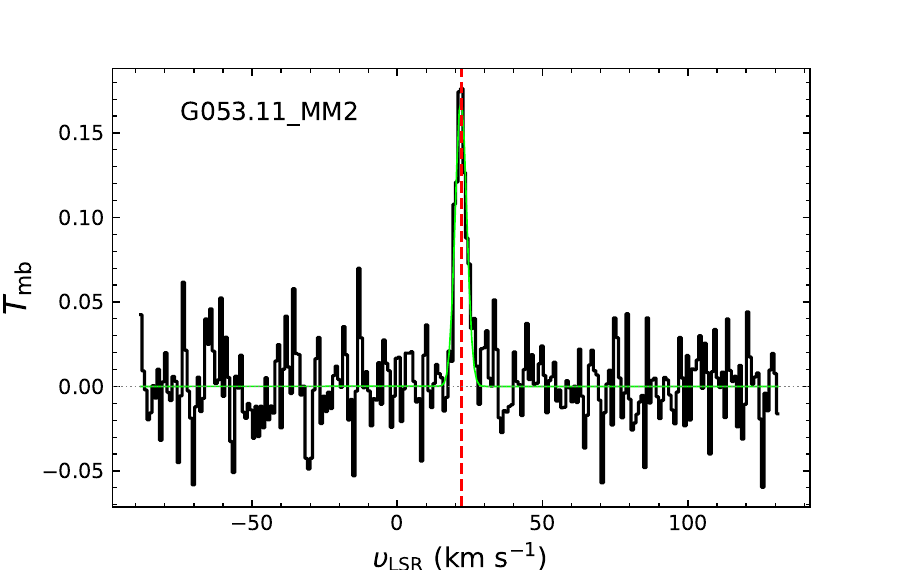}
    \includegraphics[width=0.33\textwidth]{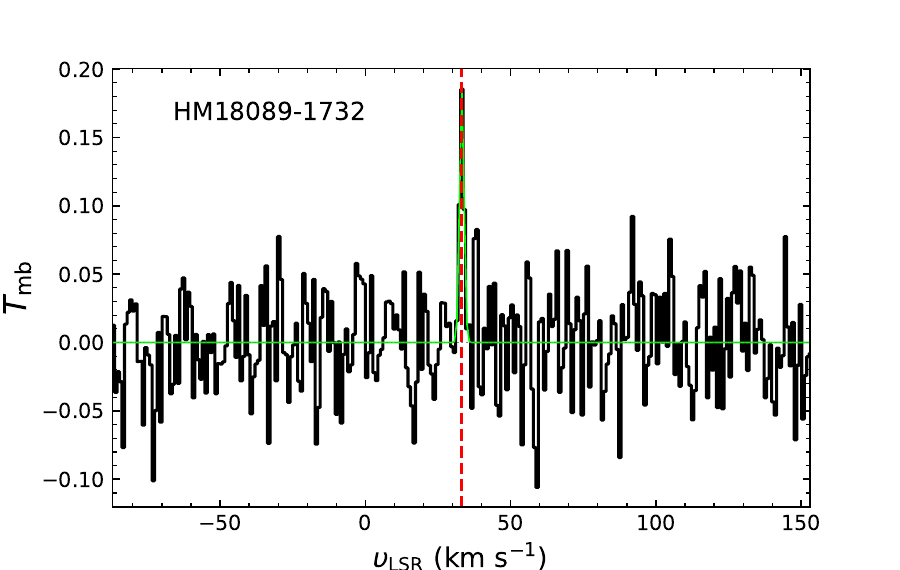}
    \includegraphics[width=0.33\textwidth]{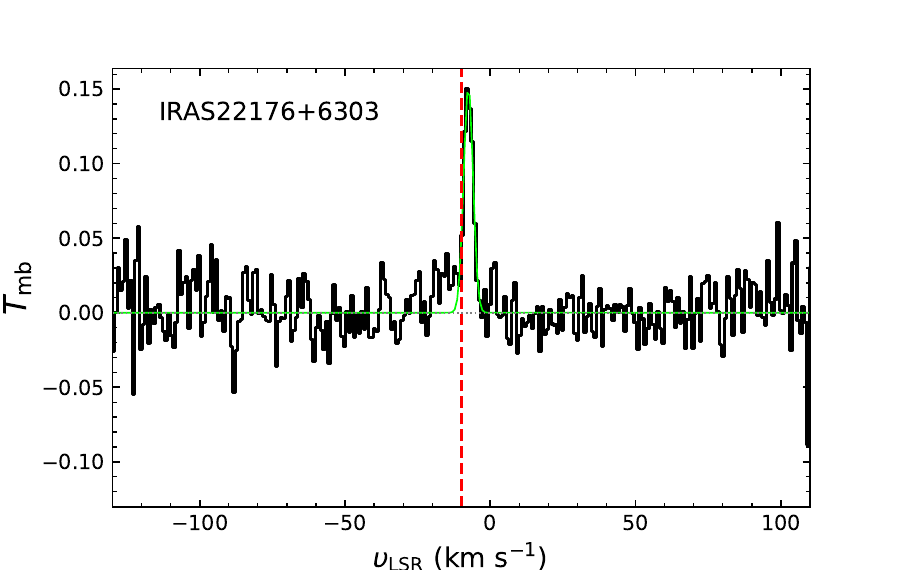}
    \includegraphics[width=0.33\textwidth]{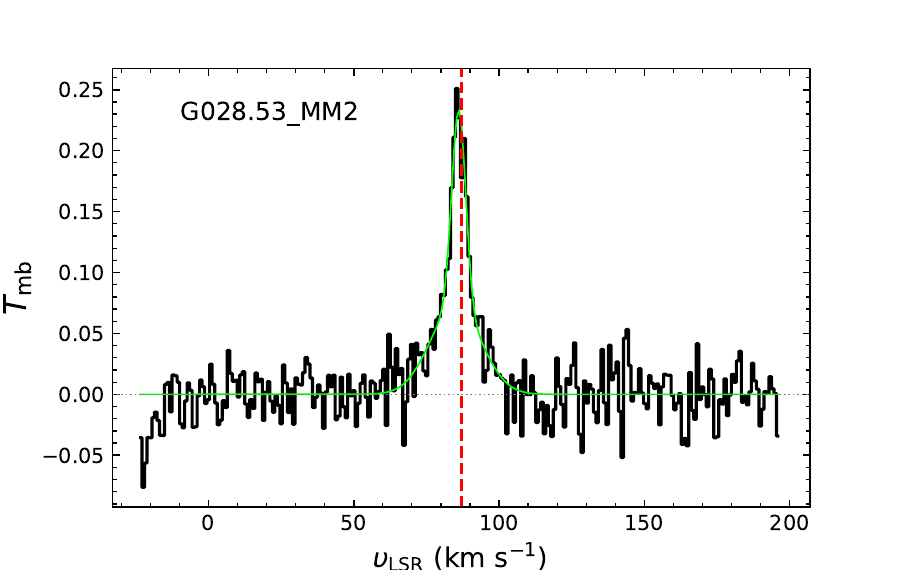}
    \includegraphics[width=0.33\textwidth]{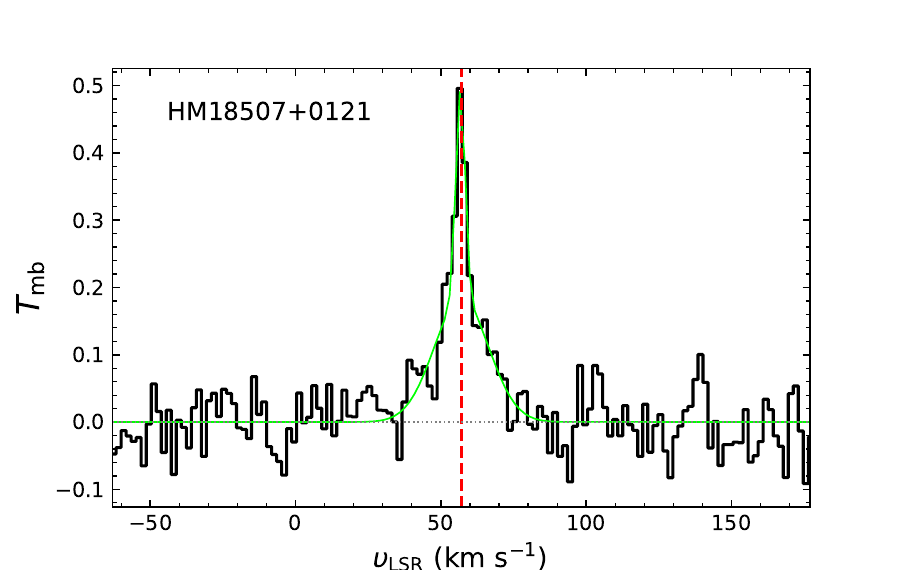}
    \includegraphics[width=0.33\textwidth]{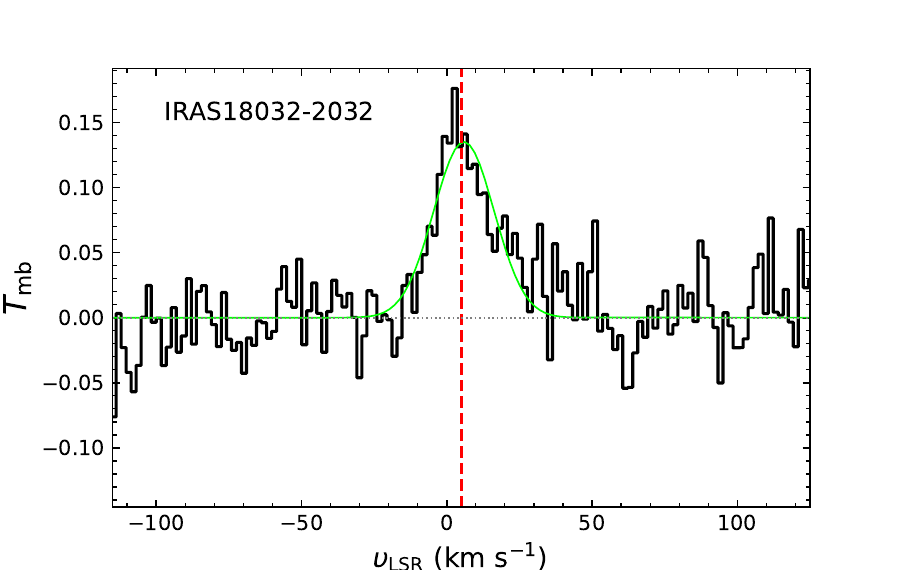}
    \caption{43\,GHz SiO $J=1-0$ spectra in $T_{\rm mb}$ scale toward two IRDCs (G053.11$+$00.05 MM2 and G28.53$-$00.25 MM2) in the left panel, two HMPOs (HM18089$-$1732 and HM18507$+$0121) in the middle panel, two \uchiis\ (IRAS22176$+$6303 and IRAS18032$-$2032) in the right panel.
    The SiO spectral plots for all other detected sources are presented in Figs.\,\ref{appedix:sio_spectra1}, \ref{appendix:sio_spectra2}, \ref{appendix:sio_spectra3},\ref{appendix:sio_spectra4}, \ref{appendix:sio_spectra5}, and \ref{appendix:sio_spectra6}. The vertical green dotted lines indicate a systemic velocity of individual sources. The purple curves are Gaussian profile fits to SiO emission.} 
    \label{fig:sio_spectra}
\end{figure*}

\begin{table*}[h!]
\centering
\caption{SiO line parameters.}
\begin{tabular}{ c l c c c c c c }
\hline \hline
ID & Source name & Fitted  & $\int{T_{\rm mb}} ~ d\varv$ & $\varv_{\rm peak}$ & $\Delta\varv_{\rm FWHM}$ & $T_{\rm mb,\,peak}$ & rms \\
 & & component & (K \kms) & (\kms) & (\kms) & (K) & (mK) \\
\hline
1&G015.05$+$00.07 MM1  &  1  &  0.604$\pm$0.143   &  24.27$\pm$0.66   &  6.45$\pm$1.34   &  0.088  &  15.82\\
&$\cdots$             &  2  &  0.374$\pm$0.157   &  14.50$\pm$1.85   &  9.21$\pm$4.50   &  0.038  &  15.82\\
2&G018.82$-$00.28 MM1  &  1  &  1.732$\pm$0.194   &  41.08$\pm$0.37   &  6.70$\pm$0.85   &  0.243  &  53.67\\
3&G018.82$-$00.28 MM3  &  1  &  0.566$\pm$0.080   &  43.53$\pm$0.56   &  7.16$\pm$1.02   &  0.074  &  16.10\\
4&G018.82$-$00.28 MM4  &  1  &  1.006$\pm$0.154   &  63.72$\pm$1.13   &  15.72$\pm$3.04   &  0.060  &  18.70\\
5&G019.27$+$00.07 MM1  &  1  &  0.962$\pm$0.142   &  26.05$\pm$0.44   &  6.26$\pm$1.15   &  0.144  &  27.29\\
6&G019.27$+$00.07 MM2  &  1  &  1.216$\pm$0.142   &  26.93$\pm$0.32   &  6.16$\pm$1.03   &  0.186  &  35.07\\
7&G022.35$+$00.41 MM1  &  1  &  0.870$\pm$0.053   &  51.48$\pm$0.86   &  4.37$\pm$0.86   &  0.187  &  45.94\\
&$\cdots$             &  2  &  1.260$\pm$0.053   &  52.44$\pm$0.86   &  20.09$\pm$0.86   &  0.059  &  45.94\\
8&G023.60$+$00.00 MM1  &  1  &  0.976$\pm$0.130   &  105.60$\pm$0.57   &  8.63$\pm$1.27   &  0.106  &  22.68\\
9&G023.60$+$00.00 MM2  &  1  &  1.452$\pm$0.218   &  50.79$\pm$0.58   &  7.53$\pm$1.38   &  0.181  &  38.91\\
10&G023.60$+$00.00 MM4  &  1  &  0.373$\pm$0.140   &  52.98$\pm$0.35   &  4.39$\pm$1.10   &  0.080  &  17.58\\
&$\cdots$             &  2  &  0.690$\pm$0.184   &  46.34$\pm$1.93   &  14.04$\pm$3.19   &  0.046  &  17.58\\
11&G023.60$+$00.00 MM7  &  1  &  0.978$\pm$0.138   &  52.09$\pm$0.49   &  7.60$\pm$1.44   &  0.121  &  32.55\\
12&G024.33$+$00.11 MM1  &  1  &  0.864$\pm$0.057   &  112.60$\pm$0.29   &  4.54$\pm$0.42   &  0.179  &  25.55\\
&$\cdots$             &  2  &  1.221$\pm$0.142   &  116.20$\pm$0.51   &  9.68$\pm$1.71   &  0.119  &  25.55\\
13&G024.33$+$00.11 MM2  &  1  &  1.026$\pm$0.110   &  116.90$\pm$0.28   &  5.15$\pm$0.66   &  0.187  &  34.18\\
14&G024.33$+$00.11 MM3  &  1  &  0.942$\pm$0.089   &  115.90$\pm$0.12   &  2.89$\pm$0.37   &  0.307  &  30.81\\
&$\cdots$             &  2  &  0.276$\pm$0.082   &  110.80$\pm$0.58   &  3.64$\pm$1.05   &  0.071  &  30.81\\
15&G024.33$+$00.11 MM5  &  1  &  0.472$\pm$0.158   &  116.10$\pm$0.31   &  3.54$\pm$0.75   &  0.125  &  23.45\\
&$\cdots$             &  2  &  2.221$\pm$0.223   &  116.20$\pm$1.04   &  15.78$\pm$2.12   &  0.132  &  23.45\\
16&G024.33$+$00.11 MM6  &  1  &  1.323$\pm$0.089   &  113.50$\pm$0.18   &  6.09$\pm$0.55   &  0.204  &  23.79\\
17&G024.33$+$00.11 MM10 &  1  &  0.487$\pm$0.102   &  116.20$\pm$0.48   &  4.94$\pm$1.36   &  0.093  &  21.63\\
\hline
\end{tabular}
\tablefoot{Extract of the table of Gaussian fit parameters the SiO spectra. The full table is available at CDS. If there are more a single Gaussian component fitted, the component number is indicated in Column 3. }
\label{tab:sio_gaussianfit}
\end{table*}

We selected 366 sources toward massive star-forming regions from five catalogs covering three different categories, there are being: infrared dark clouds (IRDCs) defined as high extinction regions against bright mid-infrared Galactic background \citep{Perault1996, Egan1998, Hennebelle2001, Rathborne2006_irdc_cata,Peretto2009_irdc}, high-mass protostellar objects (HMPOs, \citealt{Molinari1996_hmpo_cata}) considered as a precursor of ultra-compact \hii\ regions (\uchiis), and molecular clump hosting \uchiis\ \citep{Wood1989_uchii_cata, Kurtz1994_uchii_cata} considered to be the most evolved evolutionary stage of the massive star formation. The selected sources consist of 134 infrared dark cloud (IRDC) cores toward 33 IRDCs taken from \cite{Rathborne2006_irdc_cata}, 129 high-mass protostellar objects (HMPOs) taken from the catalogs of \cite{Molinari1996_hmpo_cata} and \cite{Sridharan2002_hmpo_cata}, and 103 \uchiis\ from the radio continuum catalogs of \cite{Wood1989_uchii_cata} and \cite{Kurtz1994_uchii_cata}. 

The 134 IRDC cores are identified as compact dust emission sources at 1.2 mm wavelength observed with the 117 element bolometer array MAMBO II at the Institut de Radioastronomie Millime\'trique (IRAM) 30~m telescope \citep{Rathborne2006_irdc_cata}, and based on their mid-infrared emission, the cores are characterized as associated to different evolutionary phases \citep{Chambers2009_IRDC}. The \uchii\ sources were initially selected with \textit{IRAS} colors (Log($F_{\rm 60}/F_{\rm 12}) \geq 1.30$ and Log($F_{\rm 25}/F_{\rm 12}) \geq 0.57$); \citealt{Wood1989b}) and later the 103 sources were confirmed as \uchiis\ with radio continuum emission observations at 2/6\,cm and 7\,mm wavelengths \citep{Wood1989_uchii_cata, Kurtz1994_uchii_cata, Walsh1998, carral1999} (see \citealt{Kim2019_masers} for details). The 129 HMPOs are selected from two different catalogs that are \cite{Molinari1996_hmpo_cata} and \cite{Sridharan2002_hmpo_cata}. The sources from \cite{Molinari1996_hmpo_cata} are divided into two groups, indicated as \textit{High} and \textit{Low} based on the $IRAS$ colors. The catalog from \cite{Molinari1996_hmpo_cata} has the same \textit{IRAS} color criteria used for the \uchii\ catalogs of \cite{Wood1989_uchii_cata} and \cite{Kurtz1994_uchii_cata}, and due to the same initial selection criteria, HMPO candidates referred to as \textit{High} can potentially be \uchiis. The other HMPO catalog from \cite{Sridharan2002_hmpo_cata} used the \textit{IRAS} criteria defined by \cite{Ramesh1997}, and besides, all the selected sources are satisfied by the following criteria: detection of CS (2 -- 1) line \citep{Bronfman1996_CS_uchii}, $F_{\rm 60} > 90$\,Jy and $F_{\rm 100} > 500$\,Jy, and the absences of radio continuum emission detected from the Galaxy-Wide 5\,GHz Green bank \citep{Gregory1991} and Parkes-MIT-NRAO ratio continuum surveys \citep{Griffith1994, Wright1994}. Except for the \uchii\ source group, the source classifications for the two categories (i.e., IRDC and HMPO) are not definitive. Thus instead of using the three catalogs, we use the ratio of bolometric luminosity and clump mass as well as dust temperatures for indicators of evolutionary stages. However, as supplementary information, we list the classification defined in The APEX Telescope Large Area Survey of the Galaxy (ATLASGAL) survey \citep{Urquhart2022}. For the \uchii\ sample, if there are no classifications from the ATLASGAL catalog, we refer to their classification as MSF (radio-\hii) in the classification column because we already know that all the sources have radio continuum sources indicating them as \hii\ regions \citep{Kim2019_masers}. Table\,\ref{tab:soutb} lists information on the observed sources.

\subsection{SiO (1 -- 0) observations and data reduction}
\label{sec:sio_obs}

Observations of the rotational ground state transition of SiO ($v=0, J=1-0$, 43.42376\,GHz) toward 366 sources were carried out over four observation periods (2010 October - December, 2011 February, 2012 October-November, and 2013 May) using the Korean VLBI Network (KVN) 21\,m telescopes\footnote{\url{https://radio.kasi.re.kr/kvn/main_kvn.php}} at the Yonsei, Ulsan and Tamna sites \citep{Kim2011,Lee2011}. The pointing and focus observations were performed every 2$-$3 hours by observing strong \ce{H2O} (6$_{1,6} - $5$_{2,3}$, 22.23508 GHz) and SiO ($v=1$, $J=1 - 0$, 43.12203 GHz) maser sources (i.e., V1111~Oph, R~Cas, W3(OH), and Orion~KL) and found to be within 5\arcsec. Position switching mode was used with a total ON$+$OFF integration time of $\sim$30 minutes per source, yielding a typical rms noise level up to $\sim$ 50\,mK at velocity resolutions ($\delta\varv$) of 0.86 and 1.73\,\kms\ depending on the noise of a spectrum. The main beam efficiencies for converting $T_{a}^{*}$ to $T_{\rm mb}$ and the full-width half maxima (FWHM) of the beams are listed in Table\,\ref{tab:beam_eff}. We used the Continuum and Line Analysis Single-dish Software (CLASS) software\footnote{\url{https://www.iram.fr/IRAMFR/GILDAS/doc/html/class-html/class.html}} of the Grenoble Image and Line Data Analysis Software (GILDAS) package \citep{pety2005_gildas} for the data reduction and fitting a Gaussian profile.

\begin{figure*}[h!]
    \centering
    \includegraphics[width=0.43\textwidth]{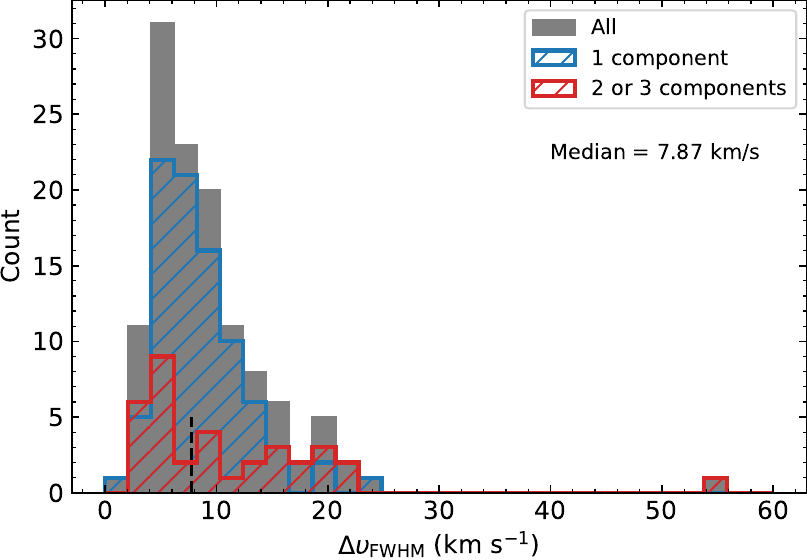}
    \includegraphics[width=0.43\textwidth]{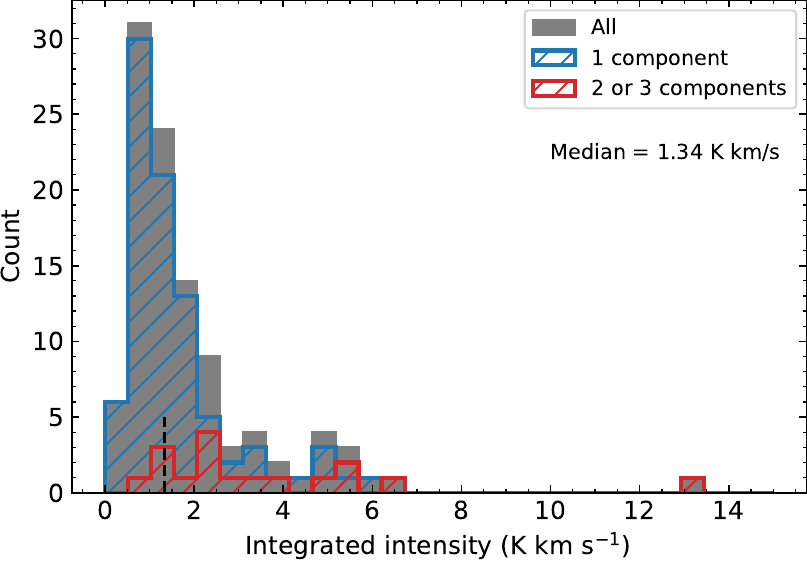}
    \caption{Histogram of Gaussian fit line widths (left panel) and full area (right panel) of SiO (1 -- 0) lines. Gray histograms indicate all the SiO detected sample. Blue and red histograms represent SiO sources fitted with one Gaussian component and with more than one Gaussian component, respectively. Black dashed-lines mark the median values of the gray histograms.}
    \label{fig:linewidth}
\end{figure*}

In addition, to investigate the association with outflows of embedded objects and surrounding environment toward our target positions, we use published observational data of 22\,GHz \ce{H2O} ($6_{1,6}-5_{2,3}$) and 44\,GHz Class I \ce{CH3OH} ($7_0 - 6_1$ A$^+$ ) masers. These maser observations were done toward the same coordinates of our SiO sources with the same three KVN 21~m telescopes. The \ce{H2O} and \ce{CH3OH} maser data toward \uchii\ sources are taken from \cite{Kim2019_masers}, but both maser line parameters for the HMPO and IRDC samples are obtained from \cite{Kim2016PhDT}. In Table\,\ref{tab:soutb}, we provide detections of both masers toward the observed sources.

\section{Results and analysis}\label{sec:result_analysis}

\begin{figure}
    \centering
    \includegraphics[width=0.43\textwidth]{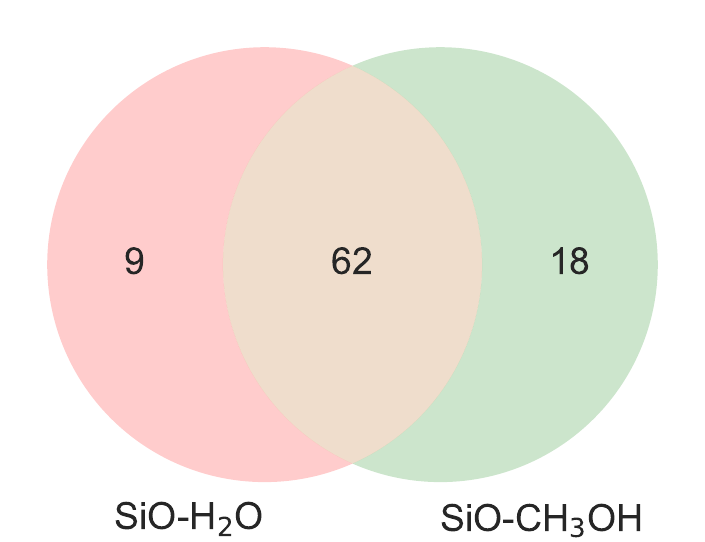}
    \caption{Venn diagram for \ce{H2O} and \ce{CH3OH} maser detections toward SiO detected sources. }
    \label{fig:venn_diagram}
\end{figure}
\begin{figure}[h!]
    \centering
    \includegraphics[width=0.48\textwidth]{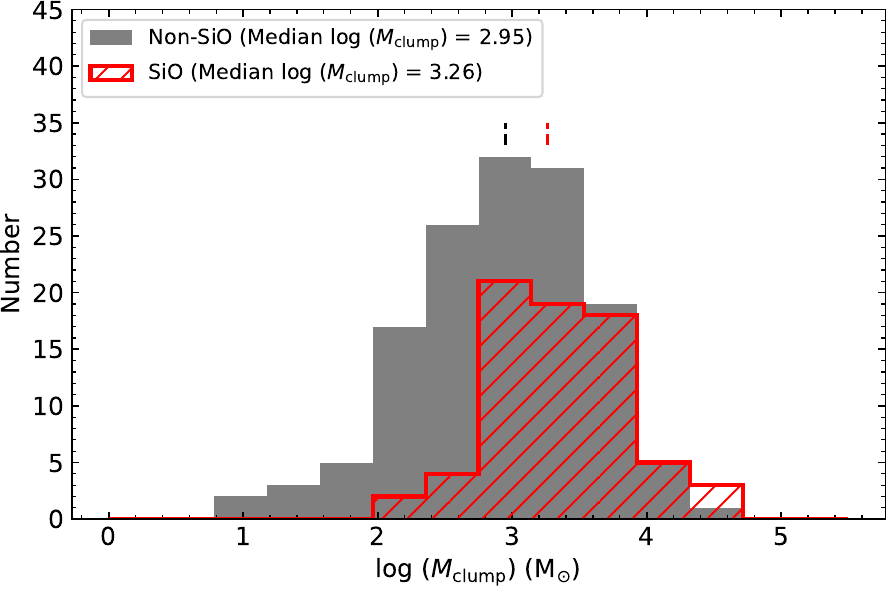}
    \includegraphics[width=0.48\textwidth]{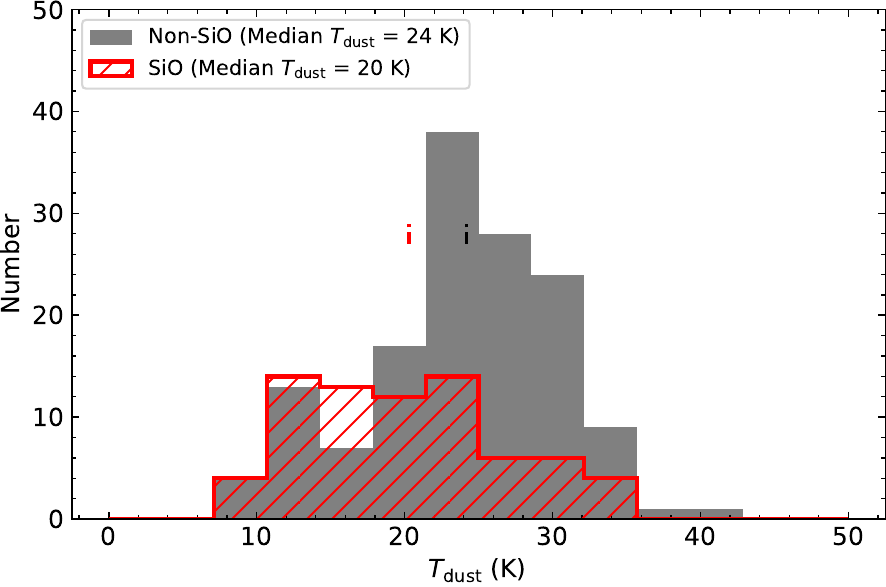}
    \includegraphics[width=0.48\textwidth]{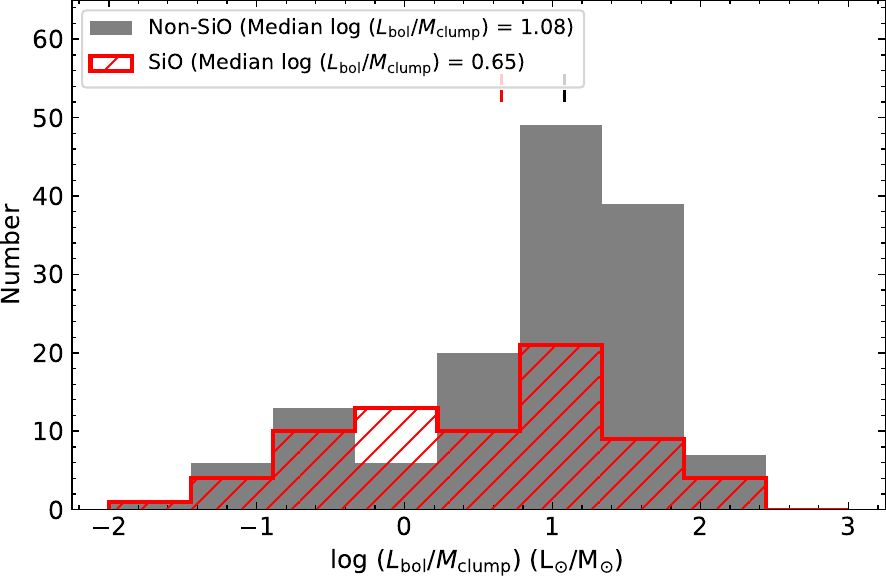}
    \caption{Histograms of dust clump mass (top panel), dust temperature (middle panel), and bolometric luminosity over dust clump mass (bottom panel) for sources with SiO detection (red) and non-detection (gray). The black and red vertical dashed lines correspond to the median values of the respective physical properties for SiO detected and non-SiO detected sources, respectively.}
    \label{fig:hist_nonsio}
\end{figure}

\subsection{Detection and SiO line properties}
We detected SiO (1 -- 0) emission lines toward 106 massive star-forming regions. Among the detected sources, 59 sources are associated with the IRDCs, and the positions of two pairs of four cores (i.e., G024.60$+$00.08 MM1/MM3 and G028.53$-$00.25 MM4/MM6) as shown in Fig.\,\ref{fig:8micro} are very close to each other within $\frac{1}{2}$ a beam width. They match a common ATLASGAL clump, respectively. The SiO detections from these sources are likely duplicates; thus, we only include the brightest spectrum for these duplicated positions toward in the following statistical analysis. The other 21 and 26 detections are found toward HMPO and \uchii\ associated sources, respectively. Figures\,\ref{fig:sio_spectra}, \ref{appedix:sio_spectra1}, \ref{appendix:sio_spectra2}, \ref{appendix:sio_spectra3}, and \ref{appendix:sio_spectra4} show SiO emission spectral lines detected from this survey. Table\,\ref{tab:sio_gaussianfit} lists Gaussian fit parameters of SiO emission lines, which are velocity-integrated intensities ($\int{T_{\rm mb}}~d\varv$) in K\,\kms, peak velocity ($\varv_{\rm peak}$) in \kms, full width at half maximum ($\Delta\varv_{\rm FWHM}$) in \kms, peak intensity ($T_{\rm mb,\,peak}$) in K, and rms level in K. For the following analyses, we exclude the beam-overlapped sources, G024.60$+$00.08 MM3 and G028.53$-$00.25 MM4, and 104 out of 106 sources is the total number of SiO emission sources confirmed as final detections. 

SiO emission lines for 86 sources are fitted with a single Gaussian profile, and in 17 cases two Gaussian components are required to fit, based on visual inspection of the residual emission after subtracting a first Gaussian component that was fitted. Only one case (G5.48$-$0.24) is fitted with three Gaussian components, and the FWHM line width of its broadest components is 54\,\kms\ (see Fig.\,\ref{appendix:sio_spectra6}). Figure\,\ref{fig:linewidth} shows the histogram of $\Delta\varv_{\rm FWHM}$ for all the SiO spectral lines, and for multicomponents, we plot all the fitted components, including a narrow component where the peak velocity is consistent with the ambient gas velocity and a broad component tracing outflows. The median line width marked as a black vertical line is 7.87\,\kms\, and a significant fraction of the Gaussian components have FWHM narrower. The components with $\Delta\varv_{\rm FWHM}$ narrower than 7.87\,\kms\ exhibit an average $\Delta\varv_{\rm FWHM}$ of 5.38\,\kms. The narrowest observed $\Delta\varv_{\rm FWHM}$ being 1.8\,\kms\ toward HM18089$-$1732, where the spectrum has a velocity resolution of 0.86\,\kms. We note that in some cases, weak SiO emission lines are smoothed to a velocity resolution of 1.73\,\kms, which can cause broadened line widths. 

The median value (7.87\,\kms) of $\Delta\varv_{\rm FWHM}$ toward our data are close to 8\,\kms. If the $\Delta\varv_{\rm FWHM}$ is narrower than 8\,\kms, we, thus, consider the SiO emission lines as low-velocity components. The same $\Delta\varv_{\rm FWHM}$ is used in \cite{Csengeri2016}. The average $\Delta\varv_{\rm FWHM}$ of the narrow components from one and multiple Gaussian fitting are 5.71 and 4.63\,\kms, respectively, and for broad components, we obtained 11.46 and 17.39\,\kms, respectively. These average values are in good agreement with the $\Delta\varv_{\rm FWHM}$ ($\sim$5 -- 6\,kms\ and 14 -- 19\,\kms) obtained from SiO 2 -- 1 emission measured toward massive dust clumps \citep{Csengeri2016}. As depicted in the upper panels of Fig.\,\ref{fig:sio_spectra}, certain sources exhibit narrower SiO profiles compared to others (lower panels), characterized by broader wings fitted with a broad Gaussian component. The histogram of the total velocity integrated intensity shows that multiple Gaussian component SiO lines have a bright total velocity integrated intensity than those with a single Gaussian component, except for some sources that are associated with a single bright components.

The Venn diagram in Figure\,\ref{fig:venn_diagram} illustrates the distribution of \ce{H2O} and Class I \ce{CH3OH} masers toward 104 SiO detected sources. We find 22\,GHz \ce{H2O} maser emission toward 71 sources (in red and orange colored areas), and 80 sources have 44\,GHz Class I \ce{CH3OH} masers (in green and orange colored areas).
Among these maser-detected sources, 62 sources (in the orange colored area) have both maser emission lines. Out of 71 \ce{SiO}-\ce{H2O} detected sources, 9 positions do not have \ce{CH3OH} maser emission lines, and 18 sources of the 80 \ce{SiO}-\ce{CH3OH} detected sources only have \ce{CH3OH} masers, not \ce{H2O} masers. Toward HMPOs and \uchiis\ associated sources, most SiO detected sources have either \ce{H2O} or \ce{CH3OH} masers, and only 2 and 5 sources toward HMPOs and \uchiis\, respectively have SiO detection without maser emission. On the other hand, toward IRDC sources, 26 and 18 sources with SiO detections have a lack of \ce{H2O} and \ce{CH3OH} maser emission lines, respectively. 

\subsection{Physical properties of SiO detected and non-detected sources}

We investigated the physical properties (bolometric luminosity, clump mass, dust temperature, and $L_{\rm bol}$/$M_{\rm clump}$) derived from \cite{Urquhart2018} toward the SiO detected and non-detected sources. Figure\,\ref{fig:hist_nonsio} displays histograms of these properties of sources showing the SiO detections (red bars) and non-SiO detections (gray bars). To qualify the statistical significance of these differences, we performed Kolmogorov-Smirnov (KS) tests to $M_{\rm clump}$ and $T_{\rm dust}$ between the two groups. The $p-$values from the KS-tests are remarkably small, $\ll 0.001$, indicating a significant distinction between the groups. The distribution of SiO emission sources in $L_{\rm bol}/M_{\rm clump}$ spread widely, while the non-SiO detected sources trend to be concentrated to associated clumps with higher $L_{\rm bol}/M_{\rm clump}$ indicating more evolved clumps. The KS-test resulted in a reasonably small $p-$value, 0.007, implying that the SiO emission clumps are likely different with the clumps without SiO emission. In addition, we do not find any apparent differences in source distances and $L_{\rm bol}$ toward sources of the two groups, with large $p-$values $>$ 0.09 and 0.8, respectively. In summary, the SiO detected sources are notably associated with massive (median Log $M_{\rm clump}=$ 3.26) and colder (a median $T_{\rm dust}=$ 20\,K) sources compared to the sources without SiO emission. In addition, there is an indication that the SiO detected sources have a lower median $L_{\rm bol}/M_{\rm dust}$ by a factor $\sim40$\% lower than the non-detected sources, suggestive of being somewhat less evolved.

\begin{table*}[h!]
\centering
\tiny
\caption{Derived column density and abundances, and physical parameters of associated dust clumps.}
\begin{tabular}{l c c c c c c c c c c}
\hline \hline
Source   &   $N_{\rm SiO}$ &  Log($N_{\rm \ce{H2}}$)$^a$ &  Log($N_{\rm \ce{H2}}$)$^b$ &  $X_{\rm SiO}$ & Dist. & $T_{\rm dust}$ & Log($L_{\rm bol}$) & Log($M_{\rm clump}$) & $L_{\rm \ce{H2O}}$ & $L_{\rm \ce{CH3OH}}$ \\
 & (cm$^{-2}$) & (cm$^{-2}$) &  (cm$^{-2}$)  & & (kpc) & (K) & (L$_{\odot}$) & (M$_{\odot}$) & (L$_{\odot}$) & (L$_{\odot}$) \\
\hline
G018.82$-$00.28MM1&1.06$\times 10^{13}$&23.055&22.936&9.34$\times 10^{-11}$&12.6&18.7&4.806&4.369&1.34$\times 10^{-3}$&9.71$\times 10^{-5}$\\
G019.27$+$00.07MM2&7.44$\times 10^{12}$&22.714&22.560&1.44$\times 10^{-10}$&2.8&12.1&2.079&2.713&2.31$\times 10^{-6}$&$\cdots$\\
G022.35$+$00.41MM1&1.30$\times 10^{13}$&22.953&22.465&1.45$\times 10^{-10}$&3.6&13.1&2.420&3.057&6.70$\times 10^{-6}$&4.23$\times 10^{-6}$\\
G023.60$+$00.00MM1&5.97$\times 10^{12}$&22.930&22.771&7.02$\times 10^{-11}$&5.9&16.9&3.782&3.519&1.21$\times 10^{-6}$&2.42$\times 10^{-5}$\\
G023.60$+$00.00MM2&8.89$\times 10^{12}$&22.806&22.673&1.39$\times 10^{-10}$&3.6&16.7&3.362&3.064&$\cdots$&1.24$\times 10^{-5}$\\
G31.41$+$0.31&5.57$\times 10^{12}$&23.696&$\cdots$&1.12$\times 10^{-11}$&4.9&22.0&4.800&3.852&1.52$\times 10^{-4}$&1.30$\times 10^{-5}$\\
G33.92$+$0.11&4.49$\times 10^{12}$&23.048&22.857&4.02$\times 10^{-11}$&6.9&27.4&5.226&3.726&7.73$\times 10^{-7}$&5.26$\times 10^{-6}$\\
G34.26$+$0.15&2.87$\times 10^{13}$&23.917&23.253&3.48$\times 10^{-11}$&2.9&29.2&5.330&3.776&3.34$\times 10^{-4}$&2.58$\times 10^{-5}$\\
G45.47$+$0.05&1.06$\times 10^{13}$&22.915&22.763&1.29$\times 10^{-10}$&7.7&29.0&5.557&3.779&2.61$\times 10^{-5}$&$\cdots$\\
HM18507$+$0121&3.40$\times 10^{13}$&23.249&22.679&1.92$\times 10^{-10}$&2.9&20.5&4.042&3.260&1.37$\times 10^{-4}$&2.25$\times 10^{-5}$\\
HM18517$+$0437&3.50$\times 10^{12}$&$\cdots$&$\cdots$&$\cdots$&1.9&$\cdots$&$\cdots$&$\cdots$&1.50$\times 10^{-6}$&7.47$\times 10^{-7}$\\
HM18530$+$0215&7.74$\times 10^{12}$&22.855&22.624&1.08$\times 10^{-10}$&4.7&26.1&4.674&3.283&$\cdots$&6.20$\times 10^{-6}$\\
HM18566$+$0408&8.28$\times 10^{12}$&22.908&22.606&1.02$\times 10^{-10}$&4.9&22.5&4.365&3.293&9.69$\times 10^{-6}$&4.64$\times 10^{-6}$\\
HM19217$+$1651&5.22$\times 10^{12}$&22.659&22.373&1.15$\times 10^{-10}$&10.9&29.1&5.035&3.457&2.37$\times 10^{-4}$&4.04$\times 10^{-5}$\\
\hline
\end{tabular}
\tablefoot{Only a fraction of the entire table is presented here, and the full table is available at CDS. The properties of dust clumps are taken from the ATLASGAL point source catalog \citep{Urquhart2018, Urquhart2022}. $N$(\ce{H2}) preceded by superscript $a$ is from the ATLASGAL catalog, and the other one marked by $b$ is measured from the Hi-GAL PPMAP over the observed KVN beam size. The listed $X$(SiO) are determined using $N$(\ce{H2}) from the ATLASGAL catalog. All other clump physical parameters (i.e., $T_{\rm dust}$, Log($L_{\rm bol}$), and Log($M_{\rm clump}$)), and distance also come from the ATLASGAL catalog. $L_{\rm \ce{H2O}}$ and $L_{\rm \ce{CH3OH}}$ are recalculated with the distance from the point source catalog and velocity-integrated intensities of \ce{H2O} and Class I \ce{CH3OH} are taken from \cite{Kim2016PhDT} and \cite{Kim2019_masers}.}
\label{tab:phy_para}
\end{table*}
\subsection{SiO (1 -- 0) column density and abundance}
We assume that the SiO (1$-$0) is optically thin (e.g., \citealt{Csengeri2016}), and so the total column density is given by,
\begin{equation}
\begin{aligned}
\label{eq:optically_thin}
& N_{\rm tot}^{\rm thin} = \left(\frac{8\pi\nu^{3}}{c^{3}A_{ul}} \right)\left(\frac{Q(T)}{g_{u}} \right)\frac{{\rm exp} \left(\frac{E_{u}}{k_{\rm B}T_{\rm ex}} \right)}{{\rm exp}\left( \frac{h\nu}{k_{\rm B}T_{\rm ex}}\right) -1} \\ &
~~~~~~~~~~~~~~~~~~~~~~~~~~~~~~~~~~\times \frac{1}{[J_{\nu}(T_{\rm ex}) - J_{\nu}(T_{\rm bg})]} \int{\frac{T_{\rm MB}}{f}}\,d{\it \varv} \\
&
~~~~~~~~\simeq~6.12 \times 10^{12} \int{T_{\rm MB}}\,d{\it \varv},
\end{aligned}
\end{equation}

\noindent  where $\nu$ is 43423.760\,MHz, the Einstein coefficient for spontaneous emission, $A_{ul}$, is $3.049\times10^{-6}$~s$^{-1}$, and the statistical weight of the upper state level, $g_{u}$ is 3. $Q(T)$ and $k_{\rm B}$ are the partition function and Boltzmann constant, respectively. The energy of the upper level, $E_{\rm u}$, is 2.084\,K. All these values are taken from the Cologne Database for Molecular Spectroscopy (CDMS, \citealt{Muller2005,Endres2016}). The partition function at 10\,K is calculated by interpolating data from the CDMS. $\int{T_{\rm MB}}~d\varv$ is the velocity integrated main beam temperature of an observed source, and $f$ is the beam filling factor, which is the fraction of the beam filled by the source. Here we consider the calculated column densities are beam-averaged values, and the medium is spatially homogeneous and larger than the size of the beam. Thus, the beam filling factor is adopted to be 1. We note that toward some sources the beam filling factor could be smaller than 1 if a SiO 1--0 emitting region is localized on scales smaller than the beam. In these cases, the beam filling factors are smaller than unity and the derived column density is a lower limit. $J_{\nu}(T)$ is the Rayleigh-Jeans temperature, $J_{\nu}(T) \equiv \frac{h\nu/k_{\rm B}}{{\rm exp}(h\nu/k_{\rm B}T) - 1}$. $T_{\rm bg}$ is the background emission temperature assumed to be 2.7\,K that is the cosmic microwave background radiation. Some sources do not have available $T_{\rm dust}$ measurements, and the SiO emitting regions are not always necessarily associated with embedded sources heating dust grains. It is also possible that the SiO and dust grains are not in thermal equilibrium. We assumed a fixed excitation temperature, $T_{\rm ex}$, of 10\,K for all the sources. Given that our data are subsequently compared with other SiO surveys that have also utilized the same fixed excitation temperature for calculating the SiO column density, this assumption remains valid within this specific context. However, to consider the potential variation in $N$(SiO) caused by an unknown $T_{\rm ex}$, we estimated uncertainties in $N$(SiO) using a range of temperatures. We applied 5\,K as the lower limit, while considering $T_{\rm dust}$ as the upper limits if $T_{\rm dust}$ is available. Thus, $N$(SiO) derived in this study only represent lower limits of the measurements. Instead of using integrated intensity over the full width at zero power (FWZP), we adopt velocity-integrated intensity from Gaussian fit to avoid overestimating SiO column density due to noise. We derived SiO (1$-$0) column density, $N$(SiO), in a range 2.18$\times$10$^{12}$ -- 7.92$\times$10$^{13}$\,cm$^{-2}$ toward the all detected positions. The median $N$(SiO) is 8.12$\times$10$^{12}$\,cm$^{-2}$. The $N$(SiO) derived in this study are comparable with the SiO column densities ($1.6\times10^{12}-7.9\times10^{13}$\,cm$^{-2}$ from SiO 2 -- 1 and $0.31-4.32\times10^{12}$\,cm$^{-2}$ from SiO 1 -- 0) determined from SiO 1 -- 0 and 2 -- 1 surveys toward other massive star-forming regions \citep{Csengeri2016,Zhu2020}.

\begin{figure}[h!]
    \centering
    \includegraphics[width=0.48\textwidth]{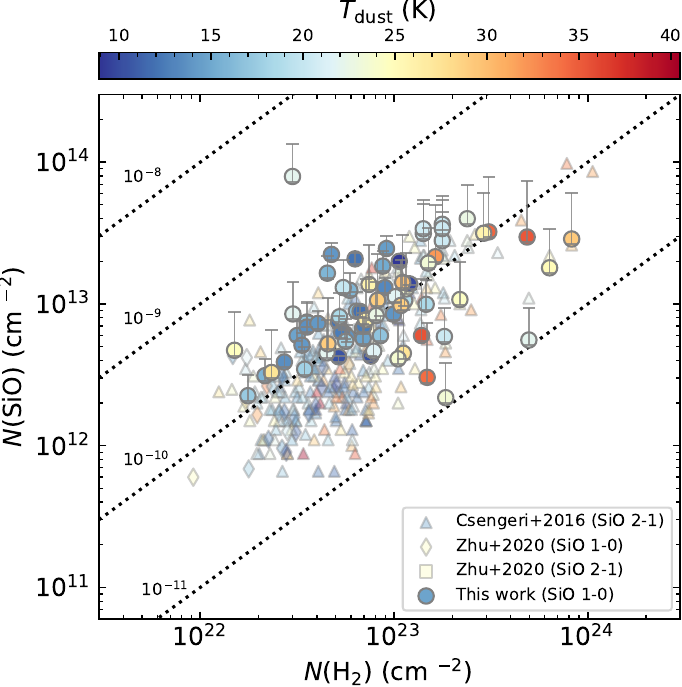}
    \vskip 0.2cm
    \includegraphics[width=0.49\textwidth]{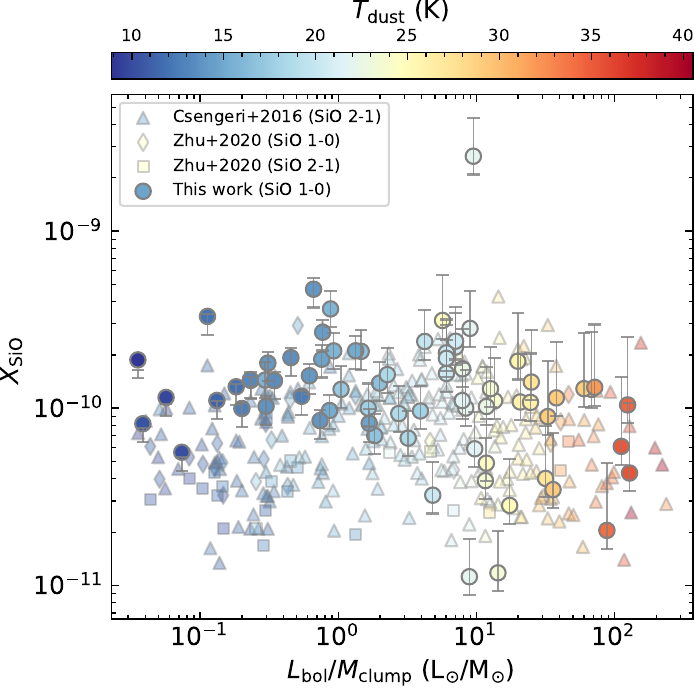}
    \caption{SiO column density and abundance versus \ce{H2} column density and $L_{\rm bol}/M_{\rm clump}$, respectively. \textit{Upper}: Molecular hydrogen column density versus SiO column density. Black diagonal dot-lines indicate SiO abundance on the y-axis relative to $N$(\ce{H2}), with its values labeled. \textit{Lower}: Comparison between $L_{\rm bol}/M_{\rm clump}$ and SiO abundance. Uncertainties depending on $T_{\rm ex}$ are indicated with error bars for both the plots.}
    \label{fig:Nsio_Nh2_Xsio}
\end{figure}

To determine the SiO abundance, $X$(SiO) relative to molecular hydrogen, we used \ce{H2} column densities from the ATLASGAL catalog \citep{Urquhart2018,Urquhart2022} but also obtained a beam-averaged $N$(\ce{H2}) from \ce{H2} column density maps produced by the point process mapping (PPMAP) to Hi-GAL data \citep{Marsh2017} using the Python package, \texttt{Photutils}. $N$(SiO), $N$(\ce{H2}), and $X$(SiO), that are derived here, and physical properties of associated dust clumps are listed in Table\,\ref{tab:phy_para}. With $N$(\ce{H2}) obtained from PPMAP, the median and mean SiO abundances are 1.95$\times10^{-10}$ and 3.00$\times10^{-10}$. The SiO abundance can change significantly depending on $N$(\ce{H2}), and thus to compare with SiO data from other surveys, we also use \ce{H2} column densities obtained from \cite{Urquhart2018,Urquhart2022} for the other catalogs, which are subsequently compared with SiO column densities of our sources. The median and mean SiO abundances determined using the \ce{H2} column densities from \cite{Urquhart2018,Urquhart2022} are 1.28$\times10^{-10}$ and 1.71$\times10^{-10}$, respectively. These abundances are close to the estimated abundances with the beam-average $N$(\ce{H2}) from PPMAP. In the following sections, we use the physical parameters of associated clumps as well as $N$(\ce{H2}) taken from the ATLASGAL catalog only for comparisons with previous surveys (i.e., \citealt{Csengeri2016, Zhu2020}). In addition, for discussion, we use the PPMAP $N$(\ce{H2}) column densities for sources that do not have ATLASGAL counterparts within the single-dish beam size. For uncertainties in $X$(SiO), we used upper and lower limits of $N$(SiO) using a range of excitation temperatures (from 5\,K to $T_{\rm dust}$) but used a fixed $N$(\ce{H2}). Thus, we note that these $X$(SiO) values and their uncertainties are probably lower limits since SiO in the gas phase likely only exists in part of the relative \ce{H2} column density. Moreover, additional uncertainties arising from Gaussian fits, calculated with $\sqrt{\sigma_{1}^2 + \sigma_2^2}$ where $\sigma_1$ and $\sigma_2$ represent uncertainties of $\int{T_{\rm mb}} ~d\varv$ for the narrow and broad Gaussian components respectively, are found to be insignificant.

\section{Discussion}\label{sec:discussion}

\subsection{Comparisons with the properties of dust clumps}

To understand the origin of SiO (1 -- 0) emission, we compare the physical parameters of dust clumps associated with the detected SiO emission. The selected categories of observed sources do not provide the evolutionary stages of the embedded sources or surroundings. For instance, even some IRDCs, considered as the earliest stage, host evolved embedded objects, such as young stellar objects or hyper/ultra-compact \hii\ regions. For this reason, we utilize dust temperature and $L_{\rm bol}/M_{\rm clump}$ as an indicator of the evolutionary stages of star formation \citep{Urquhart2018,Urquhart2022}. Thus, in the following analyses, we exclusively compare sources with ATLASGAL counterparts to use the physical parameters measured from the dust continuum. In the upper panel of Fig.\,\ref{fig:Nsio_Nh2_Xsio} shows the comparison between $N$(SiO) and $N$(\ce{H2}) for 73 sources observed in this survey (marked by circles), 261 dust clump sources hosting different evolutionary stages of massive stars (triangles) \citep{Csengeri2016}, and 24 starless clump candidates (diamonds and squares) \citep{Zhu2020} in the 1.1\,mm continuum Bolocam Galactic Plane Survey \citep{Svoboda2016_scc_catalog}. As mentioned in the previous section, the uncertainties of $N$(SiO) indicated with error bars in Fig.\,\ref{fig:Nsio_Nh2_Xsio} are determined by applying a $T_{\rm ex}$ range of 5\,K as the lower limit and $T_{\rm dust}$ as the upper limit for cases with available $T_{\rm dust}$ from the ATLASGAL catalog \citep{Urquhart2018,Urquhart2022}. $N$(SiO) trends to increase as $N$(\ce{H2}) increases, while these column densities do not show any association with the evolutionary stages as indicated by the temperatures.

The plot in the lower panel of Fig.\,\ref{fig:Nsio_Nh2_Xsio} shows the comparison of $X$(SiO) and the ratio of bolometric luminosity and clump mass ($L_{\rm bol}/M_{\rm clump}$) with dust temperature in color, for sources mentioned above. The abundance uncertainties, represented by the error bars, are derived from the upper and lower $N$(SiO) limits, while maintaining a constant $N$(\ce{H2}). Therefore, as previously indicated, these $X$(SiO) values serve as lower limits. The SiO abundances do not have any correlation with evolutionary stages. The lack of correlation between $X$(SiO) and $L_{\rm bol}/M_{\rm clump}$ is also found toward the data points from the other SiO surveys with no dependency of the observed transitions. All the surveys used here adopt the excitation temperature of 10\,K for SiO 1 -- 0 and 2 -- 1 transitions. To compare with our observations, we use the dust properties from \cite{Urquhart2018,Urquhart2022} to avoid any biased results due to applying different assumptions. \cite{Csengeri2016} and \cite{Zhu2020} estimate $N$(\ce{H2}) with different assumptions. \cite{Csengeri2016} shows a moderate correlation of $X$(SiO) with $L_{\rm bol}/M_{\rm dust}$ in their work, but that is likely due to assuming 18\,K for the dust temperature to calculate the $N$(\ce{H2}) used in their work. As a result, the $N$(\ce{H2}) estimated in \cite{Csengeri2016} are likely underestimated for infrared dark clumps, and thus SiO abundances toward the cold sources are overestimated. 

\cite{Zhu2020} also show higher abundances in their analysis toward starless clump candidates, and their $N$(\ce{H2}) column densities determined with \ce{NH3} kinetic gas temperature are smaller than the $N$(\ce{H2}) from the ATLASGAL catalog. This causes higher abundances of SiO toward the starless clump candidates. If kinetic gas temperature and dust temperatures are out of thermal equilibrium, using the gas temperature also leads to less accurate clump properties. \cite{Zhu2020} also shows higher abundances in their analysis toward starless clump candidates, and their $N$(\ce{H2}) column densities determined with \ce{NH3} kinetic gas temperature are smaller than the $N$(\ce{H2}) from the ATLASGAL catalog. This causes higher abundances of SiO toward the starless clump candidates. If the kinetic gas temperature and dust temperatures are out of thermal equilibrium, using the gas temperature also leads to less accurate clump properties. Therefore, for all sources of these three surveys (\citealt{Csengeri2016}, \citealt{Zhu2020} and this work), with ATLASGAL counterparts, we used $N$(\ce{H2}) taken from \cite{Urquhart2018,Urquhart2022} because the $N$(\ce{H2}) were estimated with dust temperatures determined by fitting spectral energy distributions of mid- and far-infrared continuum data with individual cold graybody and warm blackbody functions (see \citealt{Koenig2017} for details). The lower panel of Fig.\,\ref{fig:Nsio_Nh2_Xsio} shows that when using a uniform $N$(\ce{H2}) data the gradient in abundance disappears.

We also notice that toward sources with $T_{\rm dust}$ $<$ 20\,K, $X$(SiO) determined in our study ($J=1-0$, circle markers) and by \cite{Zhu2020} ($J=1-0$,  diamond markers) tend to be higher than \cite{Csengeri2011_b} ($J=2-1$, triangle markers) and \cite{Zhu2020} ($J=2-1$, square markers). The origin of the difference between the $J=1-0$ and $J=2-1$ is unclear. The fixed excitation temperature to 10\,K for SiO toward all the sources can affect the distribution of $X$(SiO). Until the lines attain thermalization for a given density, the $J=2-1$ transition exhibits a lower $T_{\rm ex}$ compared to the $J=1-0$ transition, though this is unlikely to be the sole explanation. Decreasing $T_{\rm ex}$ to 5\,K results in a 7\% reduction in the inferred SiO column density from the 2 -- 1 line, while further reducing $T_{\rm ex}$ to 3\,K would be required to raise the column density by 30\%. The observed discrepancy between the two transitions may be influenced by the distinct beam sizes employed for the 1 -- 0 ($\sim 61''$) and 2 -- 1 ($\sim 31''$) transitions. Consequently, the higher column densities of SiO derived from the 1 -- 0 transition could imply that the SiO emission exhibits spatial extension beyond the region probed by the 2 -- 1 transition. 

By using the same data set for the clump properties for the three SiO surveys compared here, we find that $X$(SiO) does not show any apparent correlation with $L_{\rm bol}/M_{\rm clump}$ which indicate the evolutionary stage of clumps. This is consistent with the findings from previous observations (e.g., \citealt{Liu_atoms_aca_sio_2022}). With potential uncertainty for the SiO excitation temperature, \cite{Csengeri2016} also conclude that there are no clear trends in SiO abundances compared with the evolutionary stages.
\begin{figure}
    \centering
    \includegraphics[width=0.48\textwidth]{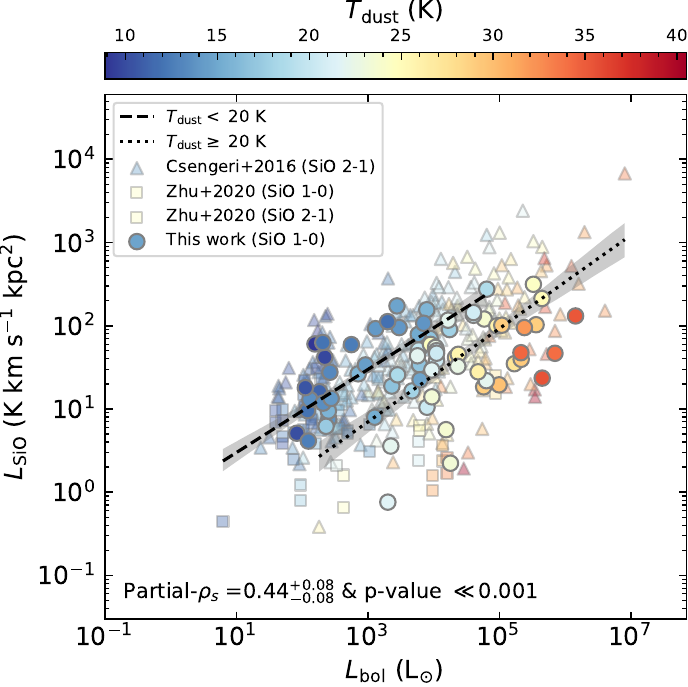}
    \vskip 0.2cm
    \includegraphics[width=0.48\textwidth]{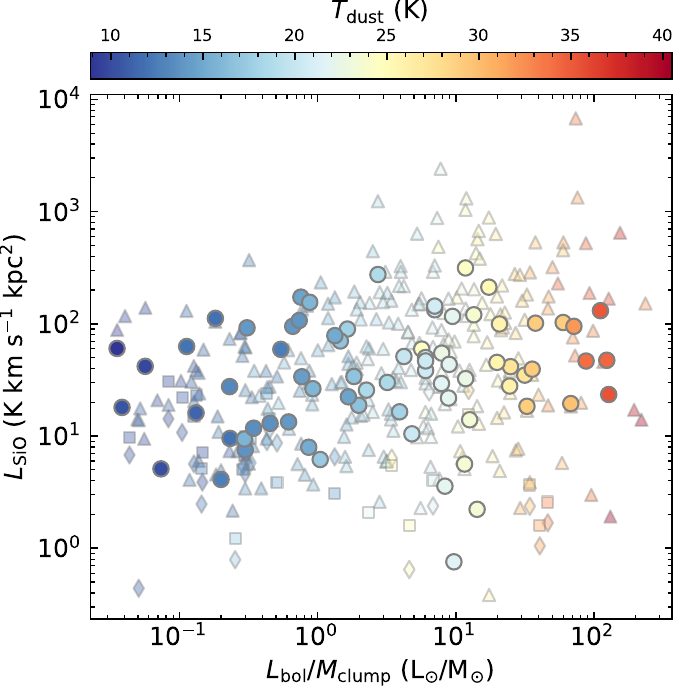}
    \caption{\textit{Upper}: Bolometric luminosity ($L_{\rm bol}$) versus SiO luminosity ($L_{\rm SiO}$). The dashed and dotted lines are the best fits for sources with temperatures $<$ 20\,K and $>$ 20\,K, respectively. The gray shaded regions represent 2$\sigma$ confidence intervals for the best fits. \textit{Lower}: The bolometric luminosity over clump mass versus SiO luminosity. The markers are as same as Fig.\ref{fig:Nsio_Nh2_Xsio}. The color bars for both the plots present dust temperature.}
    \label{fig:lbol_lsio}
\end{figure}
\begin{figure}[h!]
    \centering
    \includegraphics[width=0.48\textwidth]{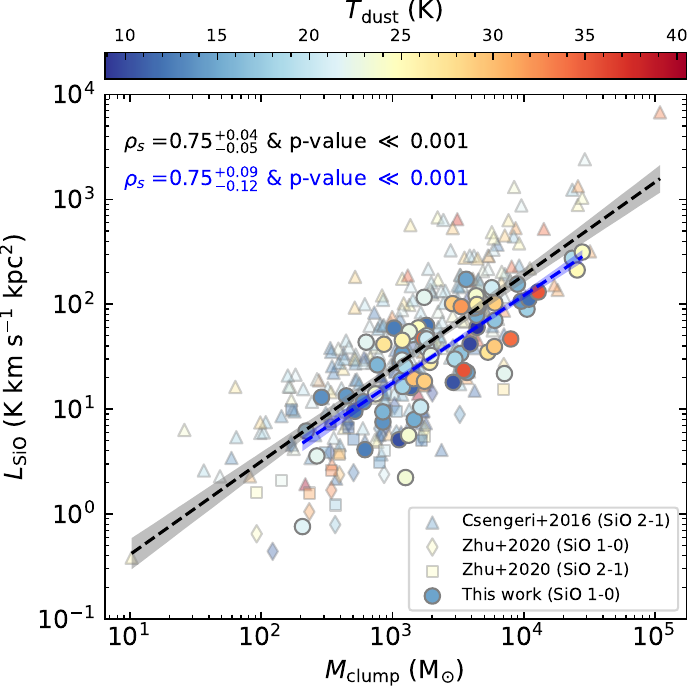}
    \vskip 0.2cm
    \includegraphics[width=0.48\textwidth]{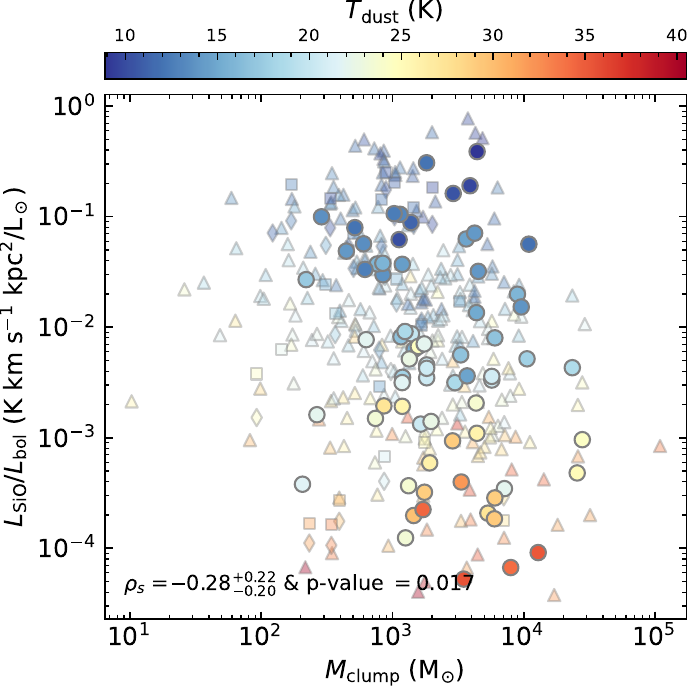}
    \caption{\textit{Upper}: Comparison of SiO luminosity and clump mass. \textit{Lower}: $L_{\rm SiO}/L_{\rm bol}$ versus $L_{\rm bol}/M_{\rm clump}$. The markers are as same as Fig.\,\ref{fig:Nsio_Nh2_Xsio}. The color bars for both the panels present dust temperature. The dashed black and blue lines in the upper panel are the best fits to data points of three surveys and only our survey, respectively. The gray and blue shaded regions present 2$\sigma$ confidence intervals for the best fits matched with the same colors. }
    \label{fig:LsioL_M}
\end{figure}
We also find that distributions of all three surveys (i.e., this survey, \citealt{Csengeri2016}, and \citealt{Zhu2020}) show significant variations in SiO abundance across the samples of objects. The median $X$(SiO) toward all the samples is about $10^{-10}$ that is consistent with SiO abundances measured toward narrow and extended SiO emission regions toward IRDCs and low-velocity ($\varv_{\rm s} \sim$ 10 -- 15\,\kms) shocked areas (e.g., \citealt{Duarte-Cabral2014,Jimenez-Serra2010_extend_sio,Cosentino_SiO_IDRC_2020}) but smaller than SiO abundances $\geq 10^{-8} - 10^{-9}$, determined with SiO emission from outflow sources and high-velocity regimes of SiO emission in low- and high-mass star-forming regions (e.g., \citealt{Jimenez-Serra2005_L1448-mm,Jimenez-Serra2010_extend_sio,Sanchez-Monge2013,Duarte-Cabral2014,Lopez-Sepulcre_2016}).

\begin{figure}[h!]
    \centering
    \includegraphics[width=0.48\textwidth]{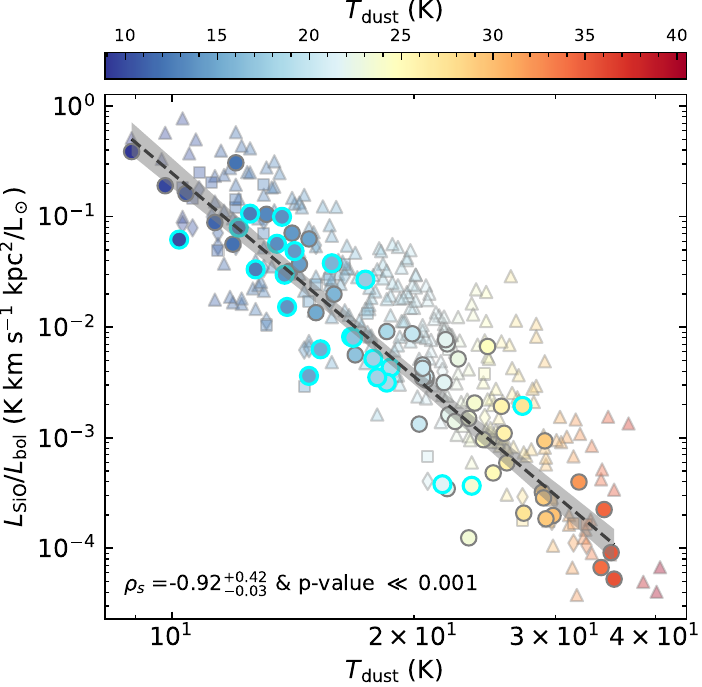}
    \caption{Comparison of $L_{\rm SiO}/L_{\rm bol}$ and $T_{\rm dust}$ toward sources with available $T_{\rm dust}$ and $L_{\rm bol}$. The color bar shows dust temperature. Black dashed line is the best fit to data points from this study, and the gray shaded region presents the 2$\sigma$ confidence intervals for the best fit. Sources without SiO wing features are indicated by open cyan circles. The symbols are same as in Figs.\ref{fig:lbol_lsio} and \ref{fig:LsioL_M}}.
    \label{fig:LsioL_Tdust}
\end{figure}

\begin{figure}[h!]
    \centering
    \includegraphics[width=0.48\textwidth]{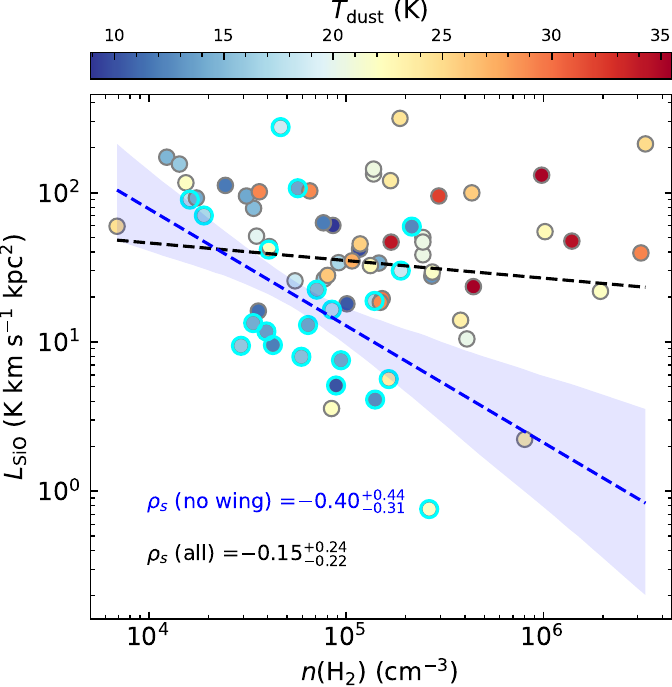}
    \includegraphics[width=0.48\textwidth]{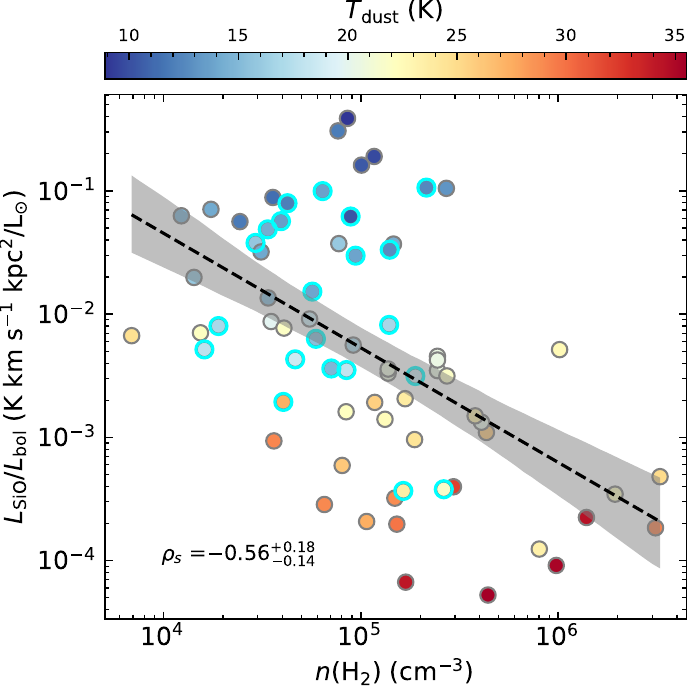}
    \caption{\textit{Upper}: Comparison of $L_{\rm SiO}$ and $n$(\ce{H2}) toward sources having $n$(\ce{H2}) parameter available. \textit{Lower}: Comparison of $L_{\rm SiO}/L_{\rm bol}$ and $n$(\ce{H2}) toward sources having $n$(\ce{H2}) and $L_{\rm bol}$ parameters. The color bars for both the panels present dust temperature. The black and blue dashed lines in the upper panel are the best fits to data points with SiO wings or without. The blue shaded region presents 2$\sigma$ confidence intervals for the best fit for non-SiO wing sources indicated by open cyan circles. The markers are same as for Figs.\ref{fig:lbol_lsio} and \ref{fig:LsioL_M}.}
    \label{fig:LsioL_volden}
\end{figure}

\begin{figure}[h!]
    \centering
    \includegraphics[width=0.48\textwidth]{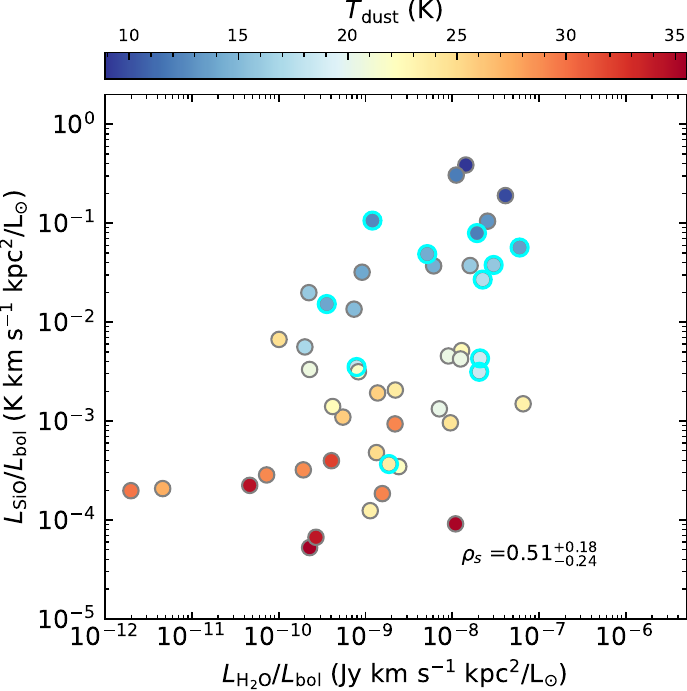}
    \includegraphics[width=0.48\textwidth]{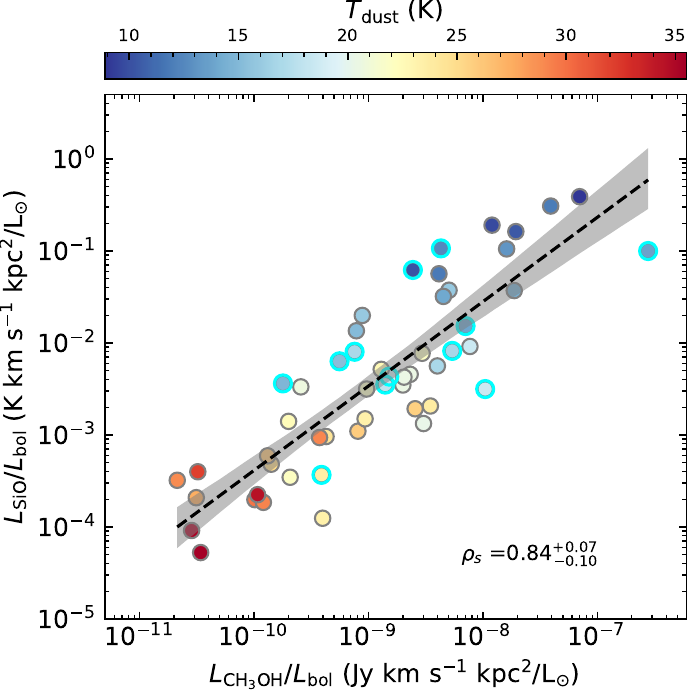}
    \caption{$L_{\rm SiO}/L_{\rm bol}$ versus ratios of 22\,GHz \ce{H2O} (\textit{upper}) maser and Class I \ce{CH3OH} maser luminosities (\textit{lower}) and $L_{\rm bol}$ toward sources with both detections of SiO and masers. The colors indicate dust temperatures. The black dashed lines and gray shade regions are the best-fit and the 95\,\% confidence intervals of fitting. The cyan open circles indicate sources considered to show no significant sign of SiO wings.}
    \label{fig:Lmaser_Lsio}
\end{figure}

The upper panel of Fig.\,\ref{fig:lbol_lsio} shows the comparison between SiO luminosity and bolometric luminosity toward the three SiO surveys. Using a partial-spearman $\rho$-rank correlation, which takes into account the influence of distance on the outcome, to all the data points, the distribution shows a moderate correlation (partial-$\rho_{\rm s}= 0.44^{+0.08}_{-0.08}$) between the two parameters as higher $L_{\rm SiO}$ sources trend to be associated with higher $L_{\rm bol}$. In addition, at a given $L_{\rm bol}$, sources with $T_{\rm dust}$ lower than 20\,K have brighter SiO luminosity compared with other sources with $T_{\rm dust}$ $\geq$ 20\,K. The dashed and dotted lines are the best fits of linear regression to the sources with $T_{\rm dust}$ colder or hotter than 20\,K, respectively. The gray areas indicate the 2$\sigma$ confidence interval of uncertainties for the best fits. These uncertainties are estimated by iteratively performing linear regression fitting randomly resampled data sets of the applied sample. It seems that SiO luminosity for sources of a given bolometric luminosity is greater by a factor of $\sim 7$ for sources with dust temperatures less than 20\,K. However, the excess SiO luminosity is not relevant to the evolutionary stage of the embedded objects or the clumps as estimated by $L_{\rm bol}/M_{\rm clump}$ as we do not find any correlation in the comparison of $L_{\rm SiO}$ and $L_{\rm bol}/M_{\rm clump}$ plotted in the lower panel of Fig.\,\ref{fig:lbol_lsio}. \cite{Liu_atoms_aca_sio_2022} also did not find a trend in the comparison between $L_{\rm SiO}$ and $L_{\rm bol}/M_{\rm clump}$ for SiO (2 -- 1) toward 171 massive star-forming clumps from the ALMA ACA observations, which have higher angular resolutions ($\sim$ 13.5$''$) than this survey. Their colder clumps also have a higher SiO luminosity at a given bolometric luminosity. This might imply that there could be additional formation mechanisms to release SiO in the gas phase from dust grains, for example, via low-velocity shocks \citep{Jimenez-Serra2008_shock_model_low_vel} apart from stronger/fast shocks from outflows or/and jets releasing SiO from dust cores \citep{Schilke1997,Gusdorf_SiO_models_2008}.

In Fig.\,\ref{fig:LsioL_M}, the upper panel shows $L_{\rm SiO}$ versus $M_{\rm clump}$. The SiO luminosities estimated over the beam size show the excellent correlations, with masses of the associated dust clumps for both cases including all the three surveys (black dashed line and gray shade region) or only our survey (blue dashed line and blue shade region) with a significant spearman $\rho$-rank coefficient of 0.75$^{+0.04}_{-0.05}$ with $p-$value $\ll$ 3$\sigma$. In contrast to the plot of $L_{\rm SiO}$ versus $L_{\rm bol}$, there is no gradient in dust temperature observed in this comparison of mass and SiO luminosity. The lower panel of Fig.\,\ref{fig:LsioL_M} shows the comparison of $L_{\rm SiO}/L_{\rm bol}$ versus $M_{\rm clump}$ indicating no obvious correlation as spearman $\rho$-rank coefficient of $-$0.28$^{+0.22}_{-0.20}$ ($p-$value of 0.017) is not significant. As found in the lower panel of Fig.\,\ref{fig:lbol_lsio} showing no trend with evolutionary stages of embedded sources, we also see no correlation between the clump mass and the dust temperature. Figure\,\ref{fig:LsioL_Tdust} shows that $L_{\rm SiO}/L_{\rm bol}$ has a significant correlation ($\rho_{\rm s} = $0.92$^{+0.42}_{-0.03}$ with $p-$value $\ll$0.001) with $T_{\rm dust}$ indicating the evolutionary stages of associated dust clumps. We find that $L_{\rm SiO}/L_{\rm bol}$ is clearly larger toward colder ($T_{\rm dust} <$ 20\,K) clumps at a given clump mass.

To distinguish SiO emission caused by ongoing star-forming activities (i.e., outflows or jets), we divide SiO emission sources into two groups based on four classifications of SiO emission features for SiO 1 -- 0 and 2 -- 1 transitions, from this survey and \cite{Csengeri2016}, respectively. In the SiO detection column of Table\,\ref{tab:soutb}, sources marked with $e$ indicating no sign of wing features in either $J=1-0$ or $J=2-1$ transitions with a narrow $\Delta \varv_{\rm FWHM} < $8\,\kms\ are classified as non-wing SiO sources, in other words, which are narrow line SiO sources. The other cases with SiO wing features or/and broad SiO line with $\Delta \varv_{\rm FWHM} \geq $8\,\kms\ are considered SiO wing sources. The narrow SiO emission marked with cyan circles in Fig.\,\ref{fig:LsioL_Tdust} are tightly associated with cold clumps, which also have large $L_{\rm SiO}/L_{\rm bol}$ ratios. We discuss the physical properties of these narrow SiO sources in detail in the following section (Sect.\,\ref{sec:association_sio}).

Figure\,\ref{fig:LsioL_volden} shows comparisons of the volume density of molecular hydrogen and SiO luminosity (in the upper panel) and $L_{\rm SiO}/L_{\rm bol}$ (in the lower panel). For all the sources in the upper panel plot, we find no correlation ($\rho_{\rm s} = -0.15^{+0.24}_{-0.22}$) between $L_{\rm SiO}$ and $n$(\ce{H2}), and the linear-regression fit is the black dashed line. However, interestingly, the narrow SiO sources marked with open cyan circles are mainly located in the lower envelope of the distribution and show a moderate correlation of $-0.40^{+0.44}_{-0.31}$. The linear regression best fit is the blue dashed line, and its 2$\sigma$ confidence intervals are indicated in the blue shaded area.

In the lower panel plot of $L_{\rm SiO}/L_{\rm bol}$ versus $n$(\ce{H2}), we see a slightly better correlation with $\rho_{\rm s} = -0.56^{+0.18}_{-0.14}$. Although the range of volume density is small, it is clear that lower volume density regions have higher SiO luminosity relative to $L_{\rm bol}$. In general, we expect lower $n$(\ce{H2}) toward early evolutionary stages due to longer free-fall times, while more evolved stages trend to have high $n$(\ce{H2}) and corresponding shorter free-fall times \citep{Urquhart2018,Urquhart2022}. The lower panel of Fig.\,\ref{fig:LsioL_volden} also shows trend as clumps with colder $T_{\rm dust}$ are mainly distributed toward $n$(\ce{H2}) $< 10^5$\,cm$^{-3}$. However, at $n$(\ce{H2}) around $10^5$\,cm$^{-3}$, both colder and hotter clumps exist. Nevertheless, it is significant that narrow SiO emission lines are often detected toward clumps with $T_{\rm dust} <$\,20\,K and $n$(\ce{H2}) $< 10^5$\,cm$^{-3}$, and these clumps tend to have $L_{\rm SiO}/L_{\rm bol}$ $> 10^{-3}$ \kms\ kpc$^2$/L$_{\odot}$. \cite{Duarte-Cabral2014} and \cite{Csengeri2016} suggest that narrow SiO emission sources without any bright infrared sources might trace the earliest phase of molecular clouds undergoing global-infalling processes or converging flows on clouds. Here we find high relative SiO luminosity to $L_{\rm bol}$ toward sources with $T_{\rm dust} < 20$\,K and $n$(\ce{H2}) $< 10^5$\,cm$^{-3}$, with most of them showing narrow SiO emission. This trend might suggest that those narrow SiO sources or high $L_{\rm SiO}/L_{\rm bol}$ sources observed in this survey are experiencing large-scale infall processes. If this suggestion is correct, it can explain why we see higher SiO luminosity toward sources with colder dust temperatures, mainly associated with IRDC sources.

\begin{table*}[h!]
\centering
\tiny
\caption{Physical properties toward sources with the presence of SiO wings.}\label{tab:siowing}
\begin{tabular}{c c c c c c c c c c}
\hline \hline
 &  &\multicolumn{3}{c}{SiO wing source} & &  & \multicolumn{3}{c}{Non-wing SiO source} \\\cline{2-5}\cline{7-10}
 &  \# &  Mean & Median & Standard deviation & & \# & Mean & Median & Standard deviation \\
\hline
$N$(\ce{H2}) [cm$^{-2}$] & 51 & $1.48\times10^{23}$ & $1.02\times10^{23}$ & $1.55\times10^{23}$ & & 22 & $6.08\times10^{22}$ & $5.68\times10^{22}$ & $3.24\times10^{22}$ \\ 
$N$(SiO) [cm$^{-2}$] & 66 & $1.49\times10^{13}$ & $1.07\times10^{13}$ & $1.23\times10^{13}$ & & 38 &$6.22\times10^{12}$ & $5.93\times10^{12}$ & $2.76\times10^{12}$  \\ 
$X$(SiO)  & 51 & $1.96\times10^{-10}$ & $1.29\times10^{-10}$  & $3.59\times10^{-10}$ & & 22 & $1.13\times10^{-10}$ & $1.01\times10^{-10}$ & $4.18\times10^{-11}$ \\ 
$T_{\rm dust}$ [K] & 51 & 21.6 & 21.9 & 6.8 & & 22 & 16.3 & 15.6 & 3.9 \\
Log ($L_{\rm bol}$) [L$_{\odot}$] & 50 &  5.03 &  4.04 & 5.38 & & 22 & 3.87 & 3.21 & 4.14 \\ 
Log ($M_{\rm clump}$) [M$_{\odot}$]  & 50 & 3.63 & 3.38 & 3.72 & & 22 & 3.50 & 3.03 & 3.72 \\ 
$n$(\ce{H2}) [cm$^{-3}$] & 51 & $5.27\times10^5$& $1.37\times10^5$ & $1.30\times10^6$& & 21 &$9.02\times10^4$ & $6.40\times10^4$ & $6.74\times10^4$  \\ 
Log ($L_{\rm bol}/M_{\rm clump}$) [L$_{\odot}$/M$_{\odot}$]  & 50 & 1.32 & 0.90 & 1.51 & & 22 & 0.49 & $-0.02$ & 0.75 \\ 
Log ($L_{\rm SiO}/L_{\rm bol}$) [K \kms\ kpc$^2$/L$_{\odot}$]& 50 & $-1.48$ & $-2.47$ & $-1.12$ & & 22 & $-1.54$ & $-1.93$ & $-1.49$ \\ 
$L_{\rm \ce{H2O}}/L_{\rm bol}$  & 37 & $7.00\times10^{-9}$& $1.34\times10^{-9}$ & $1.29\times10^{-8}$ & & 11 & $1.66\times10^{-8}$ & $1.93\times10^{-8}$ & $1.71\times10^{-8}$ \\ 
$L_{\rm \ce{CH3OH}}/L_{\rm bol}$  & 40 & $5.68\times10^
{-9}$ & $9.47\times10^{-10}$ & $1.27\times10^{-8}$ & & 12 & $2.60\times10^{-8}$ & $1.98\times10^{-9}$& $7.59\times10^{-8}$  \\ 
\hline
\end{tabular}
\end{table*}

\begin{figure}[h!]
    \centering
    \includegraphics[width=0.47\textwidth]{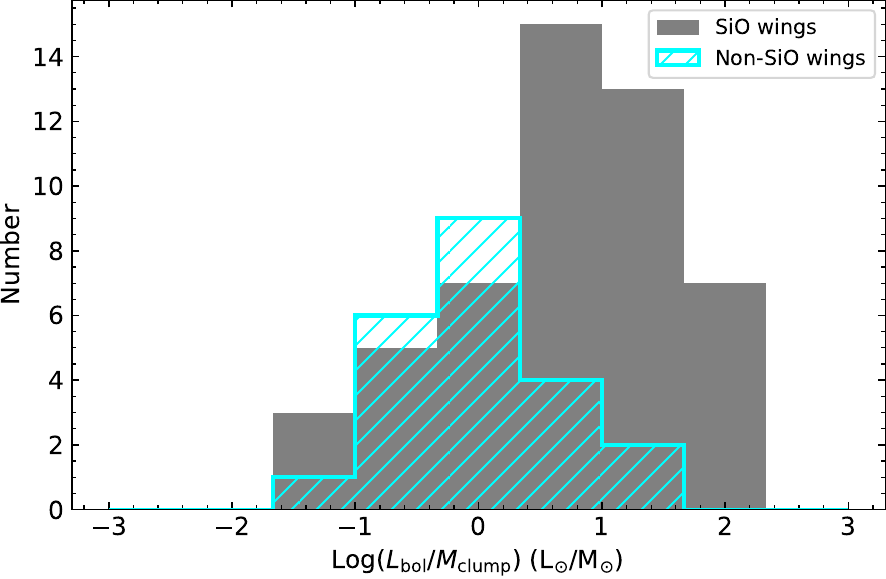}
    \includegraphics[width=0.47\textwidth]{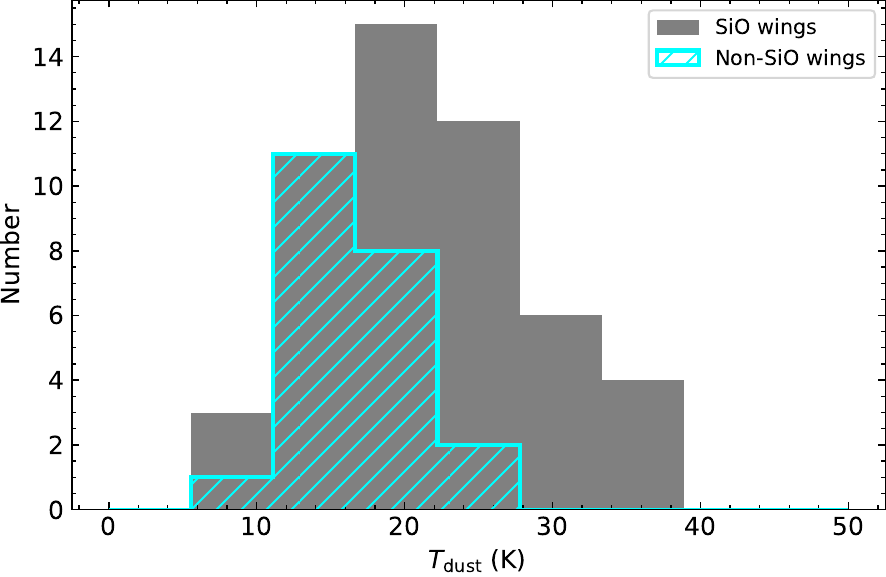}
    \includegraphics[width=0.47\textwidth]{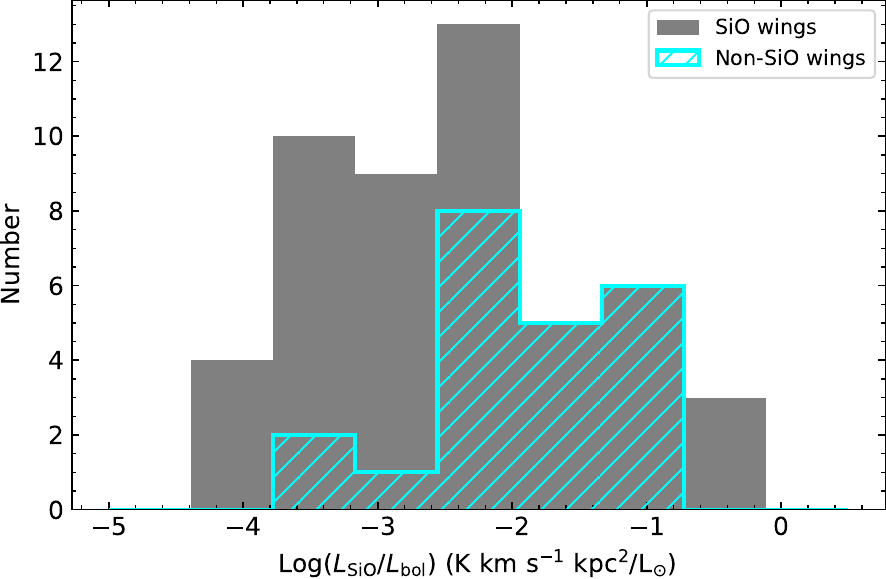}
    \caption{Histograms of Log($L_{\rm bol}/M_{\rm clump}$), $T_{\rm dust}$ Log($L_{\rm SiO}/L_{\rm bol}$) from top to bottom. The cyan histograms represent the distributions of sources without SiO wing features (non-wing sources) while the gray histograms are SiO wing sources.}
    \label{fig:hist_lm_lsiolbol}
\end{figure}

\subsection{Comparisons with 22\,GHz \ce{H2O} and 44\,GHz Class I \ce{CH3OH} maser emission}\label{sec:masers}
All of the positions observed for SiO (1 -- 0) were also observed for 22\,GHz \ce{H2O} and 44\,GHz Class I \ce{CH3OH} masers with the KVN telescopes. We recalculated the luminosities of both the maser species with the same distances used for the SiO luminosities, and these are listed in Table\,\ref{tab:phy_para}. Figure\,\ref{fig:Lmaser_Lsio} shows the comparisons of SiO and masers luminosities normalized by bolometric luminosity as $L_{\rm SiO}/L_{\rm bol}$ tends to vary with evolutionary stages (e.g., shown in the lower panel of Fig.\,\ref{fig:LsioL_volden}). 
\cite{Felli1992} show a strong correlation between $L_{\rm \ce{H2O}}$ and the mechanical luminosity of outflows in CO emission, which is proportional to the bolometric luminosity measured at submillimeter/far-infrared wavelengths. This indicates that \ce{H2O} masers are strongly associated with outflow sources. \ce{H2O} maser luminosities presented in this study, as shown, however, show a moderate correlation with SiO luminosities with the spearman correlation coefficient of 0.51$^{+0.18}_{-0.24}$ and $p$-value $\ll$ 0.001). In addition, \cite{Duarte-Cabral2014} show that $L_{\rm SiO}$ estimated for individual outflow sources correlate well with the respective CO outflow momentum flux ($F_{\rm CO}$). However, when they compare the SiO luminosities obtained over entire fields including outflows and ambient gas regions, in which the observations do not resolve individual outflow sources, with the $F_{\rm CO}$ and find no correlation between SiO emission and CO flux powered by outflows. This is consistent with the lack of a significant correlation that we see between $L_{\rm SiO}$/$L_{\rm bol}$ and $L_{\rm \ce{H2O}}$/$L_{\rm bol}$.

The scatter plot for $L_{\rm SiO}/L_{\rm bol}$ versus $L_{\rm \ce{CH3OH}}/L_{\rm bol}$ in the lower panel of Fig.\,\ref{fig:Lmaser_Lsio} shows a significant correlation. The spearman $\rho$-rank efficient is 0.84$^{+0.07}_{-0.10}$ with a small $p$-value $\ll$ 0.001. Such the strong correlation indicates that SiO and Class I 44\,GHz \ce{CH3OH} maser trace similar properties of shocked gas toward these massive star-forming regions. We also find that SiO luminosity is strongly correlated with clump mass shown in the left panel of Fig.\,\ref{fig:LsioL_M}.

The 22\,GHz \ce{H2O} and Class I 44\,GHz \ce{CH3OH} masers are known to be collisionally pumped by mainly outflows \citep{Felli1992, Kurtz2004}, but for some cases, Class I \ce{CH3OH} masers are excited by expanding ionized gas \citep{Voronkov2010,Voronkov2014,Gomez-Ruiz2016} or cloud-cloud collisions \citep{Haschick1993_CH3OHmaser_cloud_collison,Sato2000_CH3OHmaser_cloud_collison}. The processes collisionally exciting Class I \ce{CH3OH} maser can release SiO as well as possibly Si into the gas phase from icy mantles of dust grains/cores. This probably explains the tight correlation between the Class I \ce{CH3OH} maser and SiO emission shown in Fig.\,\ref{fig:Lmaser_Lsio}.

We also note that potentially the inclusion of thermal or quasi-thermal emission in the 44\,GHz \ce{CH3OH} luminosity estimate might contribute to the scatter shown in the shaded gray area in the lower panel of Fig.\,\ref{fig:Lmaser_Lsio}. For \ce{H2O} maser (the upper panel), it unexpectedly shows much larger scatters with less correlation between the \ce{H2O} maser and SiO emission. In general, outflows could be the primary origin of producing SiO and exciting \ce{H2O} maser transition but as mentioned earlier, SiO can be produced by other processes. This might cause the large scatter between the SiO and \ce{H2O} emission.

\subsection{Association with the presence of SiO wing features}\label{sec:association_sio}

Table\,\ref{tab:siowing} lists mean, median, and standard deviation values of physical parameters toward sources with SiO wings and without them. $M_{\rm clump}$ shows no differences between the two groups. On the other hand, the mean and median values for Log ($L_{\rm bol}/M_{\rm clump}$), $T_{\rm dust}$, and Log ($L_{\rm SiO}/L_{\rm bol}$), which are all of traces of evolution, are apparently different with the presence of SiO wing features. We do not find any significant differences in \ce{H2}, SiO column densities, and $X$(SiO) between SiO wing and non-wing sources. Overall sources with SiO wings have higher luminosity than non-wing sources by a factor 7 -- 14 (depending on whether considering the median or mean $L_{\rm bol}$).  

Figure\,\ref{fig:hist_lm_lsiolbol} shows histograms of those parameter ratios for sources with SiO wings (in gray color) and without them (in cyan color). To distinguish significant differences in these physical parameters as well as $T_{\rm dust}$ between SiO wing sources and non-wing sources, we performed KS-tests on $L_{\rm bol}/M_{\rm clump}$, $T_{\rm dust}$, and $L_{\rm SiO}/L_{\rm bol}$ where we find some differences in their mean and median values. According the KS-tests, we can strongly reject the null hypothesis, which is that the two group represent the same sample, for $L_{\rm bol}/M_{\rm clump}$ and $T_{\rm dust}$ with small $p-$values $\ll 0.001$. For $L_{\rm SiO}/L_{\rm bol}$, comparing the two groups gives a $p-$value of 0.04 and so can be rejected at the 0.95 confidence level (2\,$\sigma$). Therefore, the sources with no apparent SiO wing features may tend to be associated with less evolved clumps indicated by lower $L_{\rm bol}/M_{\rm clump}$ and lower $T_{\rm dust}$, whereas their $L_{\rm SiO}/L_{\rm bol}$ are higher than those of sources with SiO wings.

Several studies (e.g., \citealt{Motte2007,Sakai2010}) report bright SiO emission toward less evolved sources (i.e., mid-infrared dark). We also find the trend toward our sources showing bright SiO emission toward colder and younger sources. However, some of them do not show high-velocity SiO emission, and their SiO lines have relatively narrow line widths ($\Delta \varv_{\rm FWHM} < 8$\,\kms). This may imply that the SiO emission toward the sources with no wings are associated with low-velocity shock processes, such as cloud-cloud collision \citep{Cosentino_SiO_IDRC_2020, Armijos-Abendano_2020_SrgB2} or converging flow/global-infalling flow \citep{Jimenez-Serra2010_extend_sio,Duarte-Cabral2014,Csengeri2016}, to release the SiO into the gas-phase.

\begin{figure}[h!]
    \centering
   \includegraphics[width=0.45\textwidth]{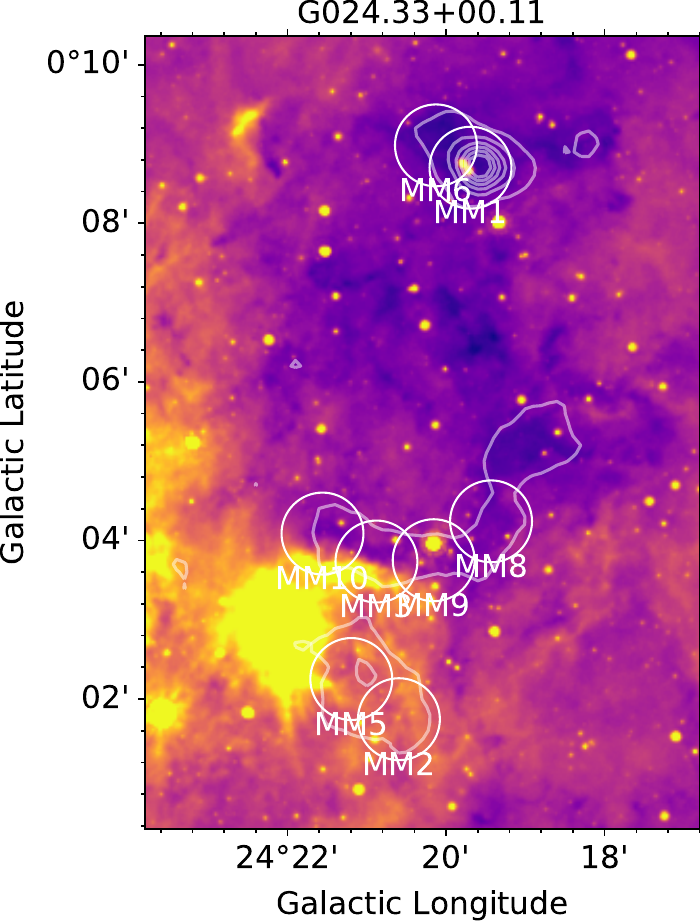}
    \caption{GLIMPSE IRAC 8\,$\mu$m image in color toward G024.33$+$00.11, with the dust emission at 870\,$\mu$m, in gray contours. The white circles indicate the beam of KVN observations toward target sources.}
    \label{fig:g24p33_8micro}
\end{figure}

\subsection{The origin of SiO emission}
Many sources observed in this survey host bright infrared objects, such as HMPOs and \uchiis\, that are selected with infrared colors \citep{Wood1989_uchii_cata, Molinari1996_hmpo_cata, Rathborne2006_irdc_cata}. Some sources of the IRDCs have compact point sources at 4.5\,$\mu$m and 8\,$\mu$m wavelengths \citep{Chambers2009_IRDC}. The detected SiO emission may originate from ongoing star-forming activity, such as outflows or jets from embedded sources. We have found 66 sources (18 \uchiis, 15 HMPOs, and 33 IRDC cores) showing SiO wing features or broad SiO lines ($\Delta\varv_{\rm FWHM} > 8$\,\kms), and most of the sources seem to be more evolved sources with a median Log($L_{\rm bol}/M_{\rm clump}$) of 0.90 compared with those showing no SiO wing features (a median Log($L_{\rm bol}/M_{\rm clump}$) $= -0.02$). Broad SiO line width and wing features apparently indicate the presence of fast shock with the typical C-shock velocity (e.g., \citealt{Gusdorf_SiO_models_2008}). However, we also find narrow SiO components toward bright infrared sources at 8\,$\mu$m or/and 24\,$\mu$m wavelengths. \cite{Lefloch1998} found widespread SiO emission toward NGC\,1333, and they proposed that the SiO emission lines trace materials shocked by outflows, being diverted and decelerated by interacting with colder ambient clumps. This can produce narrow SiO emission lines due to the processes formed by fast-shocks and surrounding gas at a systemic velocity \citep{Lefloch1998,Codella1999,Duarte-Cabral2014}.

Some SiO-detected sources, however, do not have discernible infrared objects, and so they are classified as infrared quiescent sources in the ATLASGAL catalog or unidentified. Among those sources, narrow SiO emission lines are found toward eight sources: G023.60$+$00.00 MM7, G024.33$+$00.11 MM6, G024.33$+$00.11 MM2, G024.33$+$00.11 MM3, G024.33$+$00.11 MM10, G028.37$+$00.07 MM1, G028.37$+$00.07 MM16, and G031.97$+$00.07 MM8. 

Within the IRDC G024.33$+$00.11, three sources (MM2, MM3, and MM10) showing narrow SiO lines are found in the same cloud. G024.33+00.11 MM2 is very close to another source, G024.33+00.11 MM5, which shows a slightly broad SiO emission profile. The other positions, G24.33$+$00.11 MM3 and MM10, are situated in the vicinity of a \hii\ bubble \citep{Simpson2012_bubble,Anderson2015_hii} showing bright 8\,$\mu$m emission presented in Fig.\,\ref{fig:g24p33_8micro}. It might be possible that their narrow SiO lines are produced by the photo-desorption of icy mantles of dust grains \citep{Walmsley1999_pdr,Schilke_PDR_2001}. The SiO abundances relative to $N$(\ce{H2}) obtained from PPMAP toward these four sources (MM2, MM3, MM5, and MM10 of G024.33$+$00.11) surrounding the \hii\ bubble are range from $7.1\times10^{-11}$ -- $3.85\times10^{-10}$ that are close to SiO abundances measured in translucent clouds ($\sim 10^{-11}$ -- $10^{-10}$, \citealt{Turner1998_SiO_diffuse}) and in the Orion Bar PDR ($\sim 10^{-11}$, \citealt{Schilke_PDR_2001}). However, $N$(SiO) and $X$(SiO) toward the MM2, MM3, MM5, MM10 regions also enter typical ranges of column density and abundance of SiO found in regions showing large-scale shocks \citep{Jimenez-Serra2010_extend_sio} or cloud-cloud collision by the interaction between ambient colder gas and expanding \hii\ regions \citep{Colombo2019,Zhu2023_BCGS} as all of the positions surround the 8\,$\mu$m bright \hii\ region (see Fig.\,\ref{fig:g24p33_8micro}). 

G024.33$+$00.11 MM6 is close to MM1, hosting a young stellar object that likely has outflows. Thus, the SiO emission detected toward MM6 could be tracing the shocked gas interacting with potential outflows, and this might be the same for G028.37$+$00.07 MM1, which is close to the bright 8\,$\mu$m source. However, G028.37$+$00.07 MM1 and G024.33$+$00.11 MM6 are separated by 0.64 -- 1.12 pc from the outflow candidate sources. These IRDCs are relatively isolated from the other IRDCs in IRDC G024.33$+$00.11. If these narrow SiO emission are not associated with outflows, this could be explained by shocks associated with infalling material from massive dense clumps ($\geq 0.1 -0.3$\,pc) onto the core envelope scale ($\leq 0.02$\,pc) or small-scale converging flows creating low-velocity shocks \citep{Jimenez-Serra2010_extend_sio,Csengeri2011_a,Csengeri2011_b,Nguyen-Luong_W43_2013,Duarte-Cabral2014,Louvet_2016}.

Although we adopt a $\Delta\varv_{\rm FWHM}$ of 8\,\kms\ as the threshold to divide sources into two groups, narrow (i.e., non-wing SiO) and broad (i.e., SiO wing) SiO emission sources, the value is still arbitrary, and in many previous studies, the definition of the line width for narrow SiO components are variable ($\Delta\varv_{\rm FWHM}$ $< 1$\,\kms\ to 10 -- 15\,\kms) (e.g., \citealt{Jimenez-Serra2010_extend_sio, Lefloch1998, Cosentino_SiO_IDRC_2020, Duarte-Cabral2014, Csengeri2016, Louvet_2016, Nguyen-Luong_W43_2013}). In this survey, seven sources (G015.05$+$00.07 MM1, G023.60$+$00.00 MM4, G028.53$-$00.25 MM10, G034.43$+$00.24 MM5, G034.43$+$00.24 MM6, G034.43$+$00.24 MM7, and G034.43+00.24 MM8) with $\Delta\varv_{\rm FWHM}$ $>$ 8\,\kms\ and no significant wing features of SiO do not have any infrared counterpart or are classified as embedded sources in the ATLASGAL catalog. 

\begin{figure}[h!]
    \centering
   \includegraphics[width=0.45\textwidth]{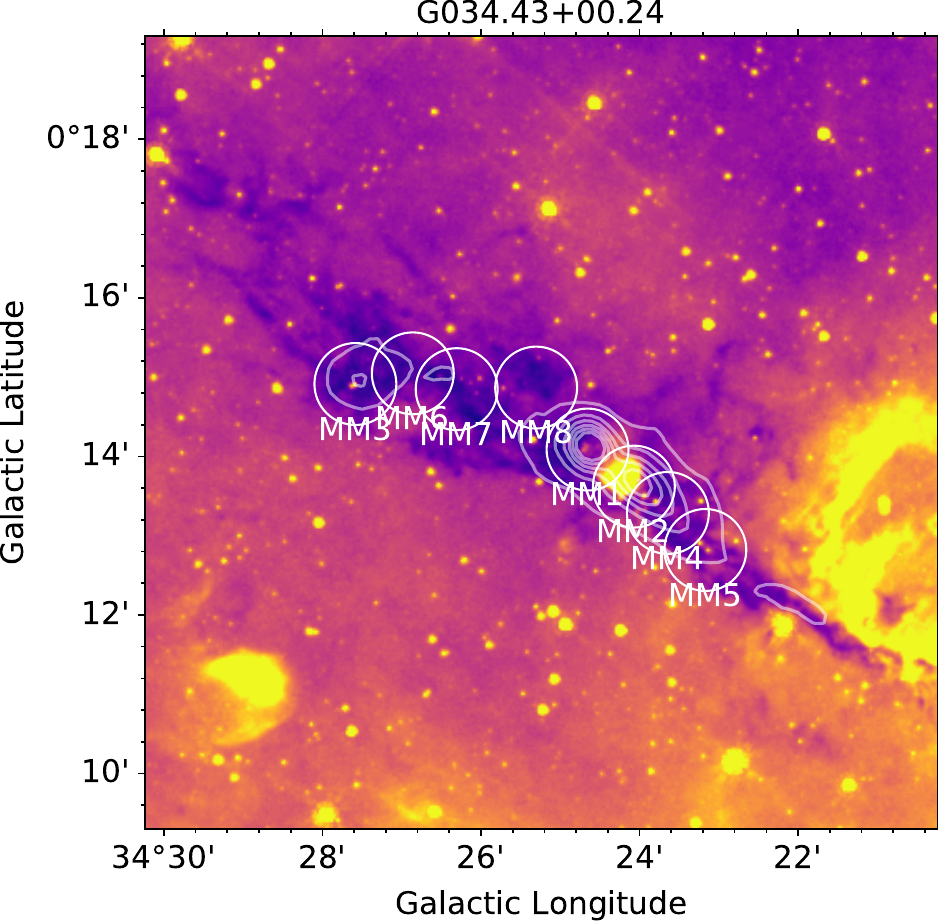}
    \caption{GLIMPSE IRAC 8\,$\mu$m image in color toward G034.43$+$00.24, with the dust emission at 870\,$\mu$m, in gray contours. The white circles indicate the beam of KVN observations toward target sources.}
    \label{fig:g34p43_8micro}
\end{figure}

Four positions out of the seven sources with SiO line width $\Delta\varv_{\rm FWHM}$ $>$ 8\,\kms\ without both SiO wing features and discernible embedded objects are located in the IRDC G034.43+00.24, these are MM5, MM6, MM7, and MM8 marked in Fig.\,\ref{fig:g34p43_8micro}. Those positions are located far from the outflow sources, that are MM1 which is known an extended green object (EGO G034.41$+$0.24, \citealt{Cyganowski2008_ego}), MM2, and MM3, with separations of 0.64--0.86\,pc. As seen in Fig.\,\ref{fig:g34p43_8micro}, these four positions clearly follow the high visual extinction region at 8\,$\mu$m along the filamentary features. We find such cares toward other IRDCs for narrow SiO sources without embedded sources; G015.05$+$00.07 MM1, G023.60$+$00.00 MM4, G023.60$+$00.00 MM7, G028.37$+$00.07 MM1, G028.37$+$00.07 MM16, G028.53$-$00.25 MM10, and G031.97$+$00.07 MM8. In cold and dense environments, Si is unlikely to be in the gas phase because of depletion back onto the icy grain mantles or oxidization forming \ce{SiO2}, which itself is also eventually depleted to the ice mantle of dust grains. This leads to the relatively short lifetime of SiO in the gas phase in about $\sim 10^{4}$ yr \citep{Martin-Pintado1992_sio,Gusdorf_SiO_models_2008}. According to \cite{Nguyen-Luong_W43_2013} and \cite{Duarte-Cabral2014}, converging flow from diffuse interstellar mediums containing preexisting SiO in the gas phase into massive dense clumps provide low-velocity shocks with $\varv_{\rm s}$ of 5\,\kms\ that could preserve the abundance of SiO above the typical abundance of $< 10^{-12}$ in low-mass dense cores \citep{Ziurys1989_sio}. Toward some diffuse and translucent clouds (e.g., \citealt{Turner1998_SiO_diffuse, Rybarczyk2023_sio_diffuseISM}), absorption observations show that SiO can already exist in the gas phase from diffuse environments. In addition, \cite{Guillet2011}, \cite{Caselli1997_sio_shock_model} and \cite{Jimenez-Serra2008_shock_model_low_vel} show that low-velocity shocks about 10 -- 15\,\kms\ can produce SiO in the gas-phase through sputtering for evaporation of the grain mantle ices caused by grain-grain collisions or desorbing Si in the gas phase if a fraction of Si exits in the grain ice mantles. Such low-velocity shocks can be created by cloud-cloud collisions, infalling materials, or converging flows. 

Our survey shows that narrow SiO without wing features are commonly detected toward IRDCs. However, this could be due to an observational bias because we observed more positions in the IRDCs than toward HMPOs and \uchiis. We note that the large number of detections toward IRDCs are not due to their distances as their detection distribution in distance is not significantly different from those of HMPOs and \uchiis. Therefore, it is hard to conclude that narrow SiO emission features appear to be related to a particular type of cloud, for example, IRDCs. If we only consider sources having ATLASGAL counterparts, we can discern that narrow SiO sources with no wings are associated with younger and colder dust clumps. This supports the suggestions proposed in previous studies that narrow and widespread SiO emission lines toward molecular clumps with a lack of star-forming activity seem to be originated from cloud formation processes producing low-velocity shocks (10 -- 15\,\kms) and to trace the early phase of molecular cloud evolution \citep{Duarte-Cabral2014, Csengeri2016}.

\section{Summary}\label{sec:summary_conclusion}

We carried out a  survey of 43\,GHz SiO ($J=1-0$) emission toward 366 massive star-forming regions consisting of different evolutionary stages of massive stars, including infrared dark cloud cores, outflow candidates of high mass young stellar objects, and ultracompact \hii\ regions, with the KVN 21~m telescopes.  
The main results are: 
\begin{itemize}
\item[$-$] 
We detected SiO (1 -- 0) emission toward 106 positions (59 IRDC cores, 21 HMPOs, and 26 \uchiis), however, two sources out of the 59 IRDC cores are excluded from further analysis as the beam significantly overlaps with two sources with stronger SiO emission. We divide the 104 sources into two groups by line profile and $\Delta\varv_{\rm FWHM}$; 66 out of 104 sources are considered as non-wing SiO group because these sources do not have wing features at high-velocities in SiO (1 -- 0) or/and (2 -- 1) or/and $\Delta\varv_{\rm FWHM}$ of their SiO spectra are narrower than 8\,\kms. The remaining 38 sources have either high-velocity wing features in SiO 1 -- 0 or 2 -- 1 or/and their $\Delta\varv_{\rm FWHM}$ is broader than 8\,\kms\ (the median width of all the fitted components) . We show that 72 (69\%) and 81 (78\%) sources have 22\,GHz \ce{H2O} and 44\,GHz Class I \ce{CH3OH} masers, respectively. By comparing physical properties of the dust clumps with SiO detection and non-detections, we show that the dust clumps with SiO emission are more massive (median Log ($M_{\rm clump}$) = 3.26) and colder (median $T_{\rm dust}= 20$\,K) than the clumps without SiO emission.

\item[$-$] 
The range of the beam-averaged SiO column density is $2.18\times10^{12} - 7.92\times10^{13}$\,cm$^{-2}$, and the median is $8.12\times10^{12}$\,cm$^{-2}$. The median and mean $X$(SiO) relative to $N$(\ce{H2}) taken from the ATLASGAL catalog are $1.28\times10^{-10}$ and $1.71\times10^{-10}$, respectively. $N$(SiO) shows a good correlation with $N$(\ce{H2}), but $X$(SiO) does not show any strong correlation with the clump evolutionary stage as measured by the clump luminosity to mass ratio. We also find that for lower dust temperature sources, the SiO column density and abundance determined from SiO 1 -- 0 are $\sim3$ times higher than those obtained from SiO 2 -- 1 surveys. We find that less evolved sources (those with lower dust temperatures) tend to have higher $L_{\rm SiO}/L_{\rm bol}$ ratio by a factor of $\sim$7 compared with higher temperature sources. $L_{\rm SiO}$ is also very well correlated with the clump dust mass.

\item[$-$] 
We have compared $L_{\rm SiO}/L_{\rm bol}$ with $L_{\rm \ce{H2O}}/L_{\rm bol}$ and $L_{\rm \ce{CH3OH}}/L_{\rm bol}$ toward \ce{H2O} and \ce{CH3OH} maser detected sources, respectively. We find that SiO does not seem to be correlated with \ce{H2O}, although both transitions are known as outflow tracers. However, the SiO luminosity correlates well with \ce{CH3OH} luminosity, which itself is also well correlated with clump mass. These two molecules occur on dust grains and are released from the dust icy mantles or grain cores. Thus, it seems that toward our sources, the SiO and Class I \ce{CH3OH} maser emission might result from the related shocks. 

\item[$-$]
Toward sources with outflow candidates, embedded young stellar objects or \uchiis, the detected SiO emission may be associated with outflows or jets as many of them show wing features at high-velocities and/or broader $\Delta\varv_{\rm FWHM}$ ($>$\,8\,\kms). However, the origin of SiO emission toward infrared dark sources and sources without clear embedded young stellar objects is unclear. The majority of these sources have narrow SiO emission, and even if a SiO line width is broader than 8\,\kms, there are no clear embedded sources, and the observed positions can be situated far from bright infrared sources by 0.64 -- 1.12\,pc. In addition, we show that the SiO emission is can be widely spread over infrared clouds, not concentrated in a small area. Such cases might be associated with other mechanisms with low-velocity shocks ($\varv_{\rm s} \sim 5 - 15$\,\kms). Cloud-cloud collisions, infalling material onto massive dense clumps, or converging flows from diffuse clouds toward dense molecular clouds are candidates for such low-velocity shock resources.  

\end{itemize}

This study demonstrates that toward colder molecular clumps without identified outflow sources, SiO $J=1-0$ traces low-velocity shocks. These shocks may be caused by various mechanisms, such as cloud-cloud collisions, expanding \hii\ regions, global infall, or converging flows. Therefore, SiO $J=1-0$ may be a more suitable tracer of shocks for such cases than a higher $J$ transition of SiO.

\begin{acknowledgements}
We would like to thank the referee for their constructive comments and suggestions that have helped to improve this article. We are grateful to all staff members in Korean VLBI Network (KVN) who helped to operate the array and to correlate the data. The KVN is a facility operated by Korea Astronomy and Space Science Institute (KASI). W.-J. Kim was supported by DLR/Verbundforschung Astronomie und Astrophysik Grant 50 OR 2007 for this work. G.A.F acknowledges support from the Universit\"at zu K\"oln and its Global Faculty Program. 
\end{acknowledgements}

%
%


\bibliographystyle{aa} 
\bibliography{arXiv_SiO_survey_WonjuKim} 

\begin{thebibliography}{90}
\expandafter\ifx\csname natexlab\endcsname\relax\def\natexlab#1{#1}\fi

\bibitem[{{Anderson} {et~al.}(2015){Anderson}, {Armentrout}, {Johnstone}, {Bania}, {Balser}, {Wenger}, \& {Cunningham}}]{Anderson2015_hii}
{Anderson}, L.~D., {Armentrout}, W.~P., {Johnstone}, B.~M., {et~al.} 2015, \apjs, 221, 26

\bibitem[{{Armijos-Abenda{\~n}o} {et~al.}(2020){Armijos-Abenda{\~n}o}, {Banda-Barrag{\'a}n}, {Mart{\'\i}n-Pintado}, {D{\'e}nes}, {Federrath}, \& {Requena-Torres}}]{Armijos-Abendano_2020_SrgB2}
{Armijos-Abenda{\~n}o}, J., {Banda-Barrag{\'a}n}, W.~E., {Mart{\'\i}n-Pintado}, J., {et~al.} 2020, \mnras, 499, 4918

\bibitem[{{Beuther} {et~al.}(2002){Beuther}, {Schilke}, {Menten}, {Motte}, {Sridharan}, \& {Wyrowski}}]{Beuther2002}
{Beuther}, H., {Schilke}, P., {Menten}, K.~M., {et~al.} 2002, \apj, 566, 945

\bibitem[{{Bronfman} {et~al.}(1996){Bronfman}, {Nyman}, \& {May}}]{Bronfman1996_CS_uchii}
{Bronfman}, L., {Nyman}, L.~A., \& {May}, J. 1996, \aaps, 115, 81

\bibitem[{{Carral} {et~al.}(1999){Carral}, {Kurtz}, {Rodr{\'\i}guez}, {Mart{\'\i}}, {Lizano}, \& {Osorio}}]{carral1999}
{Carral}, P., {Kurtz}, S., {Rodr{\'\i}guez}, L.~F., {et~al.} 1999, \rmxaa, 35, 97

\bibitem[{{Caselli} {et~al.}(1997){Caselli}, {Hartquist}, \& {Havnes}}]{Caselli1997_sio_shock_model}
{Caselli}, P., {Hartquist}, T.~W., \& {Havnes}, O. 1997, \aap, 322, 296

\bibitem[{{Chambers} {et~al.}(2009){Chambers}, {Jackson}, {Rathborne}, \& {Simon}}]{Chambers2009_IRDC}
{Chambers}, E.~T., {Jackson}, J.~M., {Rathborne}, J.~M., \& {Simon}, R. 2009, \apjs, 181, 360

\bibitem[{{Chira} {et~al.}(2013){Chira}, {Beuther}, {Linz}, {Schuller}, {Walmsley}, {Menten}, \& {Bronfman}}]{Chira2013_IRDC_NH3}
{Chira}, R.~A., {Beuther}, H., {Linz}, H., {et~al.} 2013, \aap, 552, A40

\bibitem[{{Churchwell} {et~al.}(1990){Churchwell}, {Walmsley}, \& {Cesaroni}}]{Churchwell1990_NH3}
{Churchwell}, E., {Walmsley}, C.~M., \& {Cesaroni}, R. 1990, \aaps, 83, 119

\bibitem[{{Codella} {et~al.}(1999){Codella}, {Bachiller}, \& {Reipurth}}]{Codella1999}
{Codella}, C., {Bachiller}, R., \& {Reipurth}, B. 1999, \aap, 343, 585

\bibitem[{{Colombo} {et~al.}(2019){Colombo}, {Rosolowsky}, {Duarte-Cabral}, {Ginsburg}, {Glenn}, {Zetterlund}, {Hernandez}, {Dempsey}, \& {Currie}}]{Colombo2019}
{Colombo}, D., {Rosolowsky}, E., {Duarte-Cabral}, A., {et~al.} 2019, \mnras, 483, 4291

\bibitem[{{Cosentino} {et~al.}(2019){Cosentino}, {Jim{\'e}nez-Serra}, {Caselli}, {Henshaw}, {Barnes}, {Tan}, {Viti}, {Fontani}, \& {Wu}}]{Cosentino2019_SNR_shocks_sio}
{Cosentino}, G., {Jim{\'e}nez-Serra}, I., {Caselli}, P., {et~al.} 2019, \apjl, 881, L42

\bibitem[{{Cosentino} {et~al.}(2020){Cosentino}, {Jim{\'e}nez-Serra}, {Henshaw}, {Caselli}, {Viti}, {Barnes}, {Tan}, {Fontani}, \& {Wu}}]{Cosentino_SiO_IDRC_2020}
{Cosentino}, G., {Jim{\'e}nez-Serra}, I., {Henshaw}, J.~D., {et~al.} 2020, \mnras, 499, 1666

\bibitem[{{Cosentino} {et~al.}(2022){Cosentino}, {Jim{\'e}nez-Serra}, {Tan}, {Henshaw}, {Barnes}, {Law}, {Zeng}, {Fontani}, {Caselli}, {Viti}, {Zahorecz}, {Rico-Villas}, {Meg{\'\i}as}, {Miceli}, {Orlando}, {Ustamujic}, {Greco}, {Peres}, {Bocchino}, {Fedriani}, {Gorai}, {Testi}, \& {Mart{\'\i}n-Pintado}}]{Cosentino2022_SNR_shock_sio}
{Cosentino}, G., {Jim{\'e}nez-Serra}, I., {Tan}, J.~C., {et~al.} 2022, \mnras, 511, 953

\bibitem[{{Csengeri} {et~al.}(2011{\natexlab{a}}){Csengeri}, {Bontemps}, {Schneider}, {Motte}, \& {Dib}}]{Csengeri2011_a}
{Csengeri}, T., {Bontemps}, S., {Schneider}, N., {Motte}, F., \& {Dib}, S. 2011{\natexlab{a}}, \aap, 527, A135

\bibitem[{{Csengeri} {et~al.}(2011{\natexlab{b}}){Csengeri}, {Bontemps}, {Schneider}, {Motte}, {Gueth}, \& {Hora}}]{Csengeri2011_b}
{Csengeri}, T., {Bontemps}, S., {Schneider}, N., {et~al.} 2011{\natexlab{b}}, \apjl, 740, L5

\bibitem[{{Csengeri} {et~al.}(2017){Csengeri}, {Bontemps}, {Wyrowski}, {Megeath}, {Motte}, {Sanna}, {Wienen}, \& {Menten}}]{Csengeri2017}
{Csengeri}, T., {Bontemps}, S., {Wyrowski}, F., {et~al.} 2017, \aap, 601, A60

\bibitem[{{Csengeri} {et~al.}(2016{\natexlab{a}}){Csengeri}, {Leurini}, {Wyrowski}, {Urquhart}, {Menten}, {Walmsley}, {Bontemps}, {Wienen}, {Beuther}, {Motte}, {Nguyen-Luong}, {Schilke}, {Schuller}, {Zavagno}, \& {Sanna}}]{Csengeri2016}
{Csengeri}, T., {Leurini}, S., {Wyrowski}, F., {et~al.} 2016{\natexlab{a}}, \aap, 586, A149

\bibitem[{{Csengeri} {et~al.}(2016{\natexlab{b}}){Csengeri}, {Weiss}, {Wyrowski}, {Menten}, {Urquhart}, {Leurini}, {Schuller}, {Beuther}, {Bontemps}, {Bronfman}, {Henning}, \& {Schneider}}]{csengeri2016_planck}
{Csengeri}, T., {Weiss}, A., {Wyrowski}, F., {et~al.} 2016{\natexlab{b}}, \aap, 585, A104

\bibitem[{{Cyganowski} {et~al.}(2008){Cyganowski}, {Whitney}, {Holden}, {Braden}, {Brogan}, {Churchwell}, {Indebetouw}, {Watson}, {Babler}, {Benjamin}, {Gomez}, {Meade}, {Povich}, {Robitaille}, \& {Watson}}]{Cyganowski2008_ego}
{Cyganowski}, C.~J., {Whitney}, B.~A., {Holden}, E., {et~al.} 2008, \aj, 136, 2391

\bibitem[{{Duarte-Cabral} {et~al.}(2014){Duarte-Cabral}, {Bontemps}, {Motte}, {Gusdorf}, {Csengeri}, {Schneider}, \& {Louvet}}]{Duarte-Cabral2014}
{Duarte-Cabral}, A., {Bontemps}, S., {Motte}, F., {et~al.} 2014, \aap, 570, A1

\bibitem[{{Dumas} {et~al.}(2014){Dumas}, {Vaupr{\'e}}, {Ceccarelli}, {Hily-Blant}, {Dubus}, {Montmerle}, \& {Gabici}}]{Dumas2014_SNR_sio_shocks}
{Dumas}, G., {Vaupr{\'e}}, S., {Ceccarelli}, C., {et~al.} 2014, \apjl, 786, L24

\bibitem[{{Egan} {et~al.}(1998){Egan}, {Shipman}, {Price}, {Carey}, {Clark}, \& {Cohen}}]{Egan1998}
{Egan}, M.~P., {Shipman}, R.~F., {Price}, S.~D., {et~al.} 1998, \apjl, 494, L199

\bibitem[{{Elia} {et~al.}(2021){Elia}, {Merello}, {Molinari}, {Schisano}, {Zavagno}, {Russeil}, {M{\`e}ge}, {Martin}, {Olmi}, {Pestalozzi}, {Plume}, {Ragan}, {Benedettini}, {Eden}, {Moore}, {Noriega-Crespo}, {Paladini}, {Palmeirim}, {Pezzuto}, {Pilbratt}, {Rygl}, {Schilke}, {Strafella}, {Tan}, {Traficante}, {Baldeschi}, {Bally}, {di Giorgio}, {Fiorellino}, {Liu}, {Piazzo}, \& {Polychroni}}]{Elia2021}
{Elia}, D., {Merello}, M., {Molinari}, S., {et~al.} 2021, \mnras, 504, 2742

\bibitem[{Endres {et~al.}(2016)Endres, Schlemmer, Schilke, Stutzki, \& Müller}]{Endres2016}
Endres, C.~P., Schlemmer, S., Schilke, P., Stutzki, J., \& Müller, H.~S. 2016, Journal of Molecular Spectroscopy, 327, 95, new Visions of Spectroscopic Databases, Volume II

\bibitem[{{Felli} {et~al.}(1992){Felli}, {Palagi}, \& {Tofani}}]{Felli1992}
{Felli}, M., {Palagi}, F., \& {Tofani}, G. 1992, \aap, 255, 293

\bibitem[{{Fuller} {et~al.}(2005){Fuller}, {Williams}, \& {Sridharan}}]{Fuller2005_infall}
{Fuller}, G.~A., {Williams}, S.~J., \& {Sridharan}, T.~K. 2005, \aap, 442, 949

\bibitem[{{G{\'o}mez-Ruiz} {et~al.}(2016){G{\'o}mez-Ruiz}, {Kurtz}, {Araya}, {Hofner}, \& {Loinard}}]{Gomez-Ruiz2016}
{G{\'o}mez-Ruiz}, A.~I., {Kurtz}, S.~E., {Araya}, E.~D., {Hofner}, P., \& {Loinard}, L. 2016, \apjs, 222, 18

\bibitem[{{Gregory} \& {Condon}(1991)}]{Gregory1991}
{Gregory}, P.~C. \& {Condon}, J.~J. 1991, \apjs, 75, 1011

\bibitem[{{Griffith} {et~al.}(1994){Griffith}, {Wright}, {Burke}, \& {Ekers}}]{Griffith1994}
{Griffith}, M.~R., {Wright}, A.~E., {Burke}, B.~F., \& {Ekers}, R.~D. 1994, \apjs, 90, 179

\bibitem[{{Guillet} {et~al.}(2011){Guillet}, {Pineau Des For{\^e}ts}, \& {Jones}}]{Guillet2011}
{Guillet}, V., {Pineau Des For{\^e}ts}, G., \& {Jones}, A.~P. 2011, \aap, 527, A123

\bibitem[{{Gusdorf} {et~al.}(2008){Gusdorf}, {Pineau Des For{\^e}ts}, {Cabrit}, \& {Flower}}]{Gusdorf_SiO_models_2008}
{Gusdorf}, A., {Pineau Des For{\^e}ts}, G., {Cabrit}, S., \& {Flower}, D.~R. 2008, \aap, 490, 695

\bibitem[{{Haschick} \& {Baan}(1993)}]{Haschick1993_CH3OHmaser_cloud_collison}
{Haschick}, A.~D. \& {Baan}, W.~A. 1993, \apj, 410, 663

\bibitem[{{Hennebelle} {et~al.}(2001){Hennebelle}, {P{\'e}rault}, {Teyssier}, \& {Ganesh}}]{Hennebelle2001}
{Hennebelle}, P., {P{\'e}rault}, M., {Teyssier}, D., \& {Ganesh}, S. 2001, \aap, 365, 598

\bibitem[{{Jackson} {et~al.}(2019){Jackson}, {Whitaker}, {Rathborne}, {Foster}, {Contreras}, {Sanhueza}, {Stephens}, {Longmore}, \& {Allingham}}]{Jackson2019}
{Jackson}, J.~M., {Whitaker}, J.~S., {Rathborne}, J.~M., {et~al.} 2019, \apj, 870, 5

\bibitem[{{Jim{\'e}nez-Serra} {et~al.}(2008){Jim{\'e}nez-Serra}, {Caselli}, {Mart{\'\i}n-Pintado}, \& {Hartquist}}]{Jimenez-Serra2008_shock_model_low_vel}
{Jim{\'e}nez-Serra}, I., {Caselli}, P., {Mart{\'\i}n-Pintado}, J., \& {Hartquist}, T.~W. 2008, \aap, 482, 549

\bibitem[{{Jim{\'e}nez-Serra} {et~al.}(2010){Jim{\'e}nez-Serra}, {Caselli}, {Tan}, {Hernandez}, {Fontani}, {Butler}, \& {van Loo}}]{Jimenez-Serra2010_extend_sio}
{Jim{\'e}nez-Serra}, I., {Caselli}, P., {Tan}, J.~C., {et~al.} 2010, \mnras, 406, 187

\bibitem[{{Jim{\'e}nez-Serra} {et~al.}(2005){Jim{\'e}nez-Serra}, {Mart{\'\i}n-Pintado}, {Rodr{\'\i}guez-Franco}, \& {Mart{\'\i}n}}]{Jimenez-Serra2005_L1448-mm}
{Jim{\'e}nez-Serra}, I., {Mart{\'\i}n-Pintado}, J., {Rodr{\'\i}guez-Franco}, A., \& {Mart{\'\i}n}, S. 2005, \apjl, 627, L121

\bibitem[{{Kelly} {et~al.}(2017){Kelly}, {Viti}, {Garc{\'\i}a-Burillo}, {Fuente}, {Usero}, {Krips}, \& {Neri}}]{Kelly2017}
{Kelly}, G., {Viti}, S., {Garc{\'\i}a-Burillo}, S., {et~al.} 2017, \aap, 597, A11

\bibitem[{{Kim}(2016)}]{Kim2016PhDT}
{Kim}, C.-H. 2016, PhD thesis, Seoul National University, Korea

\bibitem[{{Kim} {et~al.}(2011){Kim}, {Byun}, {Je}, {Wi}, {Bae}, {Jung}, {Lee}, {Han}, {Song}, {Jung}, {Chung}, {Kim}, \& {Kim}}]{Kim2011}
{Kim}, K.-T., {Byun}, D.-Y., {Je}, D.-H., {et~al.} 2011, Journal of Korean Astronomical Society, 44, 81

\bibitem[{{Kim} {et~al.}(2019){Kim}, {Kim}, \& {Kim}}]{Kim2019_masers}
{Kim}, W.-J., {Kim}, K.-T., \& {Kim}, K.-T. 2019, \apjs, 244, 2

\bibitem[{{K{\"o}nig} {et~al.}(2017){K{\"o}nig}, {Urquhart}, {Csengeri}, {Leurini}, {Wyrowski}, {Giannetti}, {Wienen}, {Pillai}, {Kauffmann}, {Menten}, \& {Schuller}}]{Koenig2017}
{K{\"o}nig}, C., {Urquhart}, J.~S., {Csengeri}, T., {et~al.} 2017, \aap, 599, A139

\bibitem[{{Kurtz} {et~al.}(1994){Kurtz}, {Churchwell}, \& {Wood}}]{Kurtz1994_uchii_cata}
{Kurtz}, S., {Churchwell}, E., \& {Wood}, D.~O.~S. 1994, \apjs, 91, 659

\bibitem[{{Kurtz} {et~al.}(2004){Kurtz}, {Hofner}, \& {{\'A}lvarez}}]{Kurtz2004}
{Kurtz}, S., {Hofner}, P., \& {{\'A}lvarez}, C.~V. 2004, \apjs, 155, 149

\bibitem[{{Lee} {et~al.}(2011){Lee}, {Byun}, {Oh}, {Han}, {Je}, {Kim}, {Wi}, {Cho}, {Sohn}, {Kim}, {Lee}, {Oh}, {Song}, {Kang}, {Chung}, {Lee}, {Oh}, {Bae}, {Yun}, {Lee}, {Kim}, {Chung}, {Roh}, {Lee}, {Kim}, {Ryoung Kim}, {Yeom}, {Kurayama}, {Jung}, {Park}, {Kim}, {Yoon}, \& {Kim}}]{Lee2011}
{Lee}, S.-S., {Byun}, D.-Y., {Oh}, C.~S., {et~al.} 2011, \pasp, 123, 1398

\bibitem[{{Lefloch} {et~al.}(1998){Lefloch}, {Castets}, {Cernicharo}, \& {Loinard}}]{Lefloch1998}
{Lefloch}, B., {Castets}, A., {Cernicharo}, J., \& {Loinard}, L. 1998, \apjl, 504, L109

\bibitem[{{Liu} {et~al.}(2022){Liu}, {Liu}, {Chen}, {Liu}, {Wang}, {Li}, {Lee}, {Liu}, {Juvela}, {Garay}, {Dewangan}, {Soam}, {Bronfman}, {He}, {Eswaraiah}, {Zhang}, {Zhang}, {Xu}, {T{\'o}th}, {Shen}, {Li}, {Wu}, {Qin}, {Ren}, {Zhang}, {Tej}, {Goldsmith}, {Baug}, {Luo}, {Zhou}, \& {Zhang}}]{Liu_atoms_aca_sio_2022}
{Liu}, R., {Liu}, T., {Chen}, G., {et~al.} 2022, \mnras, 511, 3618

\bibitem[{{L{\'o}pez-Sepulcre} {et~al.}(2016){L{\'o}pez-Sepulcre}, {Watanabe}, {Sakai}, {Furuya}, {Saruwatari}, \& {Yamamoto}}]{Lopez-Sepulcre_2016}
{L{\'o}pez-Sepulcre}, A., {Watanabe}, Y., {Sakai}, N., {et~al.} 2016, \apj, 822, 85

\bibitem[{{Louvet} {et~al.}(2016){Louvet}, {Motte}, {Gusdorf}, {Nguy{\^e}n Luong}, {Lesaffre}, {Duarte-Cabral}, {Maury}, {Schneider}, {Hill}, {Schilke}, \& {Gueth}}]{Louvet_2016}
{Louvet}, F., {Motte}, F., {Gusdorf}, A., {et~al.} 2016, \aap, 595, A122

\bibitem[{{Marsh} {et~al.}(2017){Marsh}, {Whitworth}, {Lomax}, {Ragan}, {Becciani}, {Cambr{\'e}sy}, {Di Giorgio}, {Eden}, {Elia}, {Kacsuk}, {Molinari}, {Palmeirim}, {Pezzuto}, {Schneider}, {Sciacca}, \& {Vitello}}]{Marsh2017}
{Marsh}, K.~A., {Whitworth}, A.~P., {Lomax}, O., {et~al.} 2017, \mnras, 471, 2730

\bibitem[{{Martin-Pintado} {et~al.}(1992){Martin-Pintado}, {Bachiller}, \& {Fuente}}]{Martin-Pintado1992_sio}
{Martin-Pintado}, J., {Bachiller}, R., \& {Fuente}, A. 1992, \aap, 254, 315

\bibitem[{{Molinari} {et~al.}(1996){Molinari}, {Brand}, {Cesaroni}, \& {Palla}}]{Molinari1996_hmpo_cata}
{Molinari}, S., {Brand}, J., {Cesaroni}, R., \& {Palla}, F. 1996, \aap, 308, 573

\bibitem[{{Motte} {et~al.}(2007){Motte}, {Bontemps}, {Schilke}, {Schneider}, {Menten}, \& {Brogui{\`e}re}}]{Motte2007}
{Motte}, F., {Bontemps}, S., {Schilke}, P., {et~al.} 2007, \aap, 476, 1243

\bibitem[{{M{\"u}ller} {et~al.}(2005){M{\"u}ller}, {Schl{\"o}der}, {Stutzki}, \& {Winnewisser}}]{Muller2005}
{M{\"u}ller}, H. S.~P., {Schl{\"o}der}, F., {Stutzki}, J., \& {Winnewisser}, G. 2005, Journal of Molecular Structure, 742, 215

\bibitem[{{Nguyen-Lu'o'ng} {et~al.}(2013){Nguyen-Lu'o'ng}, {Motte}, {Carlhoff}, {Louvet}, {Lesaffre}, {Schilke}, {Hill}, {Hennemann}, {Gusdorf}, {Didelon}, {Schneider}, {Bontemps}, {Duarte-Cabral}, {Menten}, {Martin}, {Wyrowski}, {Bendo}, {Roussel}, {Bernard}, {Bronfman}, {Henning}, {Kramer}, \& {Heitsch}}]{Nguyen-Luong_W43_2013}
{Nguyen-Lu'o'ng}, Q., {Motte}, F., {Carlhoff}, P., {et~al.} 2013, \apj, 775, 88

\bibitem[{{Perault} {et~al.}(1996){Perault}, {Omont}, {Simon}, {Seguin}, {Ojha}, {Blommaert}, {Felli}, {Gilmore}, {Guglielmo}, {Habing}, {Price}, {Robin}, {de Batz}, {Cesarsky}, {Elbaz}, {Epchtein}, {Fouque}, {Guest}, {Levine}, {Pollock}, {Prusti}, {Siebenmorgen}, {Testi}, \& {Tiphene}}]{Perault1996}
{Perault}, M., {Omont}, A., {Simon}, G., {et~al.} 1996, \aap, 315, L165

\bibitem[{{Peretto} \& {Fuller}(2009)}]{Peretto2009_irdc}
{Peretto}, N. \& {Fuller}, G.~A. 2009, \aap, 505, 405

\bibitem[{{Peretto} {et~al.}(2013){Peretto}, {Fuller}, {Duarte-Cabral}, {Avison}, {Hennebelle}, {Pineda}, {Andr{\'e}}, {Bontemps}, {Motte}, {Schneider}, \& {Molinari}}]{Peretto2013}
{Peretto}, N., {Fuller}, G.~A., {Duarte-Cabral}, A., {et~al.} 2013, \aap, 555, A112

\bibitem[{{Pety}(2005)}]{pety2005_gildas}
{Pety}, J. 2005, in SF2A-2005: Semaine de l'Astrophysique Francaise, ed. F.~{Casoli}, T.~{Contini}, J.~M. {Hameury}, \& L.~{Pagani}, 721

\bibitem[{{Pillai} {et~al.}(2023){Pillai}, {Urquhart}, {Leurini}, {Zhang}, {Traficante}, {Colombo}, {Wang}, {Gomez}, \& {Wyrowski}}]{Pillai2023_infall}
{Pillai}, T.~G.~S., {Urquhart}, J.~S., {Leurini}, S., {et~al.} 2023, \mnras, 522, 3357

\bibitem[{{Ramesh} \& {Sridharan}(1997)}]{Ramesh1997}
{Ramesh}, B. \& {Sridharan}, T.~K. 1997, \mnras, 284, 1001

\bibitem[{{Rathborne} {et~al.}(2006){Rathborne}, {Jackson}, \& {Simon}}]{Rathborne2006_irdc_cata}
{Rathborne}, J.~M., {Jackson}, J.~M., \& {Simon}, R. 2006, \apj, 641, 389

\bibitem[{{Rybarczyk} {et~al.}(2023){Rybarczyk}, {Stanimirovic}, \& {Gusdorf}}]{Rybarczyk2023_sio_diffuseISM}
{Rybarczyk}, D.~R., {Stanimirovic}, S., \& {Gusdorf}, A. 2023, arXiv e-prints, arXiv:2304.06741

\bibitem[{{Sakai} {et~al.}(2010){Sakai}, {Sakai}, {Hirota}, \& {Yamamoto}}]{Sakai2010}
{Sakai}, T., {Sakai}, N., {Hirota}, T., \& {Yamamoto}, S. 2010, \apj, 714, 1658

\bibitem[{{S{\'a}nchez-Monge} {et~al.}(2013){S{\'a}nchez-Monge}, {L{\'o}pez-Sepulcre}, {Cesaroni}, {Walmsley}, {Codella}, {Beltr{\'a}n}, {Pestalozzi}, \& {Molinari}}]{Sanchez-Monge2013}
{S{\'a}nchez-Monge}, {\'A}., {L{\'o}pez-Sepulcre}, A., {Cesaroni}, R., {et~al.} 2013, \aap, 557, A94

\bibitem[{{Sato} {et~al.}(2000){Sato}, {Hasegawa}, {Whiteoak}, \& {Miyawaki}}]{Sato2000_CH3OHmaser_cloud_collison}
{Sato}, F., {Hasegawa}, T., {Whiteoak}, J.~B., \& {Miyawaki}, R. 2000, \apj, 535, 857

\bibitem[{{Schilke} {et~al.}(2001){Schilke}, {Pineau des For{\^e}ts}, {Walmsley}, \& {Mart{\'\i}n-Pintado}}]{Schilke_PDR_2001}
{Schilke}, P., {Pineau des For{\^e}ts}, G., {Walmsley}, C.~M., \& {Mart{\'\i}n-Pintado}, J. 2001, \aap, 372, 291

\bibitem[{{Schilke} {et~al.}(1997){Schilke}, {Walmsley}, {Pineau des Forets}, \& {Flower}}]{Schilke1997}
{Schilke}, P., {Walmsley}, C.~M., {Pineau des Forets}, G., \& {Flower}, D.~R. 1997, \aap, 321, 293

\bibitem[{{Schuller} {et~al.}(2009){Schuller}, {Menten}, {Contreras}, {Wyrowski}, \& {Schilke}}]{schuller2009}
{Schuller}, F., {Menten}, K.~M., {Contreras}, Y., {Wyrowski}, F., \& {Schilke}, e.~a. 2009, \aap, 504, 415

\bibitem[{{Simpson} {et~al.}(2012){Simpson}, {Povich}, {Kendrew}, {Lintott}, {Bressert}, {Arvidsson}, {Cyganowski}, {Maddison}, {Schawinski}, {Sherman}, {Smith}, \& {Wolf-Chase}}]{Simpson2012_bubble}
{Simpson}, R.~J., {Povich}, M.~S., {Kendrew}, S., {et~al.} 2012, \mnras, 424, 2442

\bibitem[{{Sridharan} {et~al.}(2002){Sridharan}, {Beuther}, {Schilke}, {Menten}, \& {Wyrowski}}]{Sridharan2002_hmpo_cata}
{Sridharan}, T.~K., {Beuther}, H., {Schilke}, P., {Menten}, K.~M., \& {Wyrowski}, F. 2002, \apj, 566, 931

\bibitem[{{Svoboda} {et~al.}(2016){Svoboda}, {Shirley}, {Battersby}, {Rosolowsky}, {Ginsburg}, {Ellsworth-Bowers}, {Pestalozzi}, {Dunham}, {Evans}, {Bally}, \& {Glenn}}]{Svoboda2016_scc_catalog}
{Svoboda}, B.~E., {Shirley}, Y.~L., {Battersby}, C., {et~al.} 2016, \apj, 822, 59

\bibitem[{{Traficante} {et~al.}(2018){Traficante}, {Fuller}, {Smith}, {Billot}, {Duarte-Cabral}, {Peretto}, {Molinari}, \& {Pineda}}]{Traficante2018}
{Traficante}, A., {Fuller}, G.~A., {Smith}, R.~J., {et~al.} 2018, \mnras, 473, 4975

\bibitem[{{Turner}(1998)}]{Turner1998_SiO_diffuse}
{Turner}, B.~E. 1998, \apj, 495, 804

\bibitem[{{Urquhart} {et~al.}(2018){Urquhart}, {K{\"o}nig}, {Giannetti}, {Leurini}, {Moore}, {Eden}, {Pillai}, {Thompson}, {Braiding}, {Burton}, {Csengeri}, {Dempsey}, {Figura}, {Froebrich}, {Menten}, {Schuller}, {Smith}, \& {Wyrowski}}]{Urquhart2018}
{Urquhart}, J.~S., {K{\"o}nig}, C., {Giannetti}, A., {et~al.} 2018, \mnras, 473, 1059

\bibitem[{{Urquhart} {et~al.}(2022){Urquhart}, {Wells}, {Pillai}, {Leurini}, {Giannetti}, {Moore}, {Thompson}, {Figura}, {Colombo}, {Yang}, {K{\"o}nig}, {Wyrowski}, {Menten}, {Rigby}, {Eden}, \& {Ragan}}]{Urquhart2022}
{Urquhart}, J.~S., {Wells}, M.~R.~A., {Pillai}, T., {et~al.} 2022, \mnras, 510, 3389

\bibitem[{{Vaupr{\'e}} {et~al.}(2014){Vaupr{\'e}}, {Hily-Blant}, {Ceccarelli}, {Dubus}, {Gabici}, \& {Montmerle}}]{Vaupre2014_SNRs_shock}
{Vaupr{\'e}}, S., {Hily-Blant}, P., {Ceccarelli}, C., {et~al.} 2014, \aap, 568, A50

\bibitem[{{Voronkov} {et~al.}(2014){Voronkov}, {Caswell}, {Ellingsen}, {Green}, \& {Breen}}]{Voronkov2014}
{Voronkov}, M.~A., {Caswell}, J.~L., {Ellingsen}, S.~P., {Green}, J.~A., \& {Breen}, S.~L. 2014, \mnras, 439, 2584

\bibitem[{{Voronkov} {et~al.}(2010){Voronkov}, {Caswell}, {Ellingsen}, \& {Sobolev}}]{Voronkov2010}
{Voronkov}, M.~A., {Caswell}, J.~L., {Ellingsen}, S.~P., \& {Sobolev}, A.~M. 2010, \mnras, 405, 2471

\bibitem[{{Walmsley} {et~al.}(1999){Walmsley}, {Pineau des For{\^e}ts}, \& {Flower}}]{Walmsley1999_pdr}
{Walmsley}, C.~M., {Pineau des For{\^e}ts}, G., \& {Flower}, D.~R. 1999, \aap, 342, 542

\bibitem[{{Walsh} {et~al.}(1998){Walsh}, {Burton}, {Hyland}, \& {Robinson}}]{Walsh1998}
{Walsh}, A.~J., {Burton}, M.~G., {Hyland}, A.~R., \& {Robinson}, G. 1998, \mnras, 301, 640

\bibitem[{{Widmann} {et~al.}(2016){Widmann}, {Beuther}, {Schilke}, \& {Stanke}}]{Widmann2016}
{Widmann}, F., {Beuther}, H., {Schilke}, P., \& {Stanke}, T. 2016, \aap, 589, A29

\bibitem[{{Wood} \& {Churchwell}(1989{\natexlab{a}})}]{Wood1989b}
{Wood}, D. O.~S. \& {Churchwell}, E. 1989{\natexlab{a}}, \apj, 340, 265

\bibitem[{{Wood} \& {Churchwell}(1989{\natexlab{b}})}]{Wood1989_uchii_cata}
{Wood}, D. O.~S. \& {Churchwell}, E. 1989{\natexlab{b}}, \apjs, 69, 831

\bibitem[{{Wright} {et~al.}(1994){Wright}, {Griffith}, {Burke}, \& {Ekers}}]{Wright1994}
{Wright}, A.~E., {Griffith}, M.~R., {Burke}, B.~F., \& {Ekers}, R.~D. 1994, \apjs, 91, 111

\bibitem[{{Wyrowski} {et~al.}(2016){Wyrowski}, {G{\"u}sten}, {Menten}, {Wiesemeyer}, {Csengeri}, {Heyminck}, {Klein}, {K{\"o}nig}, \& {Urquhart}}]{Wyrowski2016_infall}
{Wyrowski}, F., {G{\"u}sten}, R., {Menten}, K.~M., {et~al.} 2016, \aap, 585, A149

\bibitem[{{Zhu} {et~al.}(2023){Zhu}, {Wang}, {Yan}, {Zhu}, \& {Li}}]{Zhu2023_BCGS}
{Zhu}, F.-Y., {Wang}, J., {Yan}, Y., {Zhu}, Q.-F., \& {Li}, J. 2023, \mnras, 523, 2770

\bibitem[{{Zhu} {et~al.}(2020){Zhu}, {Wang}, {Liu}, {Kim}, {Zhu}, \& {Li}}]{Zhu2020}
{Zhu}, F.-Y., {Wang}, J.-Z., {Liu}, T., {et~al.} 2020, \mnras, 499, 6018

\bibitem[{{Ziurys} {et~al.}(1989){Ziurys}, {Friberg}, \& {Irvine}}]{Ziurys1989_sio}
{Ziurys}, L.~M., {Friberg}, P., \& {Irvine}, W.~M. 1989, \apj, 343, 201

\end{thebibliography}

\newpage
\onecolumn

\begin{appendix}

\section{SiO 1 -- 0 spectral lines}
\begin{figure*}[h!]
    \centering
    \includegraphics[width=0.33\textwidth]{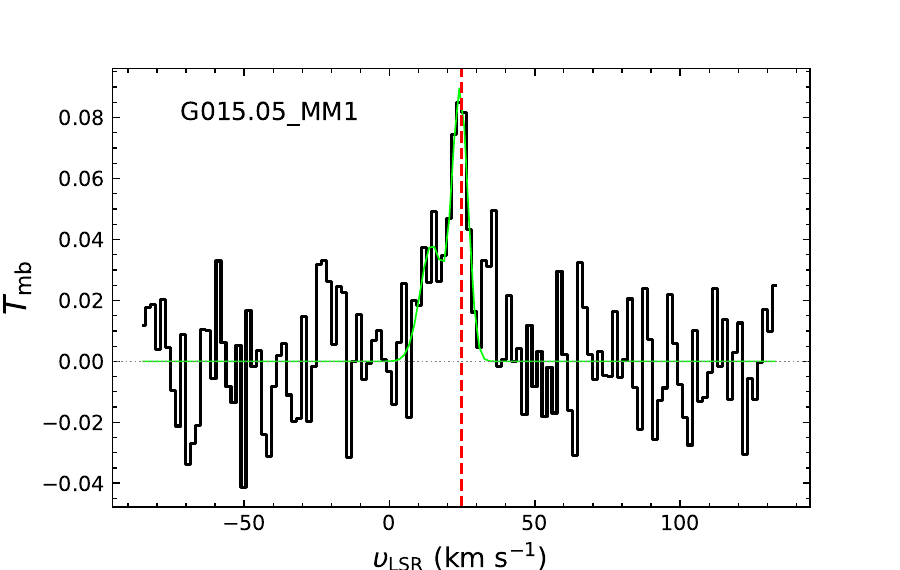}
    \includegraphics[width=0.33\textwidth]{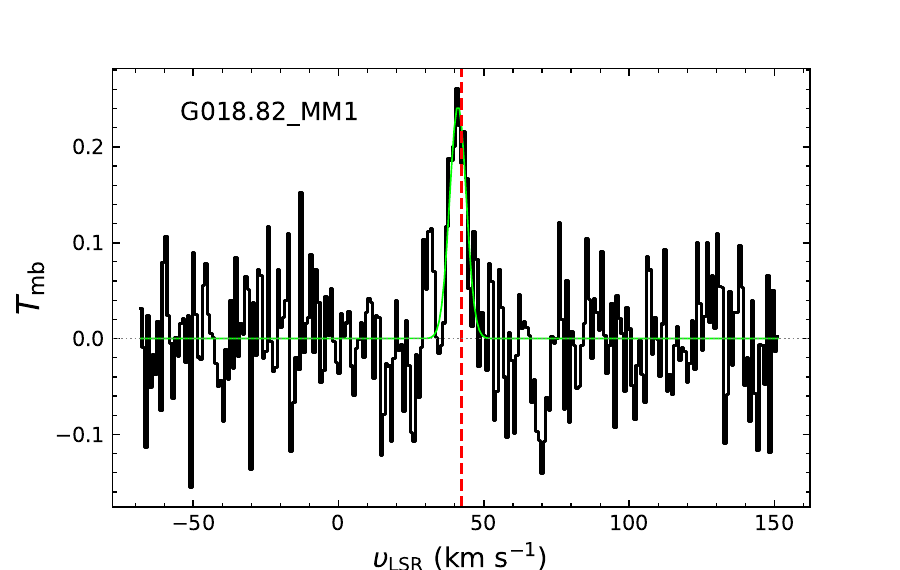}
    \includegraphics[width=0.33\textwidth]{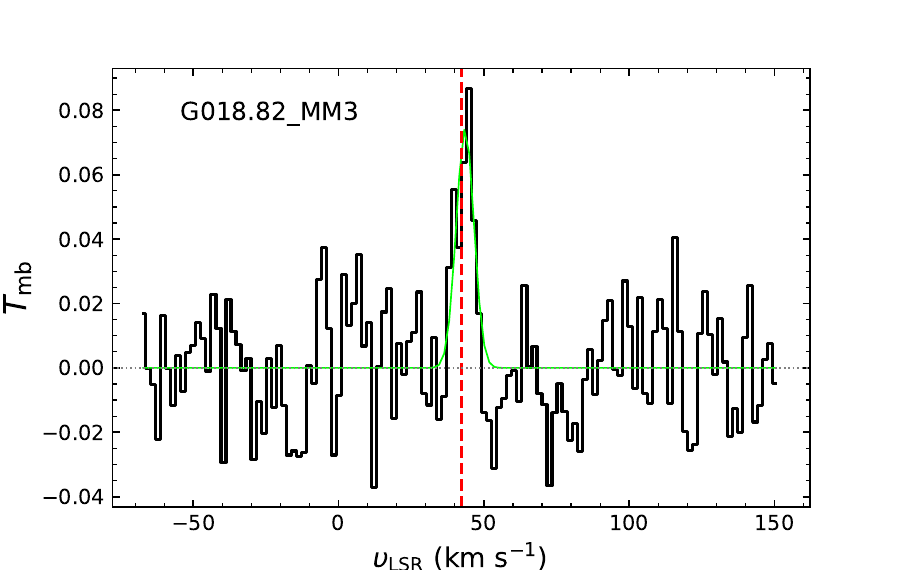}
    \includegraphics[width=0.33\textwidth]{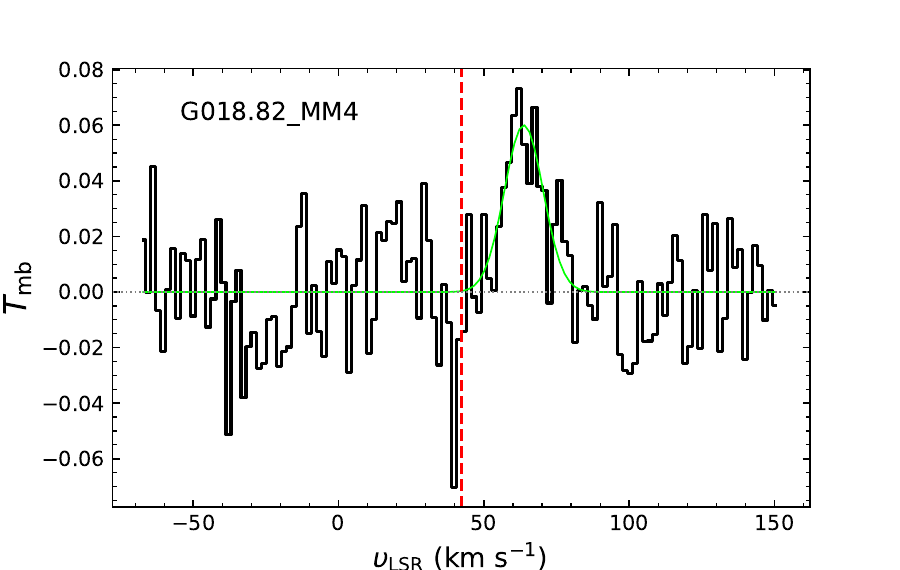} 
    \includegraphics[width=0.33\textwidth]{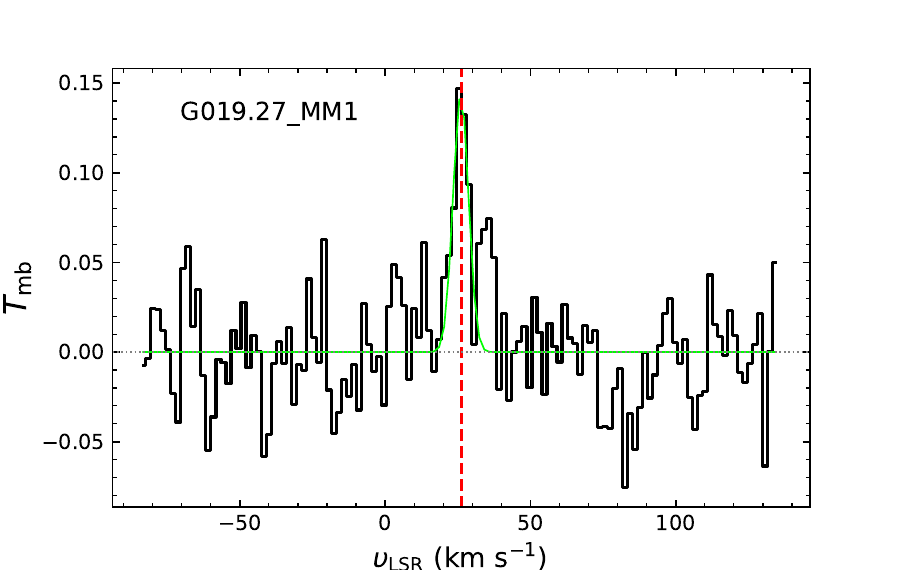}
    \includegraphics[width=0.33\textwidth]{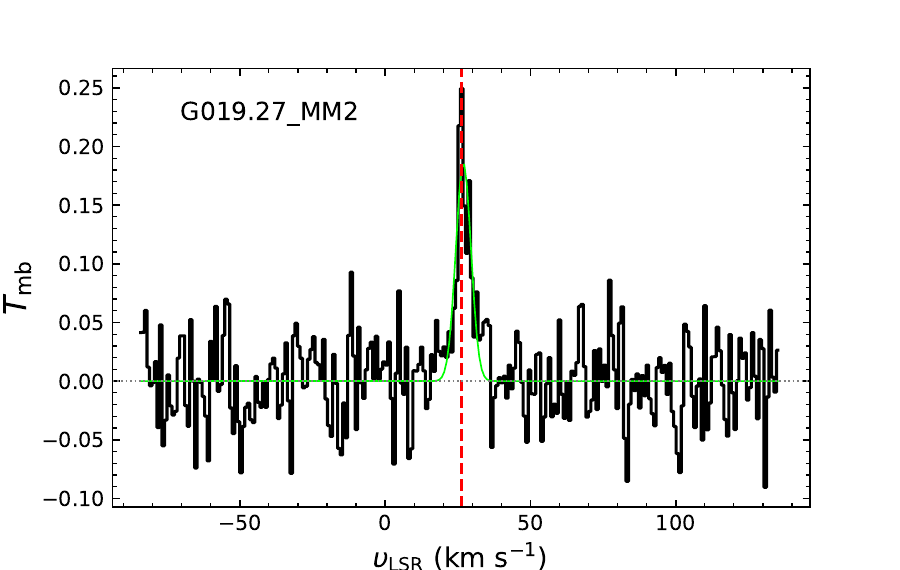}
    \includegraphics[width=0.33\textwidth]{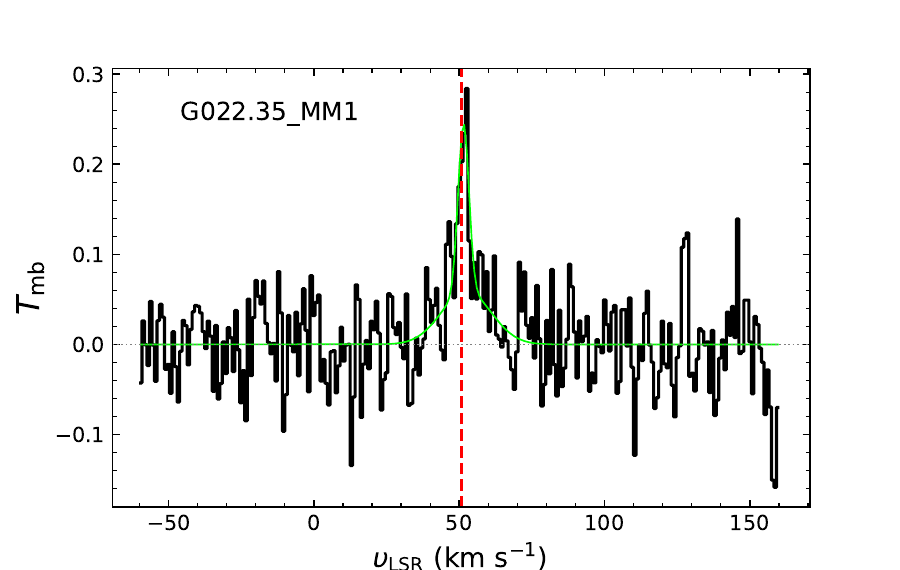}
    \includegraphics[width=0.33\textwidth]{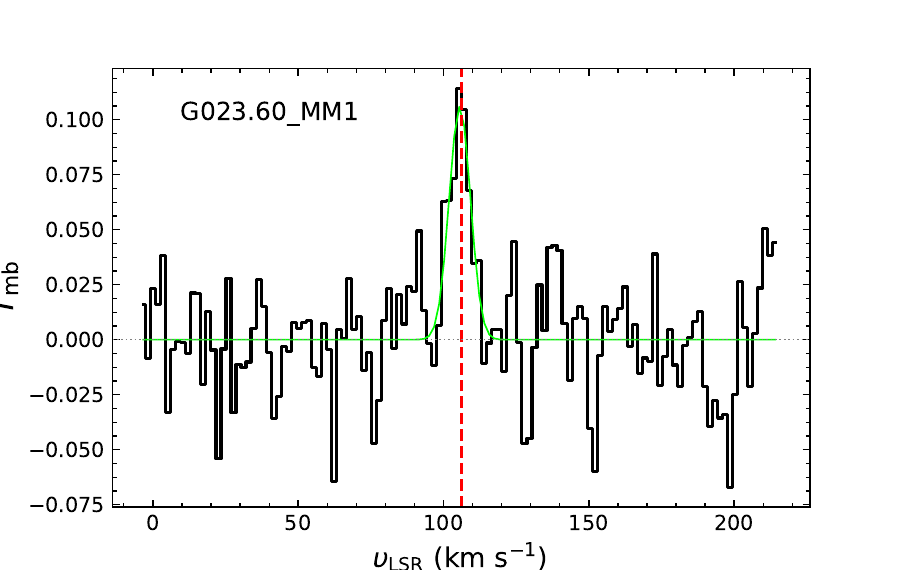}
    \includegraphics[width=0.33\textwidth]{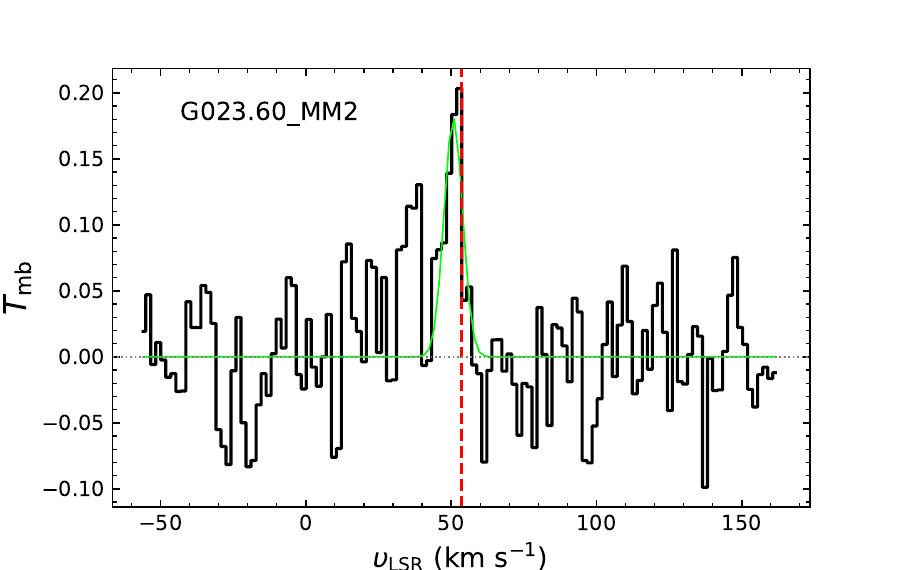}
    \includegraphics[width=0.33\textwidth]{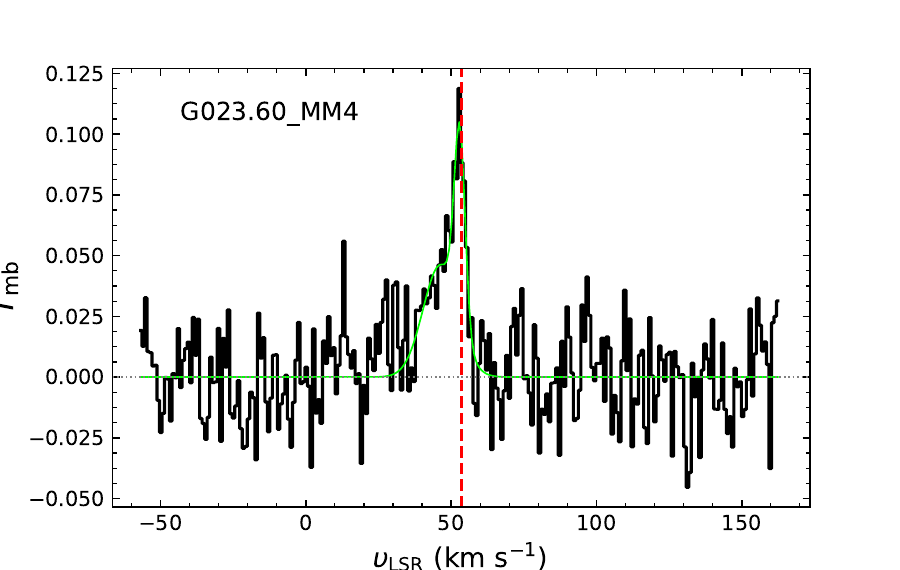}
    \includegraphics[width=0.33\textwidth]{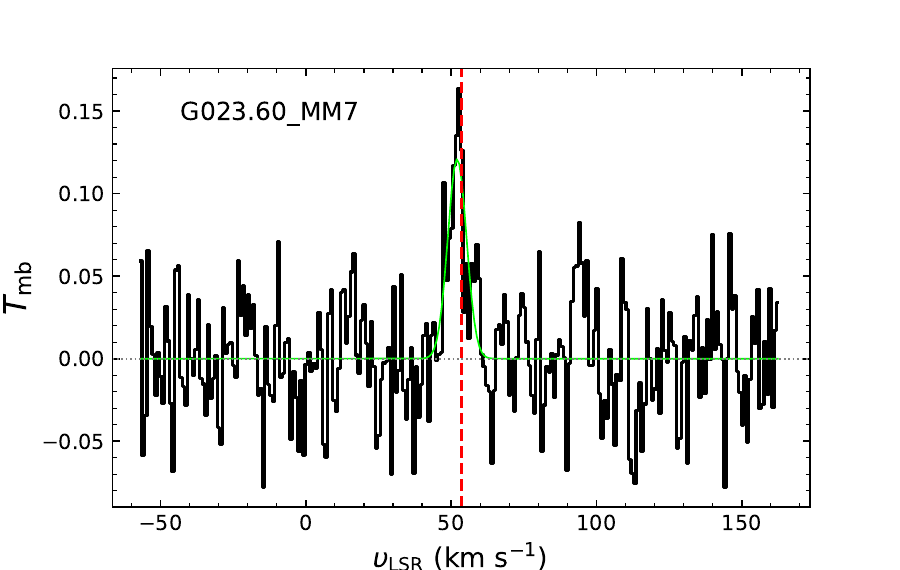} 
    \includegraphics[width=0.33\textwidth]{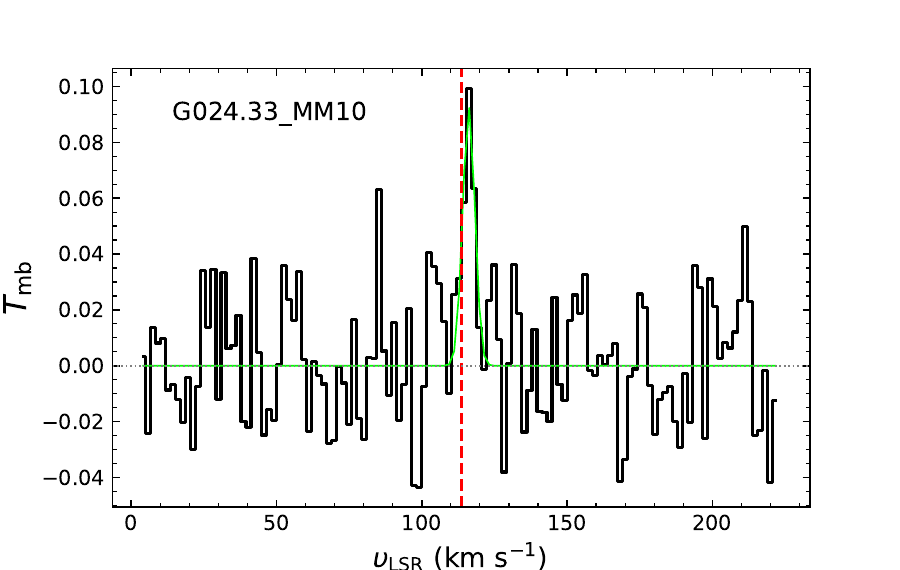} 
    \includegraphics[width=0.33\textwidth]{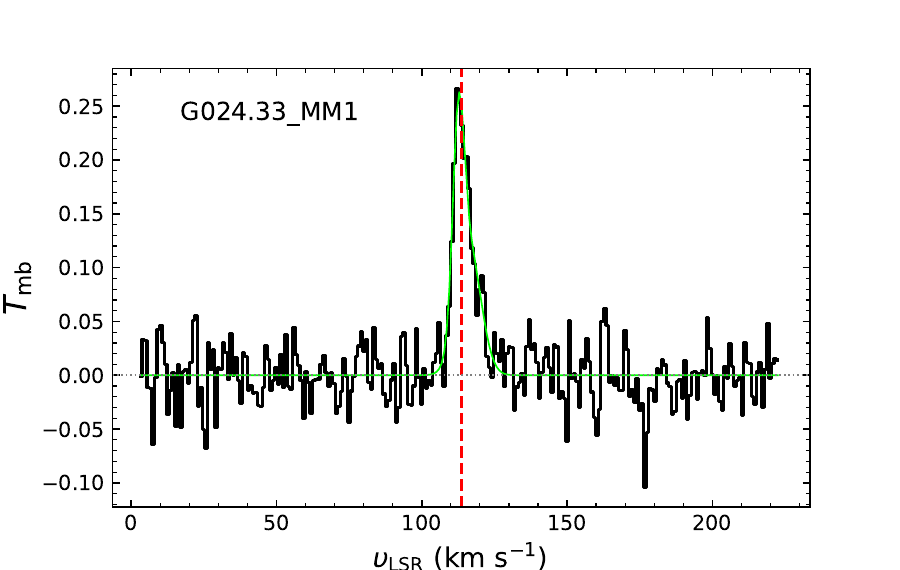}     
    \includegraphics[width=0.33\textwidth]{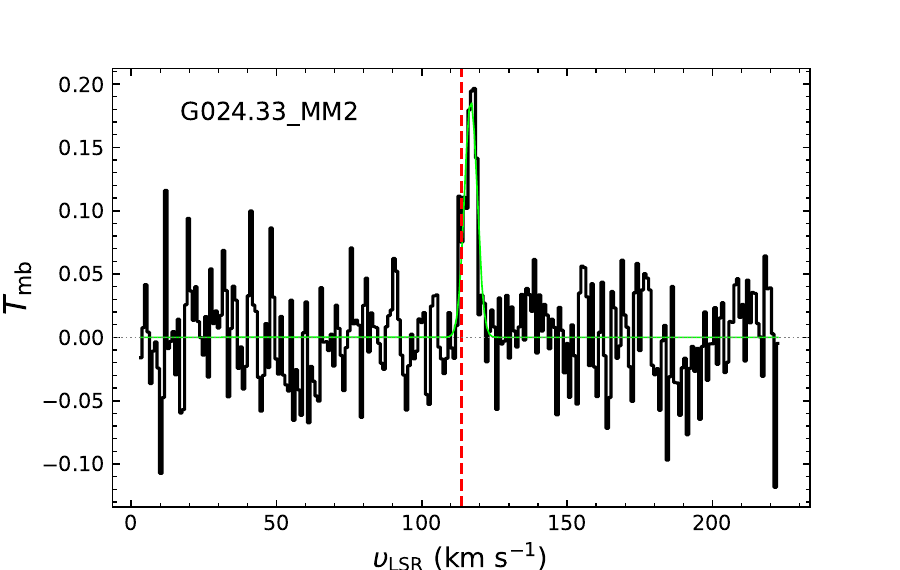}
    \includegraphics[width=0.33\textwidth]{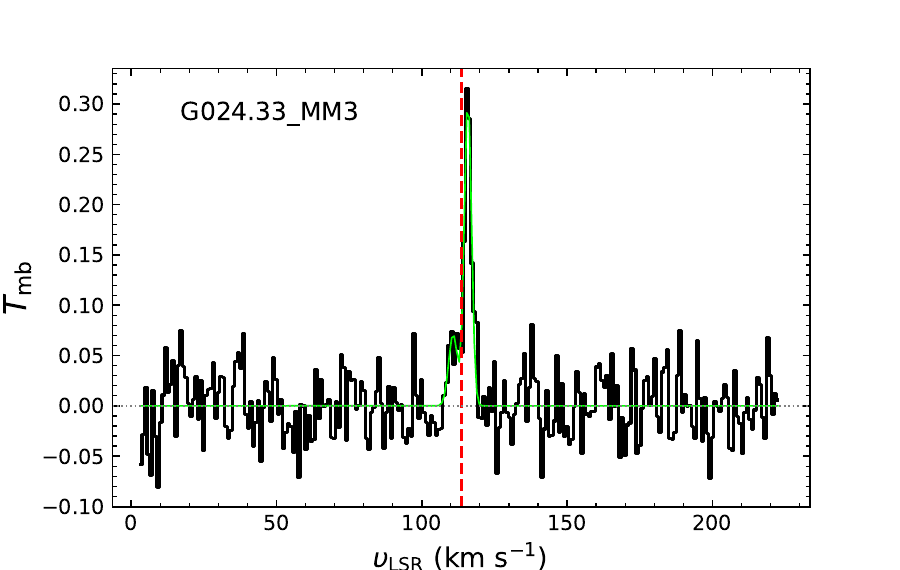}  
    \caption{43\,GHz SiO $J=1-0$ spectra in a $T_{\rm mb}$ scale, toward sources from the three different catalogs that are IRDC, HMPO, and \uchii.  }
    \label{appedix:sio_spectra1}
\end{figure*}

\begin{figure*}[h!]
    \centering
    \includegraphics[width=0.33\textwidth]{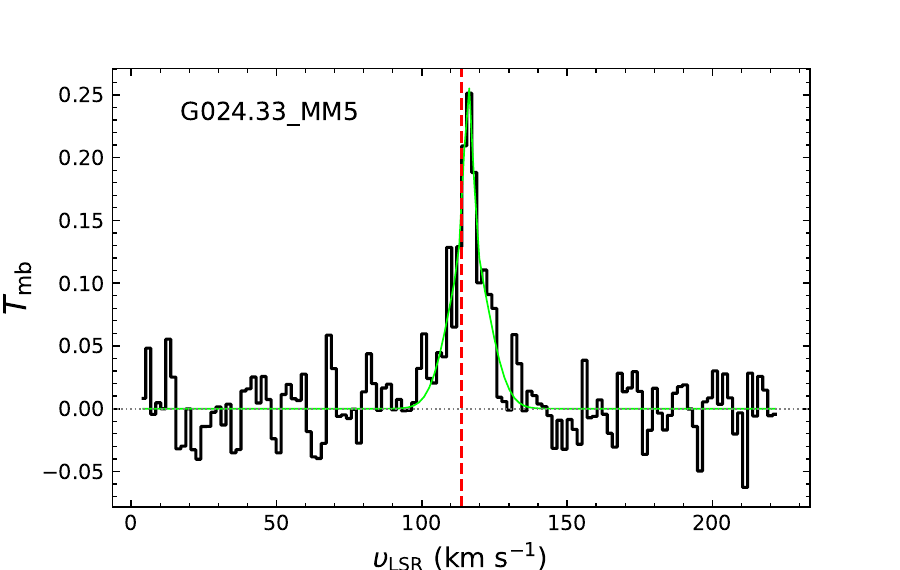}
    \includegraphics[width=0.33\textwidth]{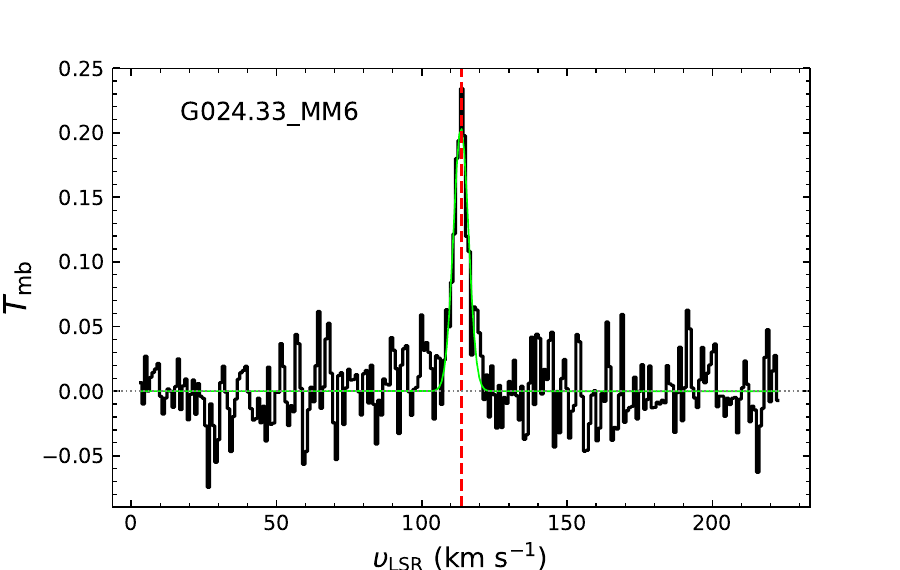}
    \includegraphics[width=0.33\textwidth]{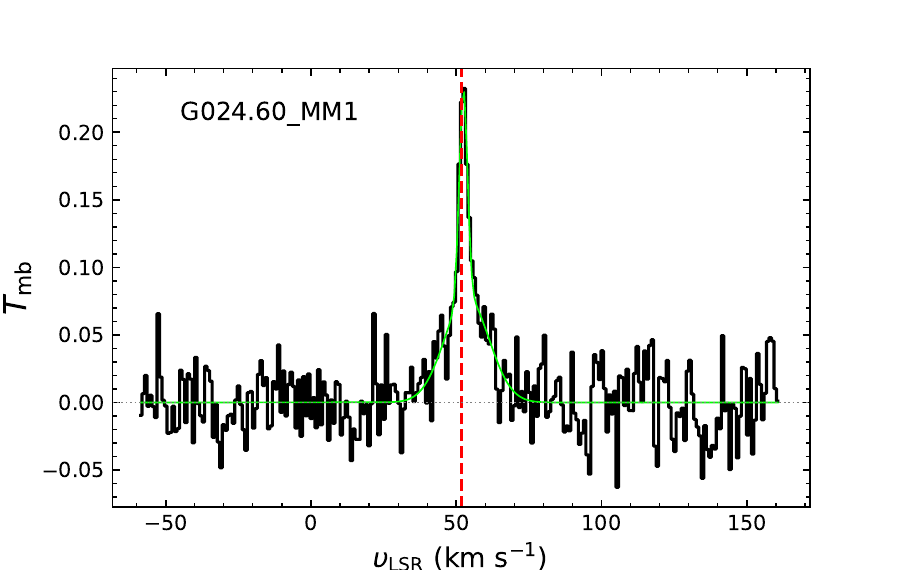}
    \includegraphics[width=0.33\textwidth]{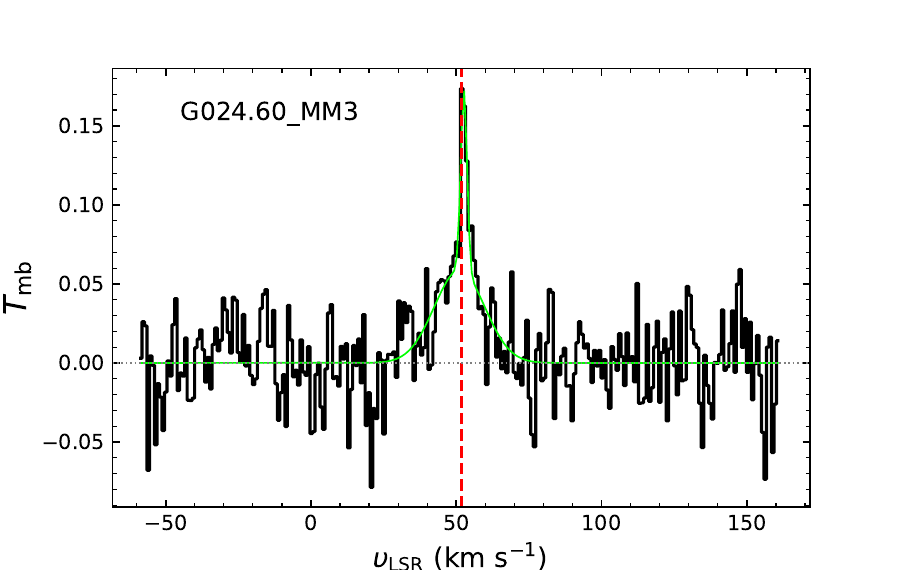}
    \includegraphics[width=0.33\textwidth]{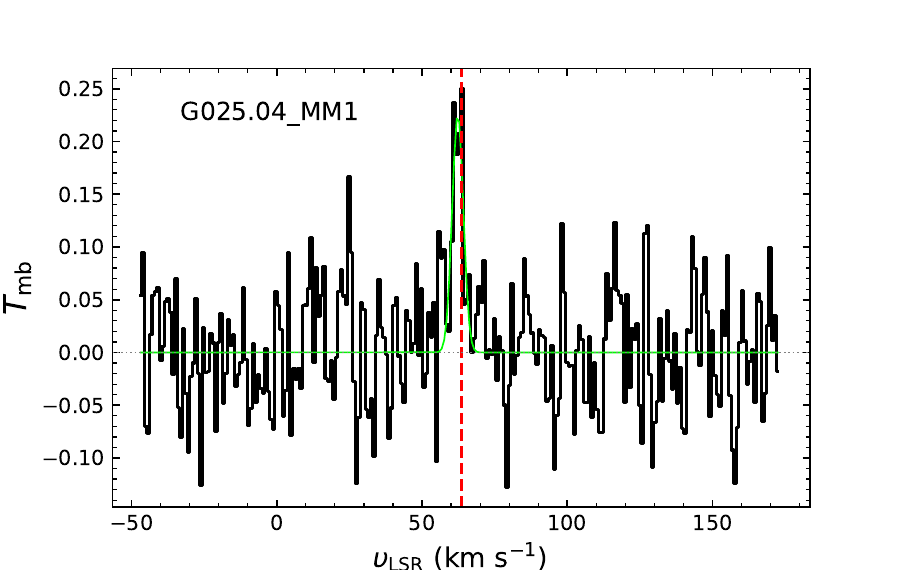}
    \includegraphics[width=0.33\textwidth]{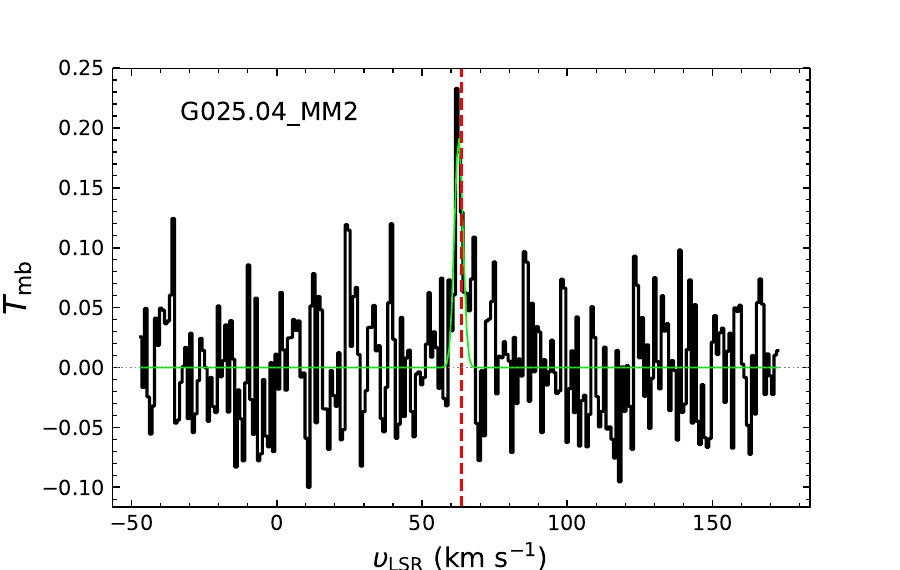}
    \includegraphics[width=0.33\textwidth]{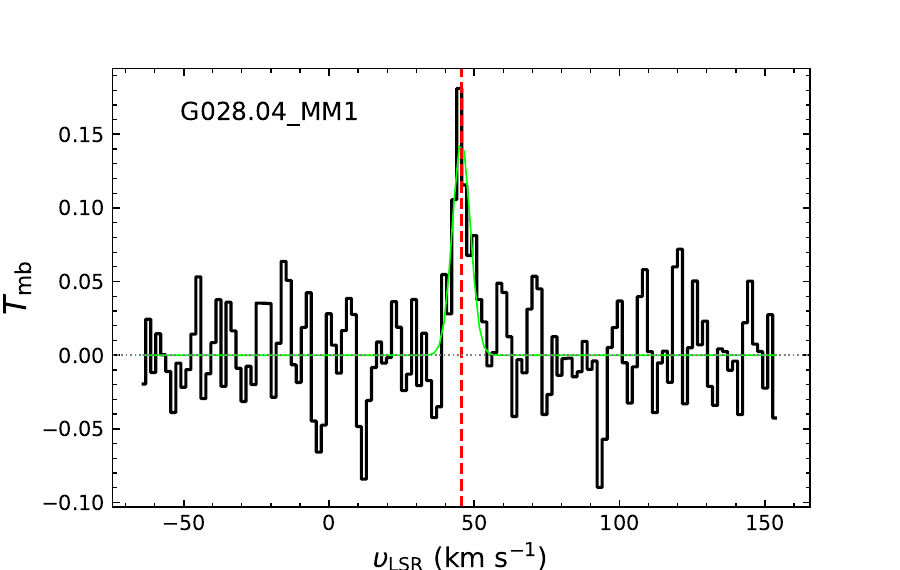}
    \includegraphics[width=0.33\textwidth]{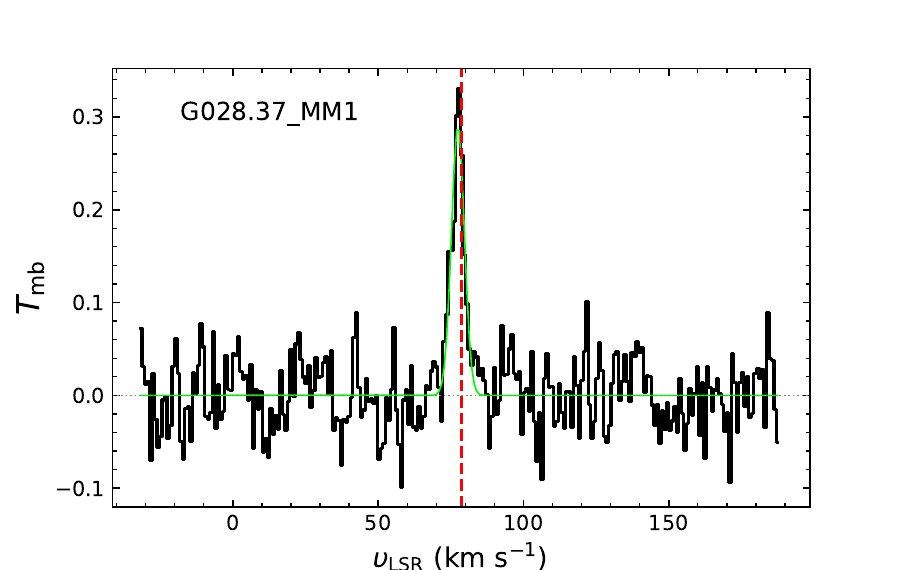}
    \includegraphics[width=0.33\textwidth]{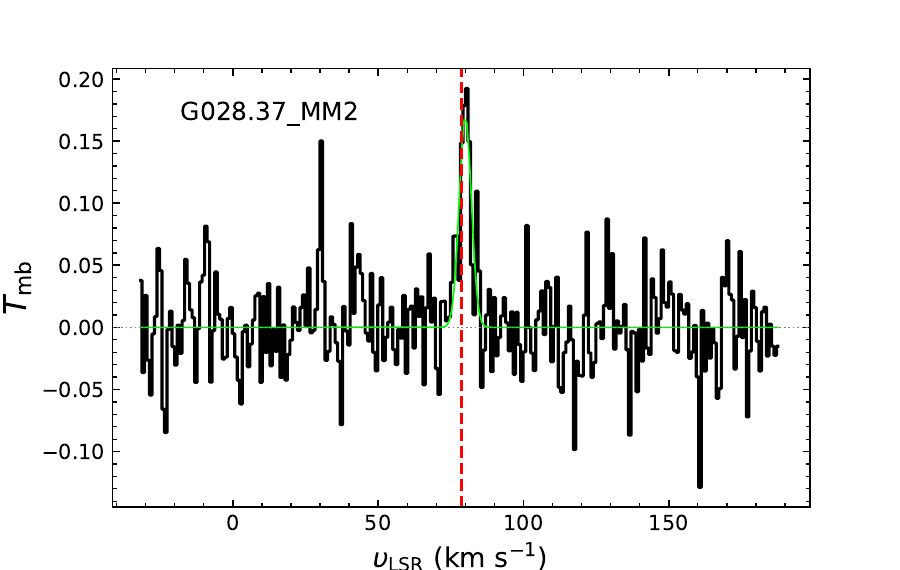}
    \includegraphics[width=0.33\textwidth]{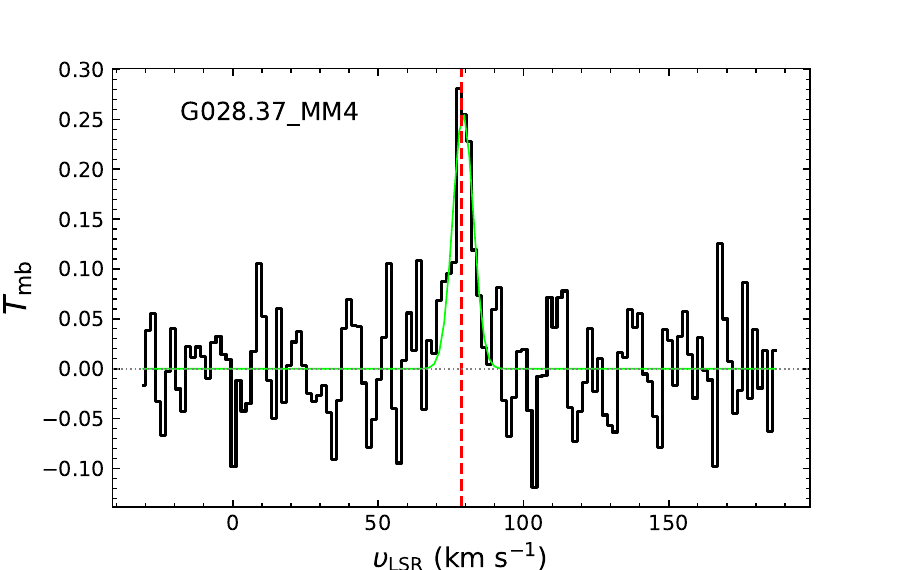}
    \includegraphics[width=0.33\textwidth]{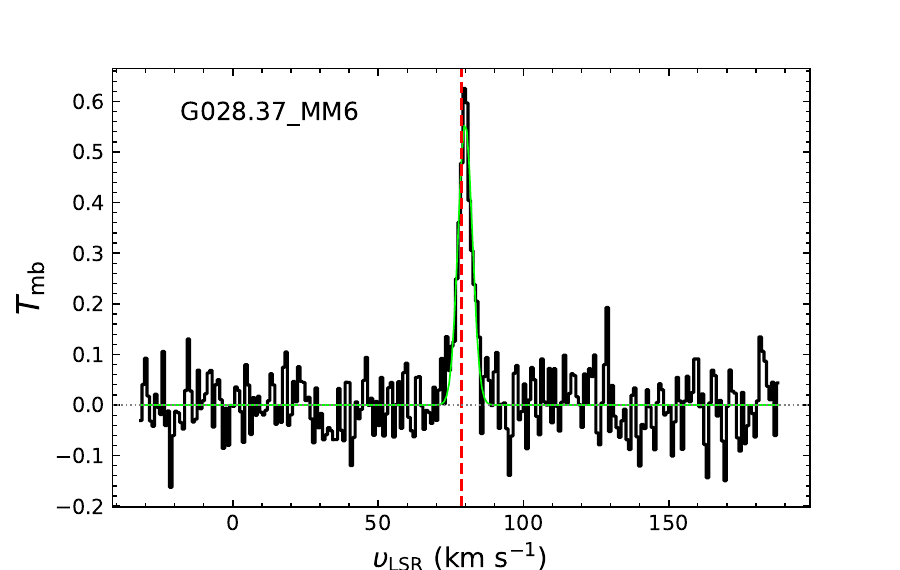}
    \includegraphics[width=0.33\textwidth]{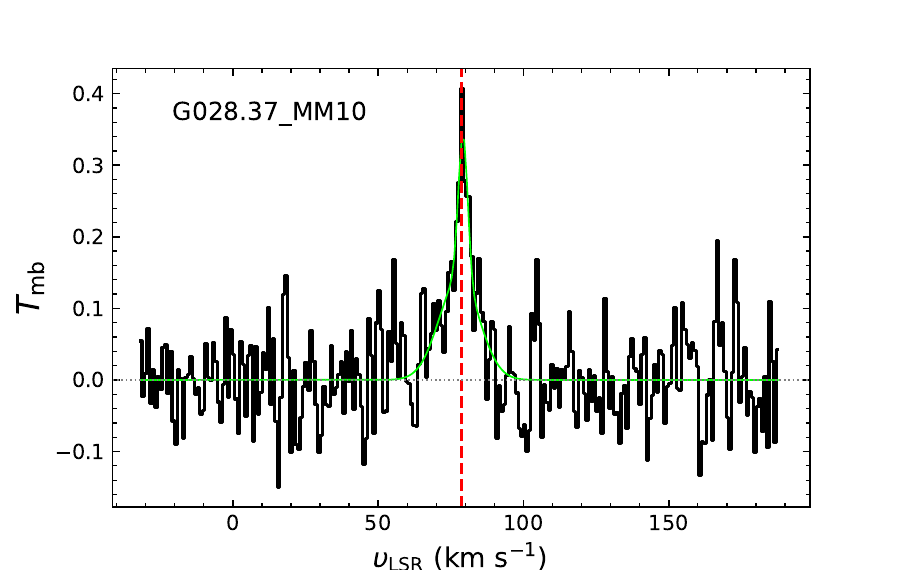}
    \includegraphics[width=0.33\textwidth]{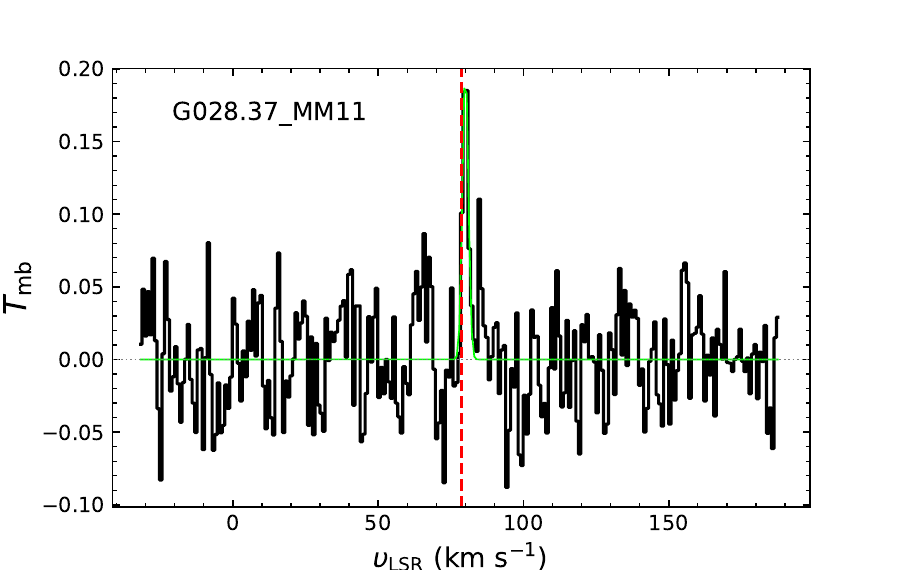}
    \includegraphics[width=0.33\textwidth]{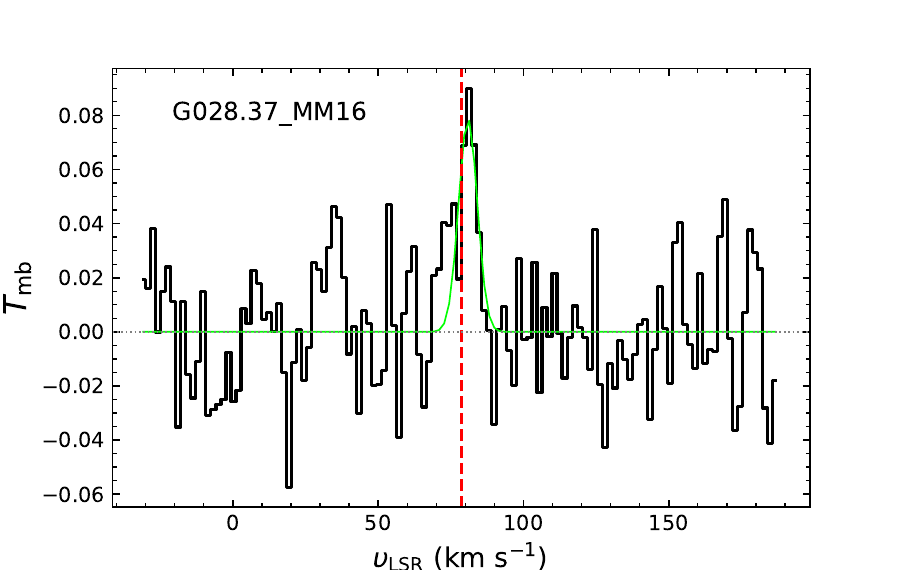}
    \includegraphics[width=0.33\textwidth]{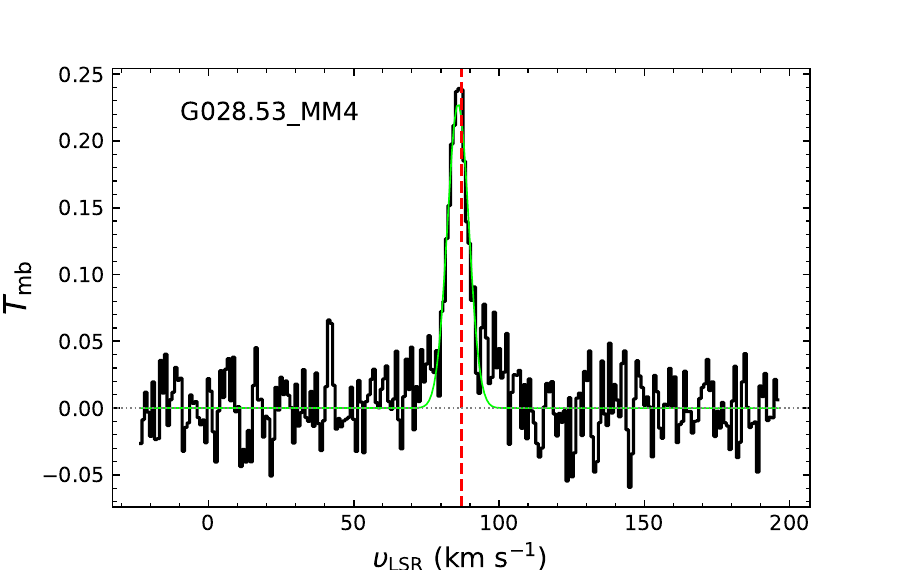}
    \includegraphics[width=0.33\textwidth]{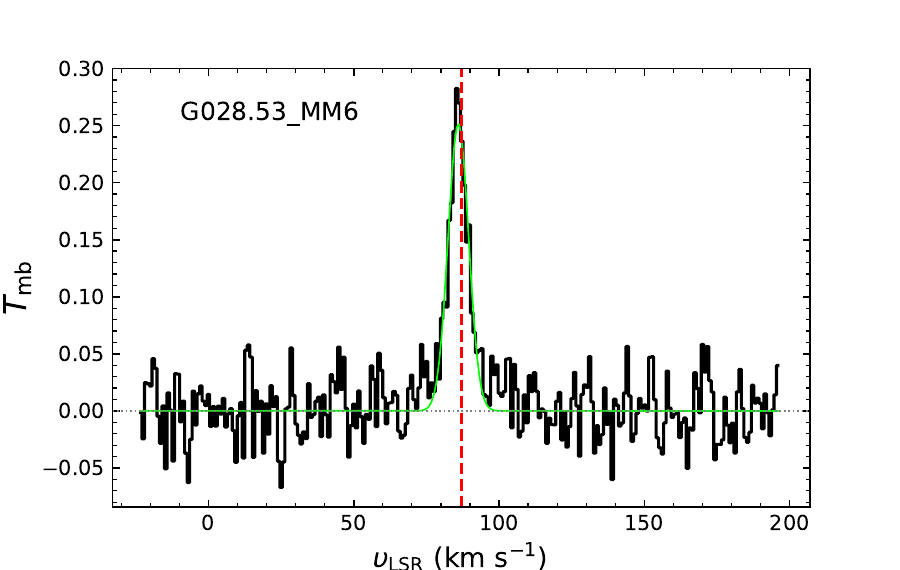}
    \includegraphics[width=0.33\textwidth]{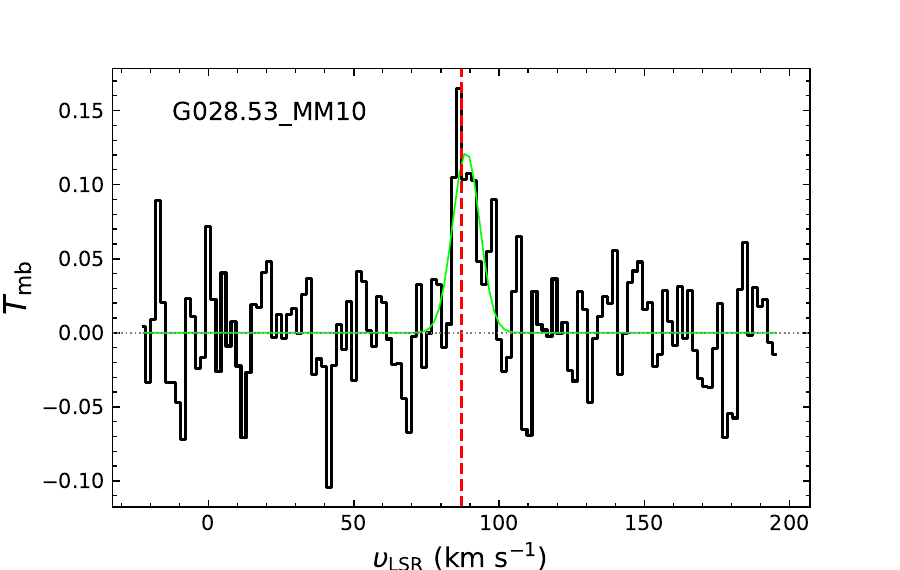}
    \includegraphics[width=0.33\textwidth]{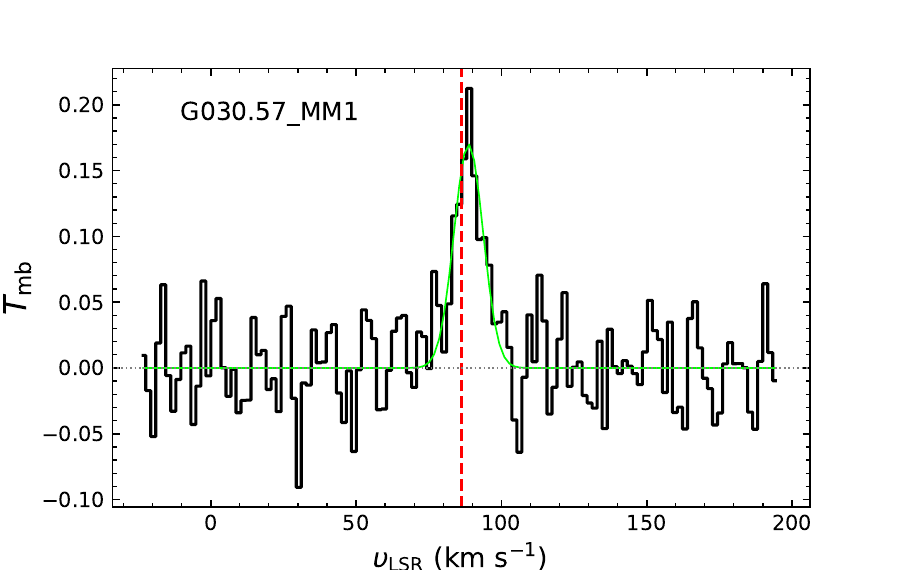}
    \caption{Continuation of Fig.\,\ref{appedix:sio_spectra1} }
    \label{appendix:sio_spectra2}
\end{figure*}

\begin{figure*}[h!]
    \centering
    \includegraphics[width=0.33\textwidth]{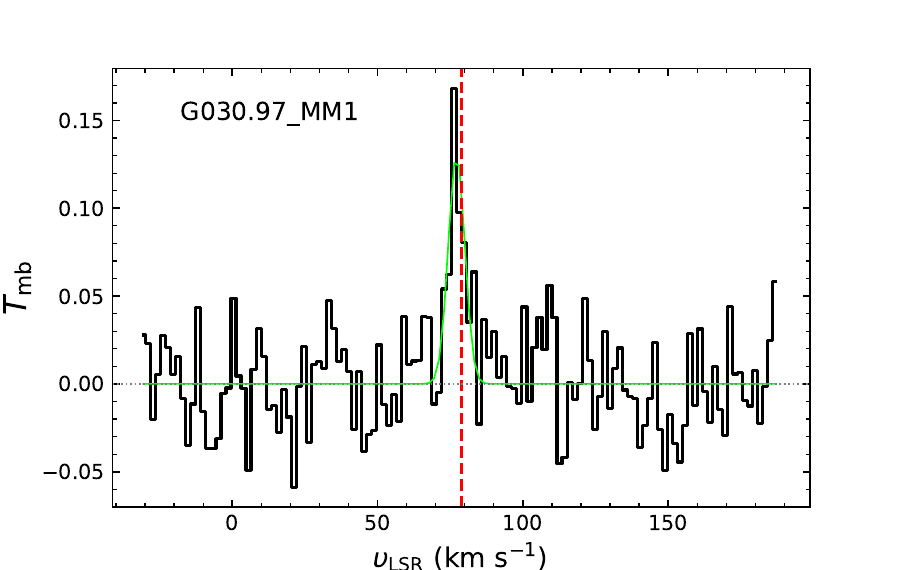}
    \includegraphics[width=0.33\textwidth]{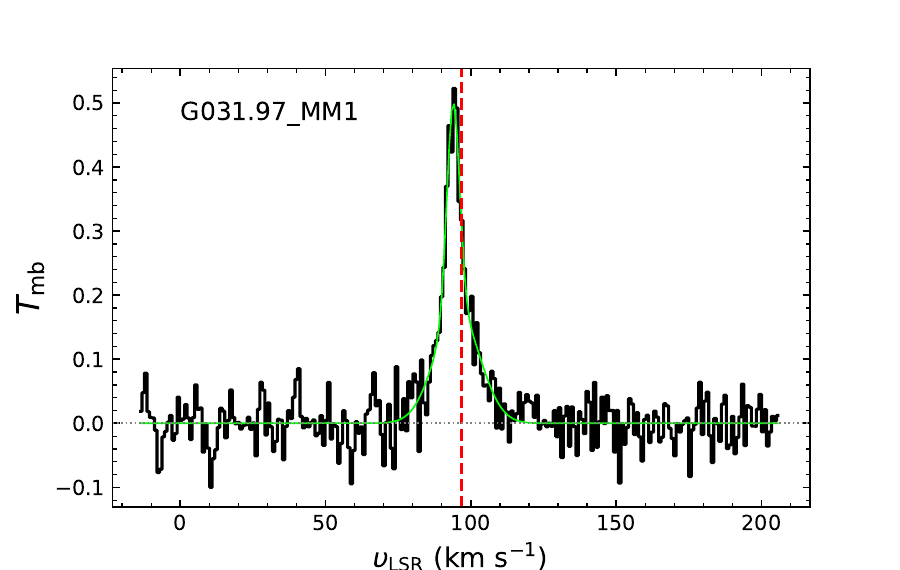}
    \includegraphics[width=0.33\textwidth]{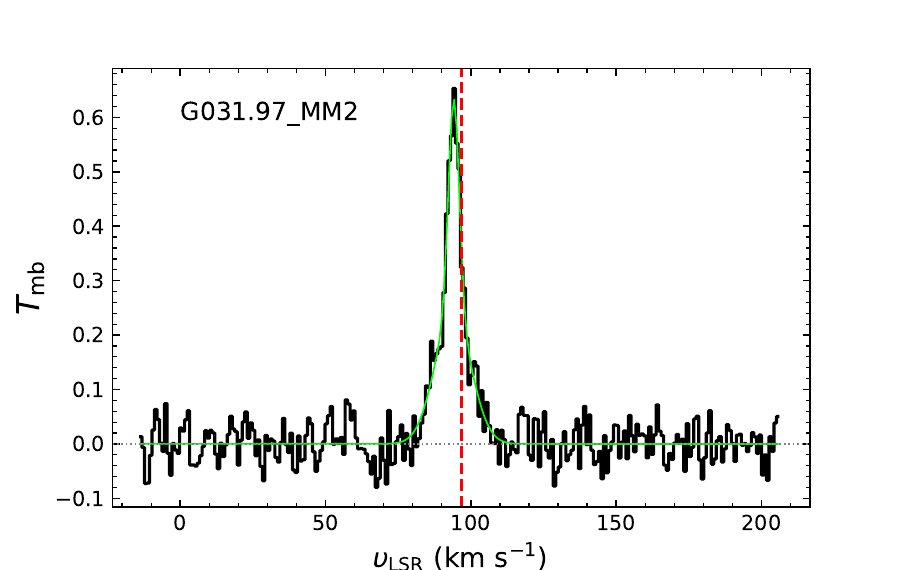}
    \includegraphics[width=0.33\textwidth]{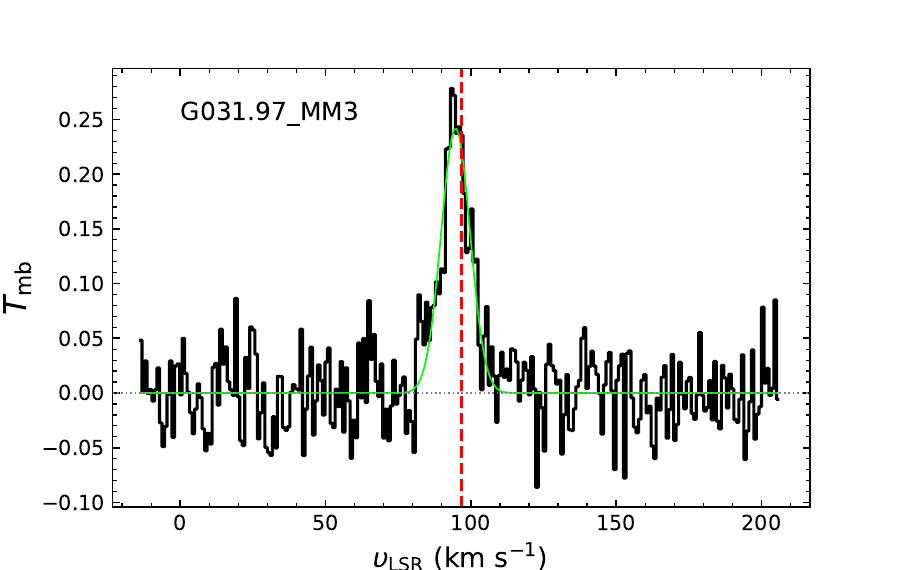}
    \includegraphics[width=0.33\textwidth]{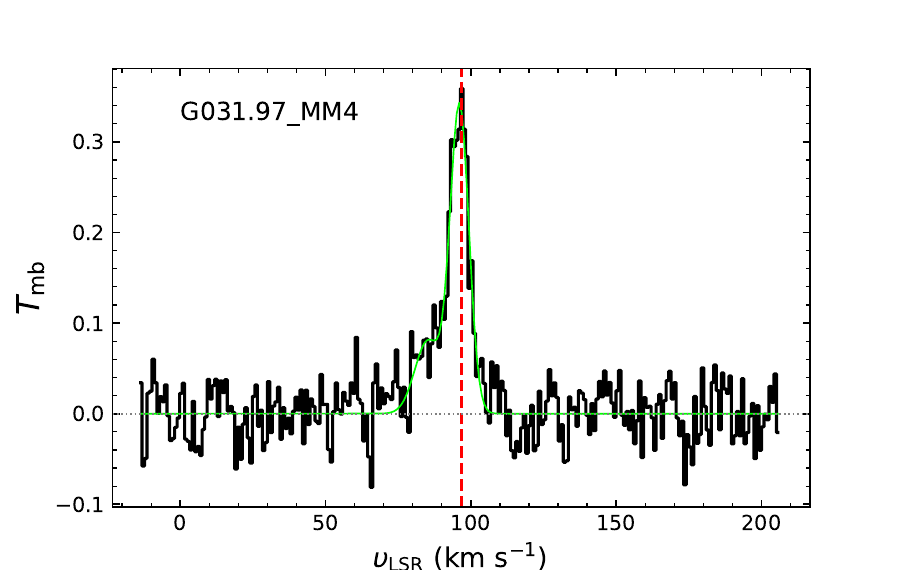}
    \includegraphics[width=0.33\textwidth]{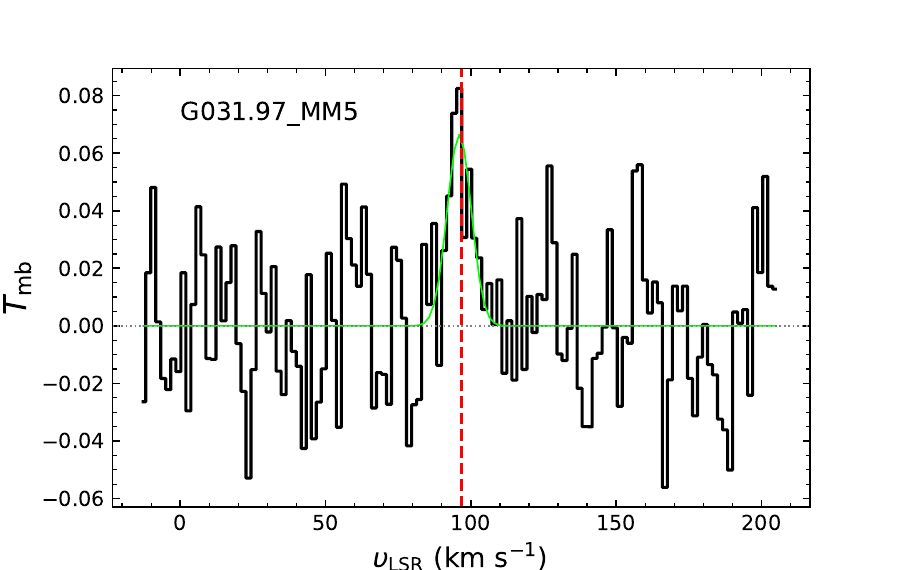}
    \includegraphics[width=0.33\textwidth]{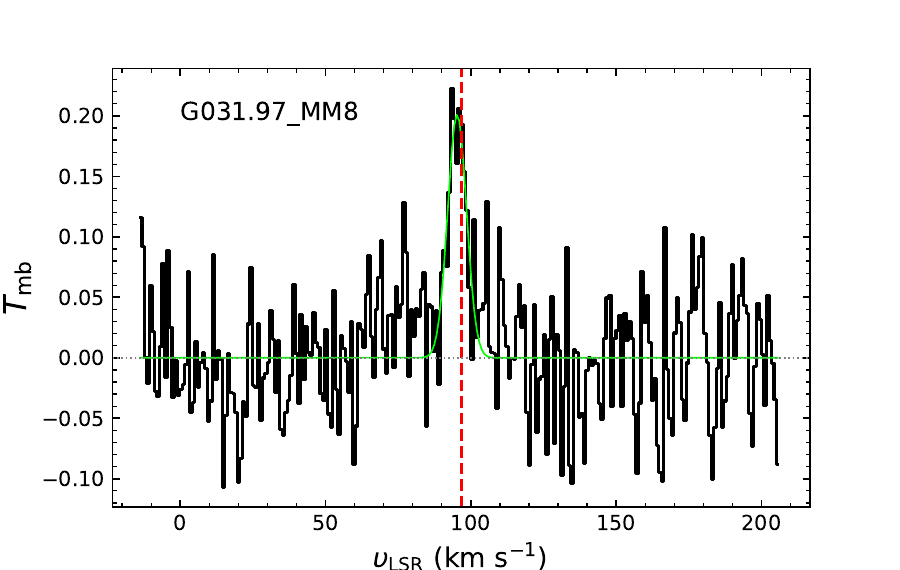}
    \includegraphics[width=0.33\textwidth]{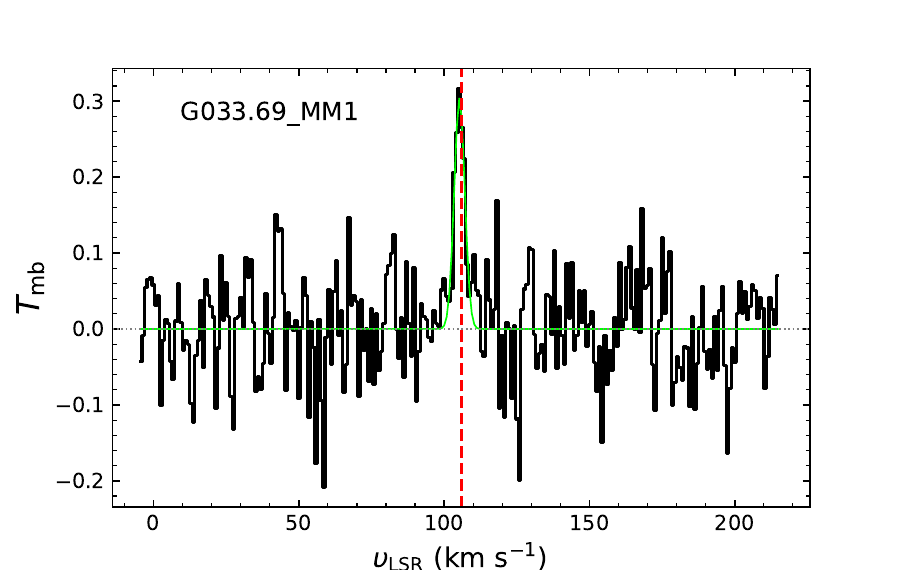}
    \includegraphics[width=0.33\textwidth]{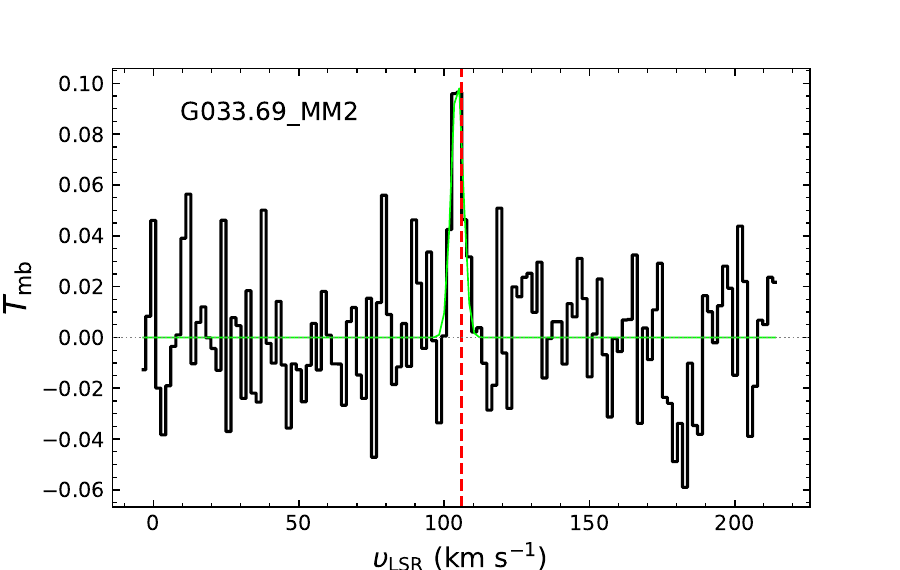}
    \includegraphics[width=0.33\textwidth]{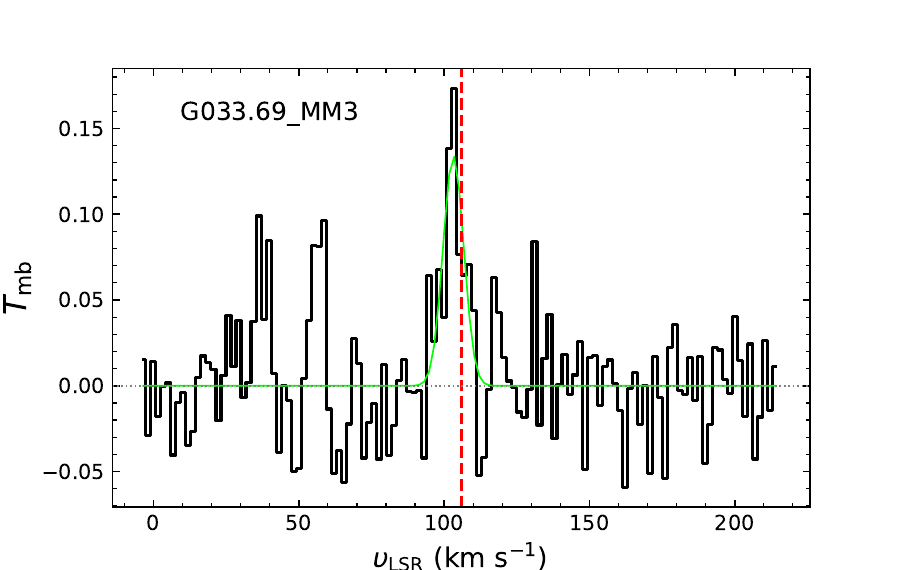}
    \includegraphics[width=0.33\textwidth]{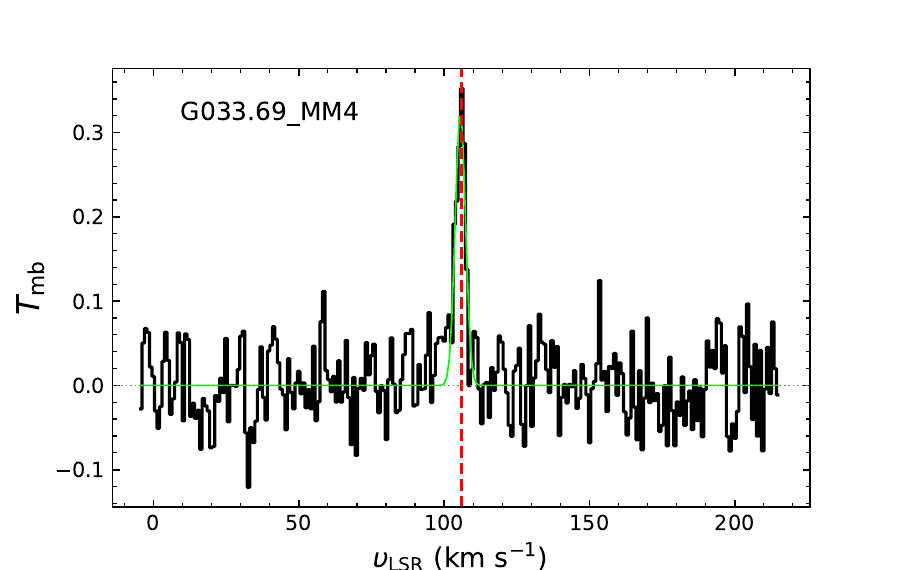}
    \includegraphics[width=0.33\textwidth]{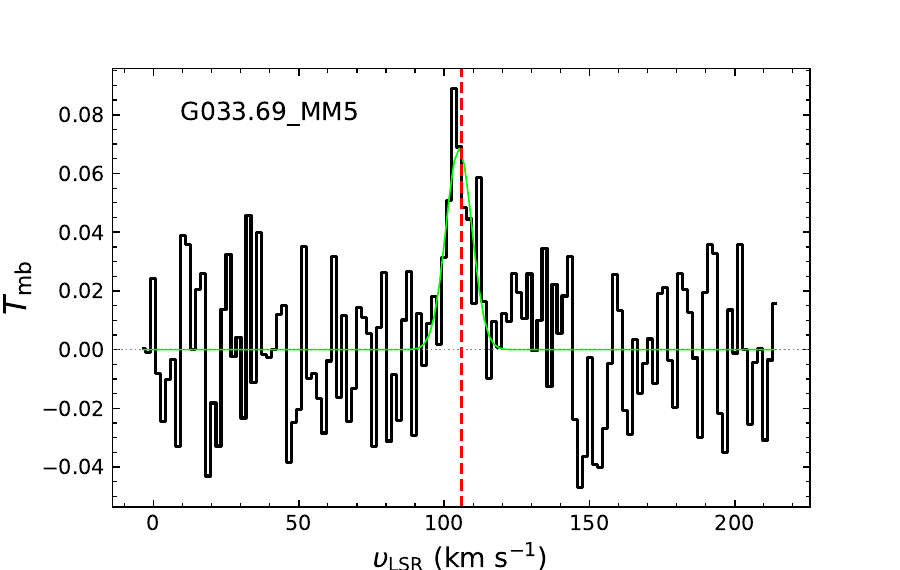}
    \includegraphics[width=0.33\textwidth]{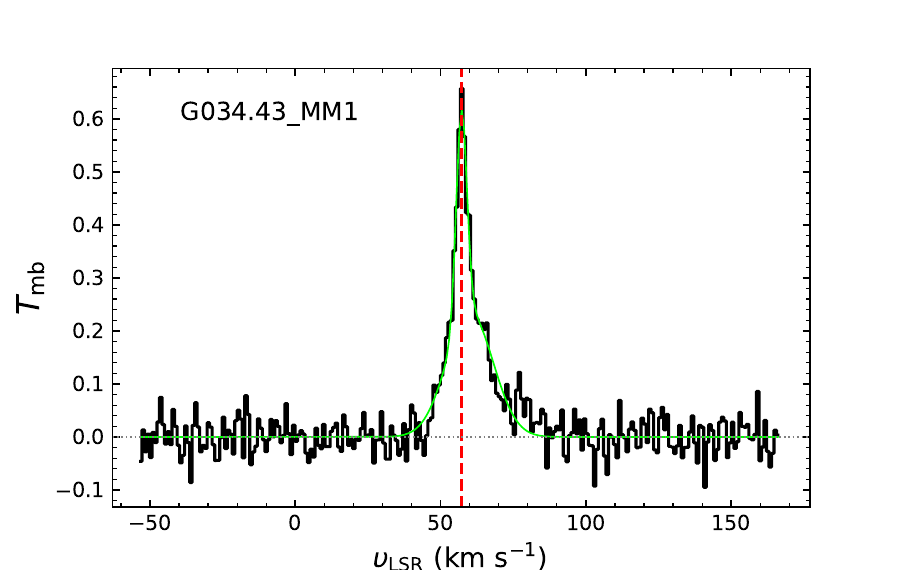}
    \includegraphics[width=0.33\textwidth]{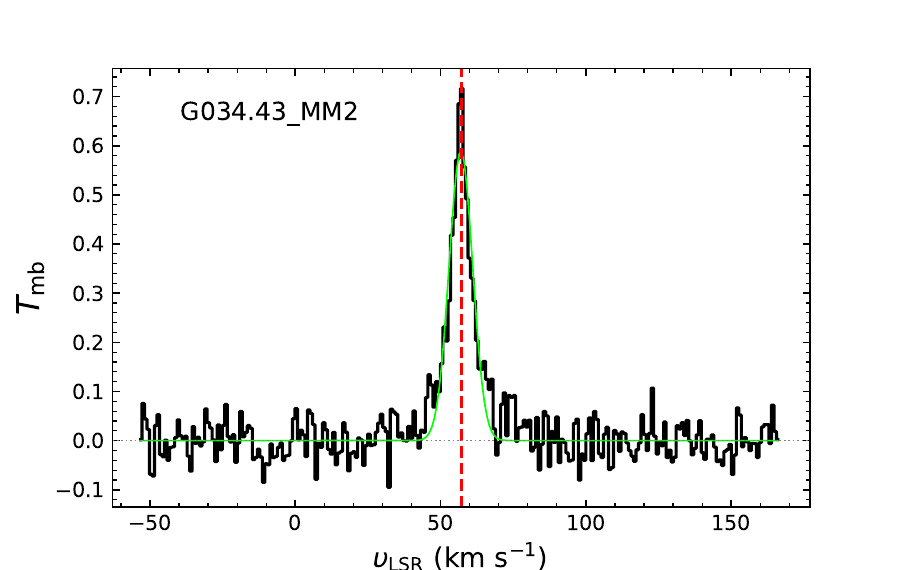}
    \includegraphics[width=0.33\textwidth]{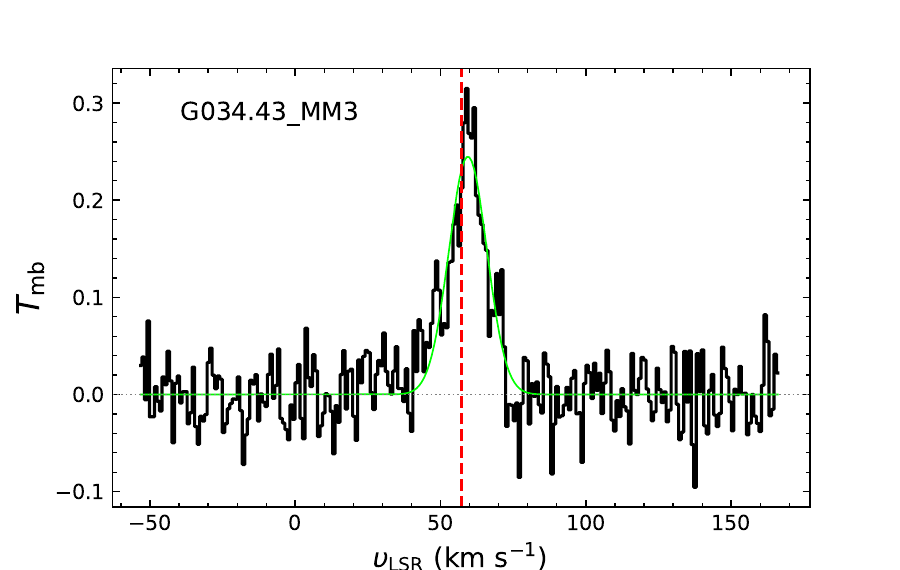}
    \includegraphics[width=0.33\textwidth]{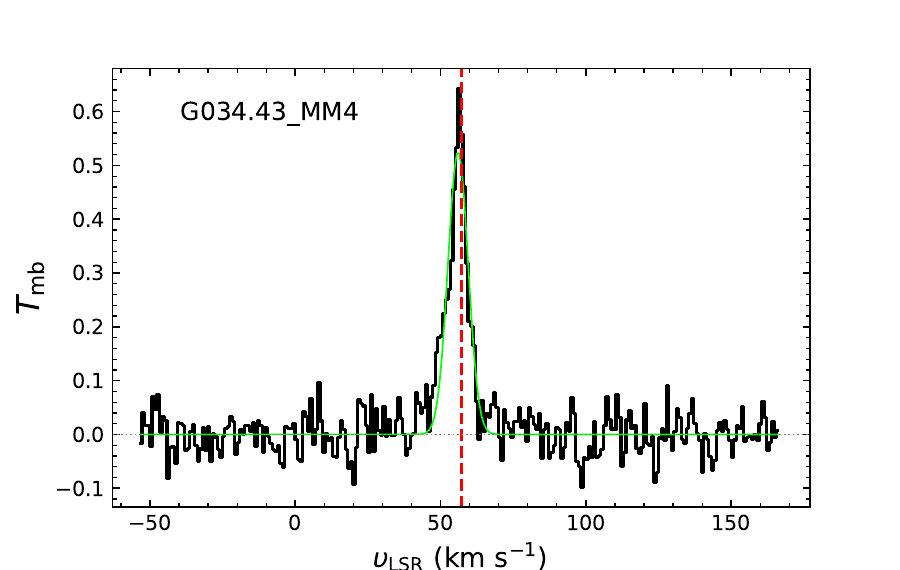}
    \includegraphics[width=0.33\textwidth]{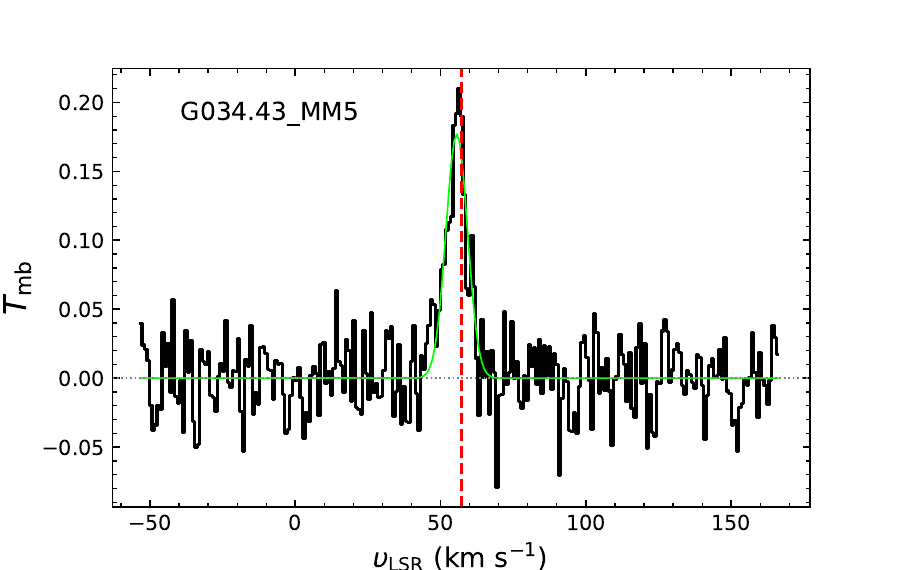}
    \includegraphics[width=0.33\textwidth]{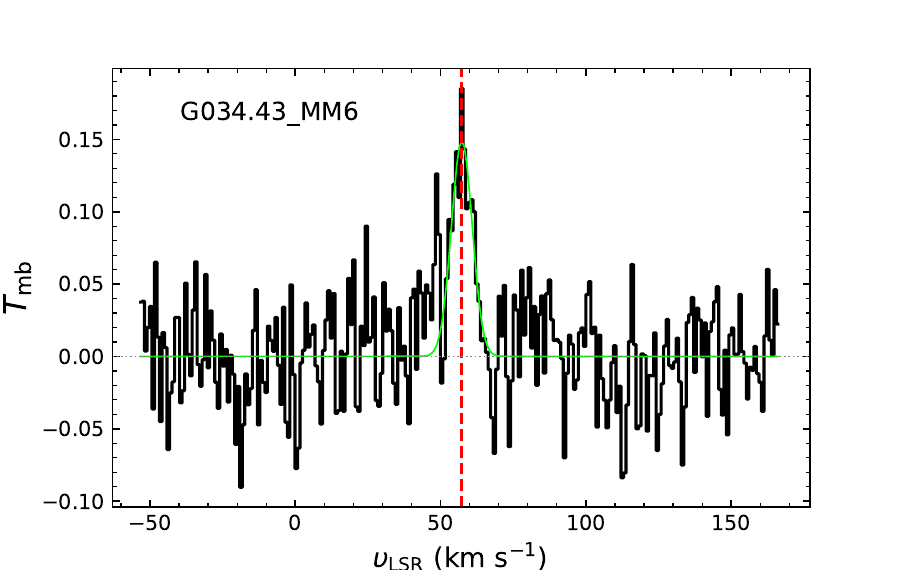}
    \caption{Continuation of Fig.\,\ref{appedix:sio_spectra1}} 
    \label{appendix:sio_spectra3}
\end{figure*}

\begin{figure*}
    \centering
    \includegraphics[width=0.33\textwidth]{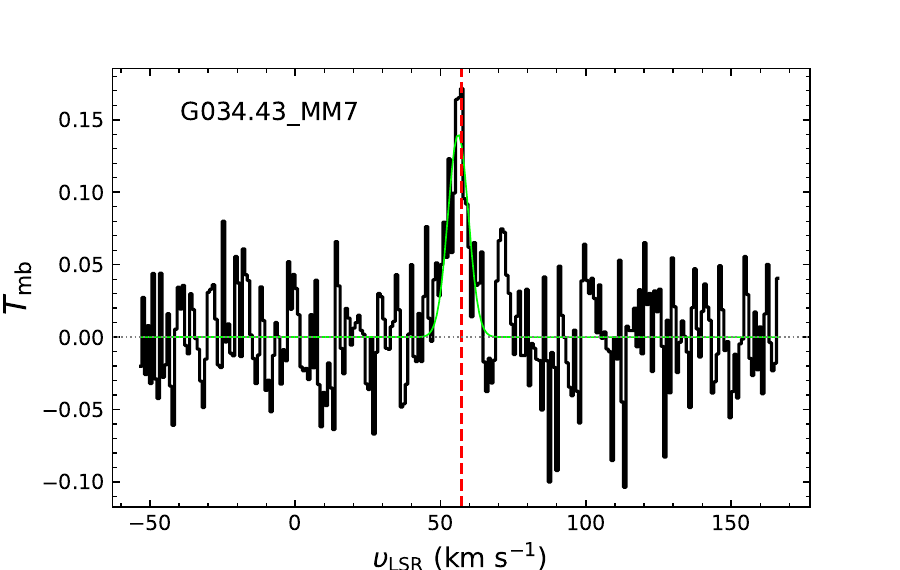}
    \includegraphics[width=0.33\textwidth]{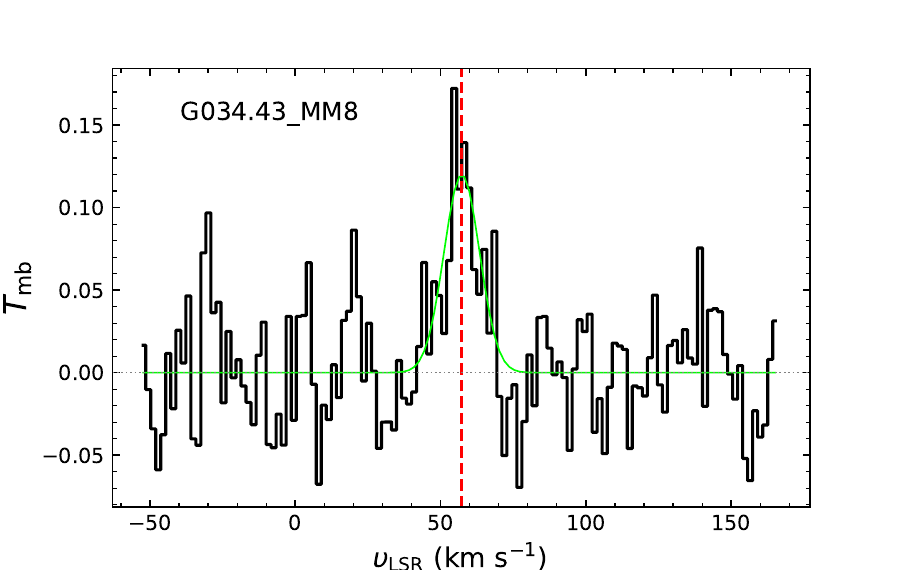}
    \includegraphics[width=0.33\textwidth]{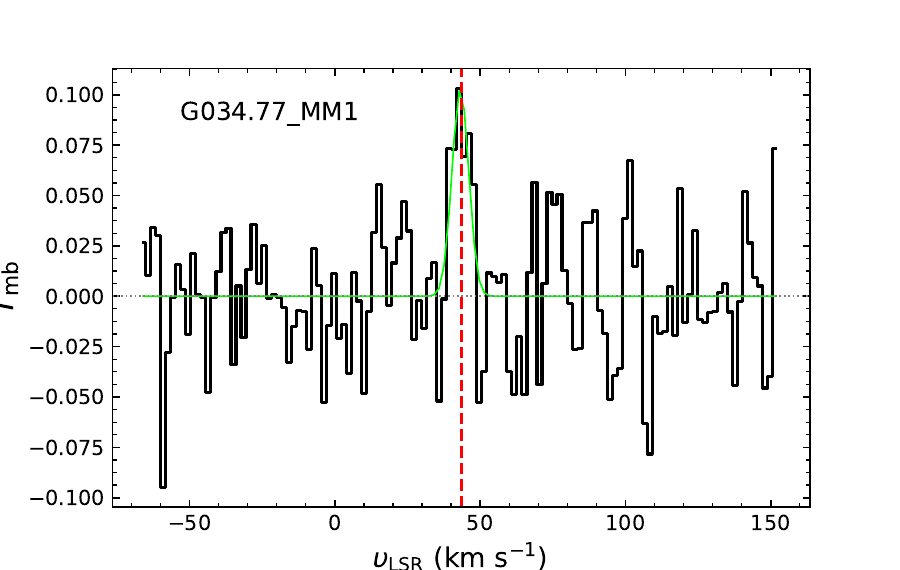}
    \includegraphics[width=0.33\textwidth]{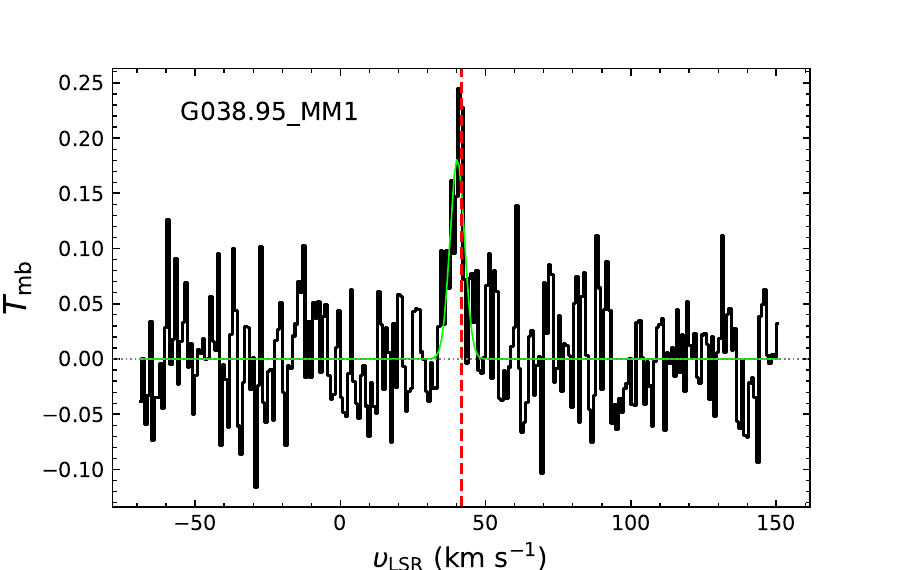}
    \includegraphics[width=0.33\textwidth]{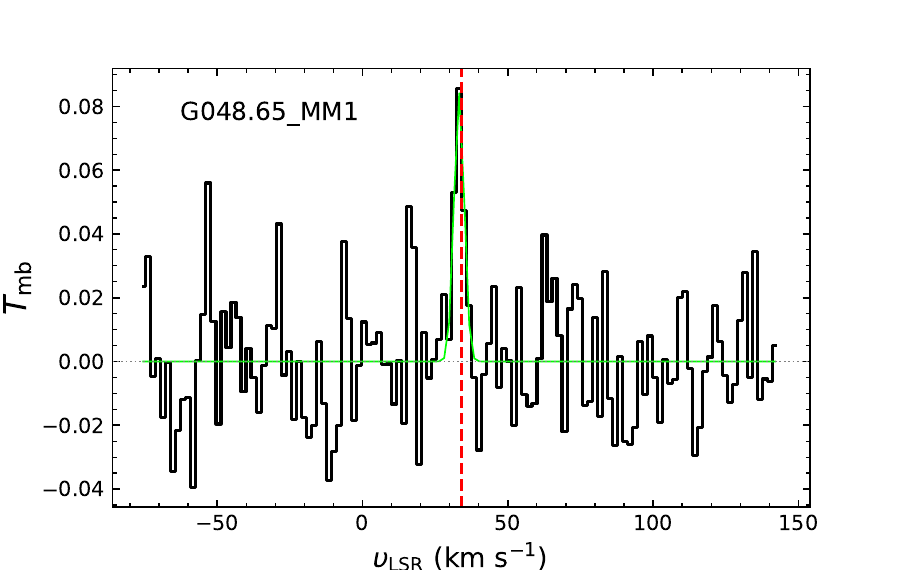}
    \includegraphics[width=0.33\textwidth]{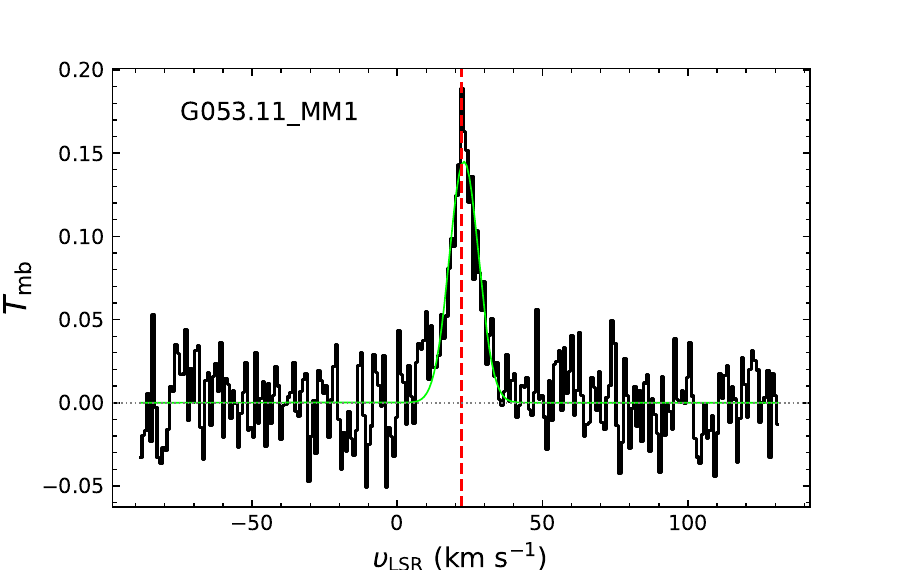}
    \includegraphics[width=0.33\textwidth]{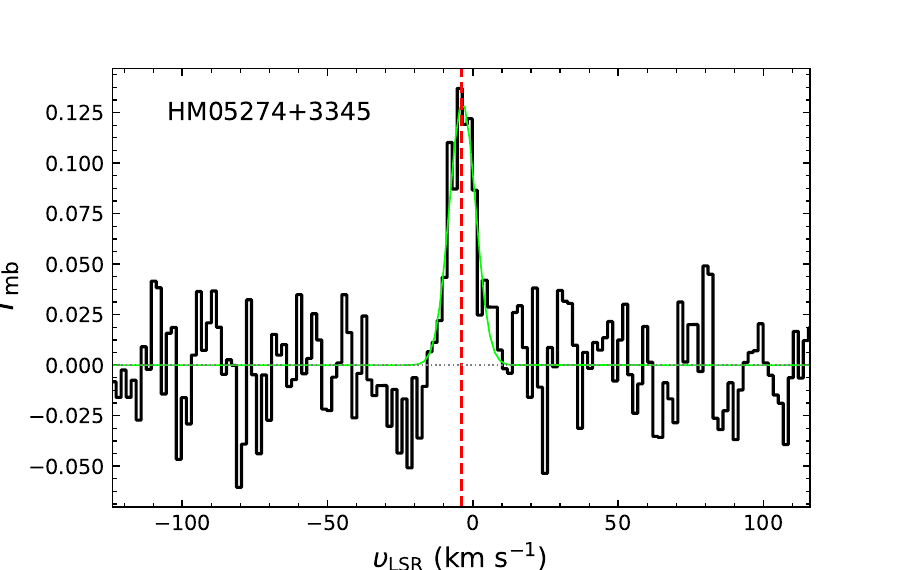}
    \includegraphics[width=0.33\textwidth]{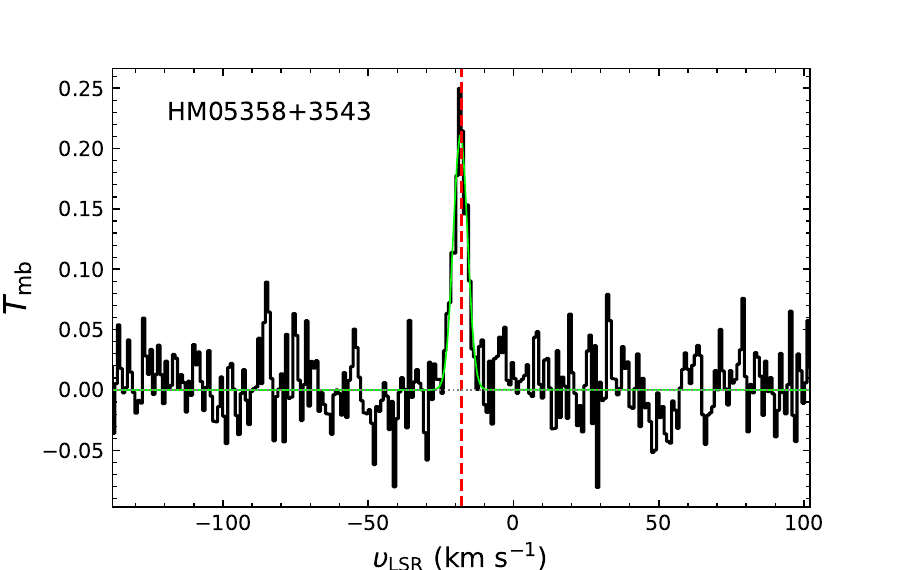}
    \includegraphics[width=0.33\textwidth]{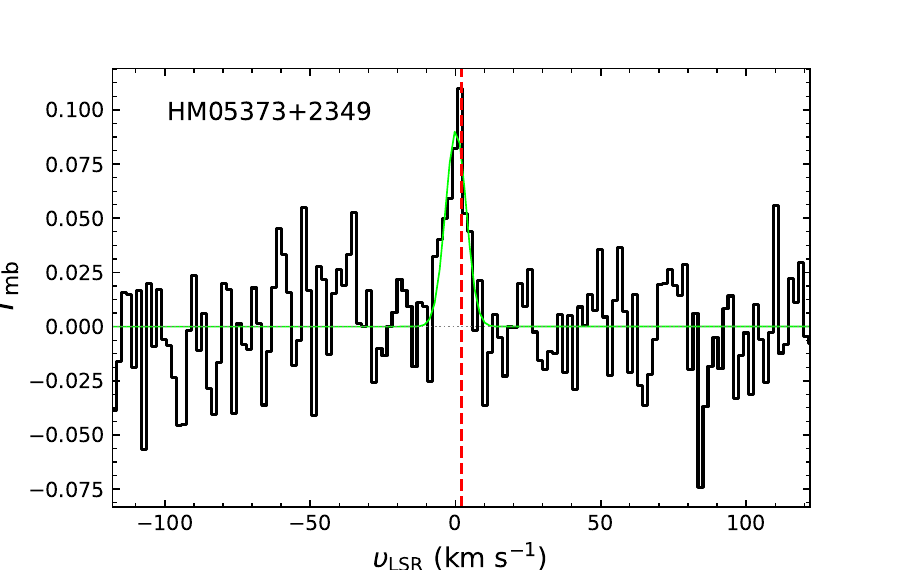}
    \includegraphics[width=0.33\textwidth]{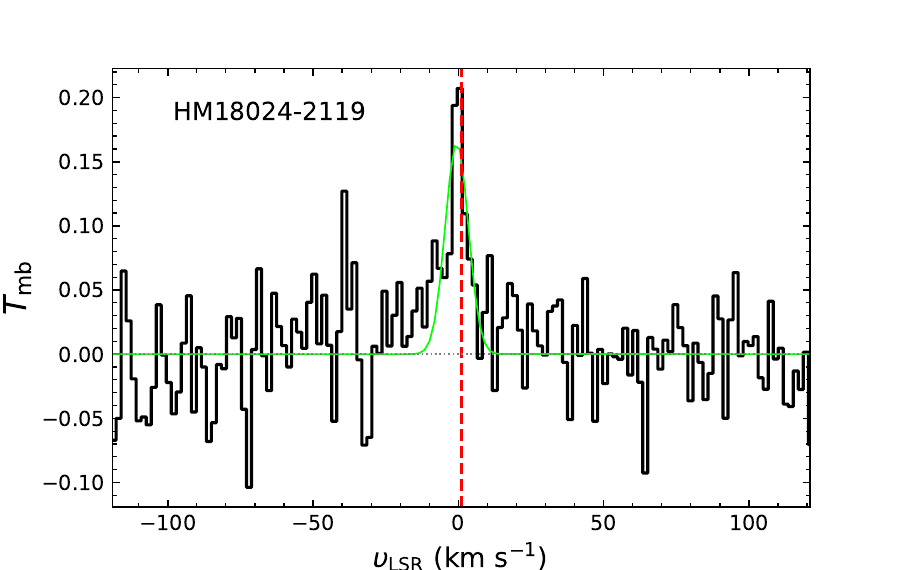}
    \includegraphics[width=0.33\textwidth]{./figures/HM18089-1732_SiO_tmb.pdf}
    \includegraphics[width=0.33\textwidth]{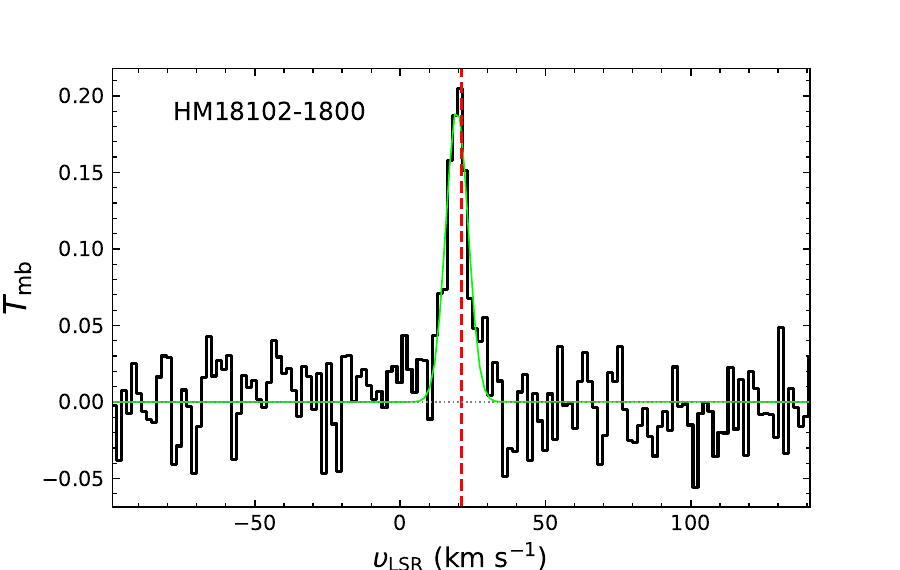}
    \includegraphics[width=0.33\textwidth]{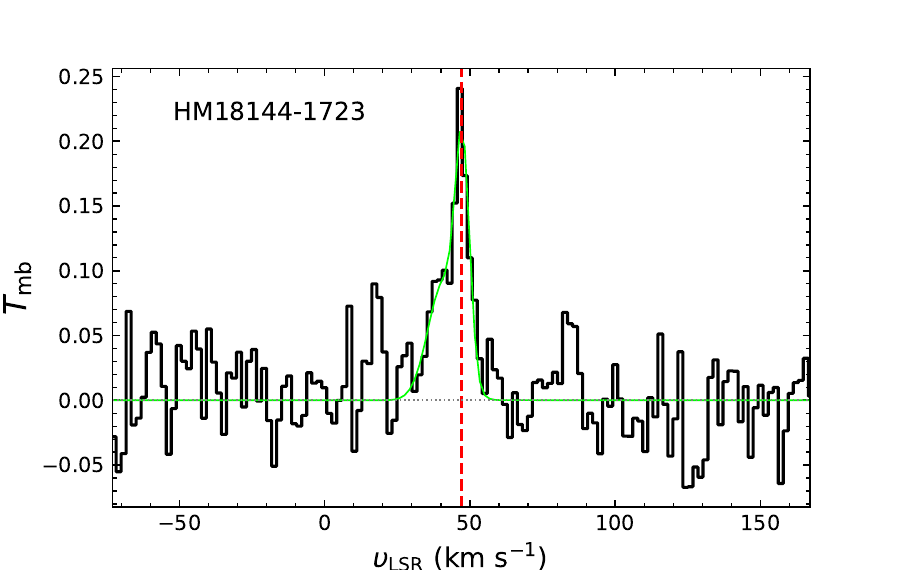}
    \includegraphics[width=0.33\textwidth]{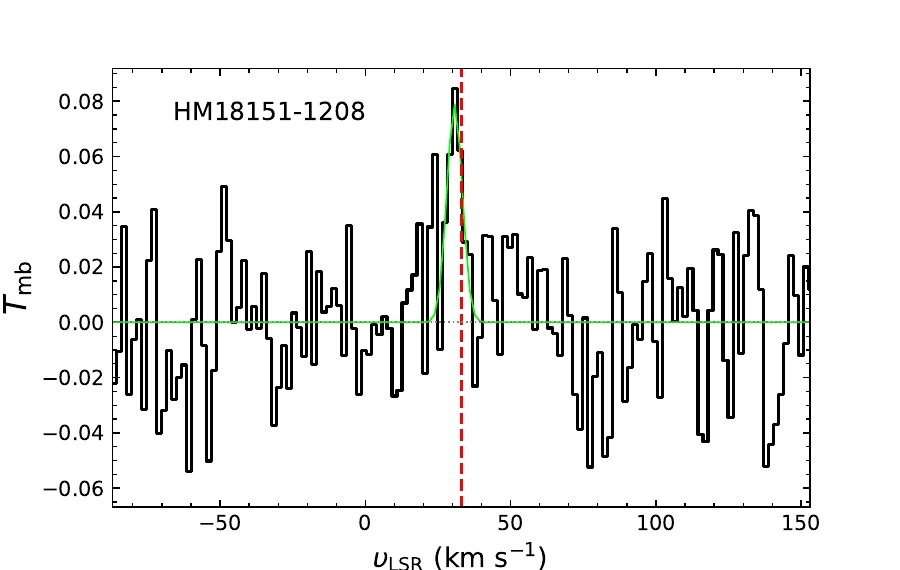}
    \includegraphics[width=0.33\textwidth]{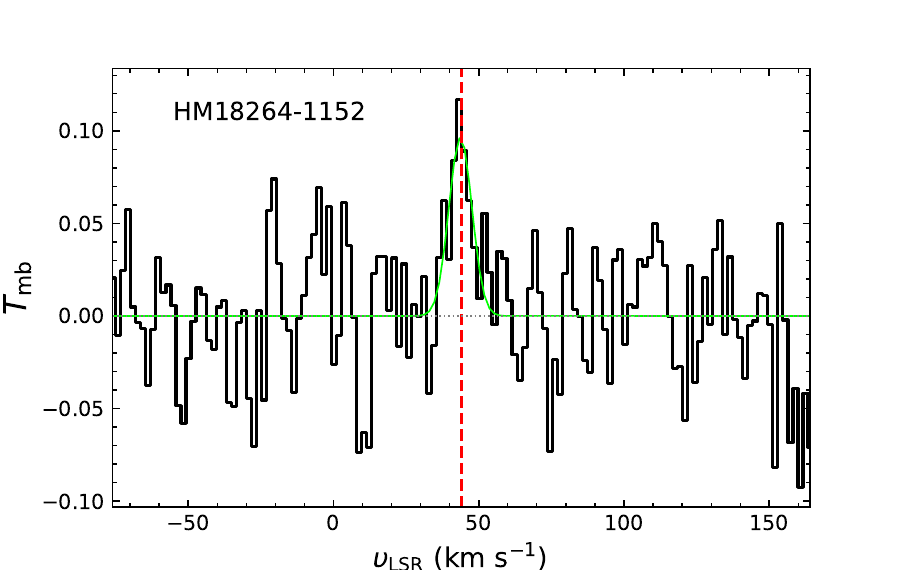}
    \includegraphics[width=0.33\textwidth]{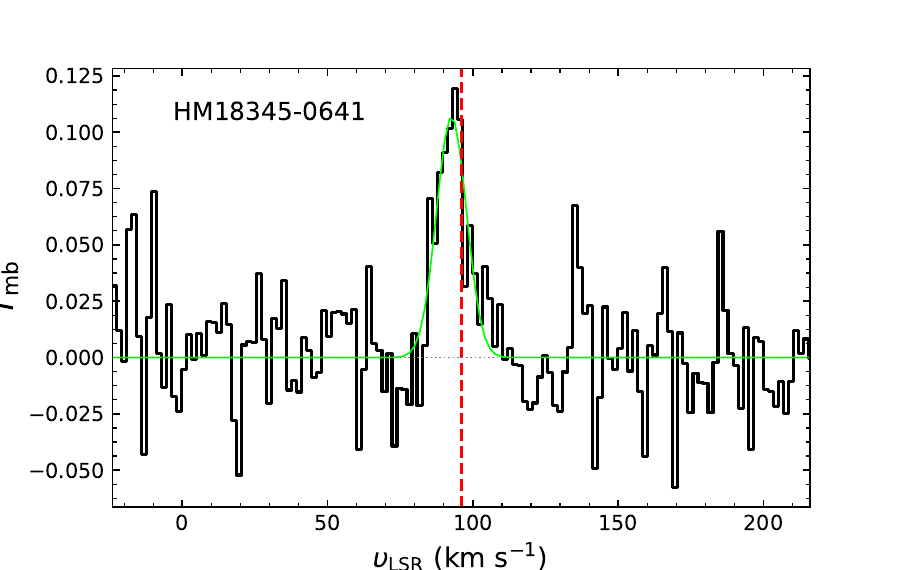}
    \includegraphics[width=0.33\textwidth]{./figures/HM18507+0121_SiO_tmb.pdf}
    \includegraphics[width=0.33\textwidth]{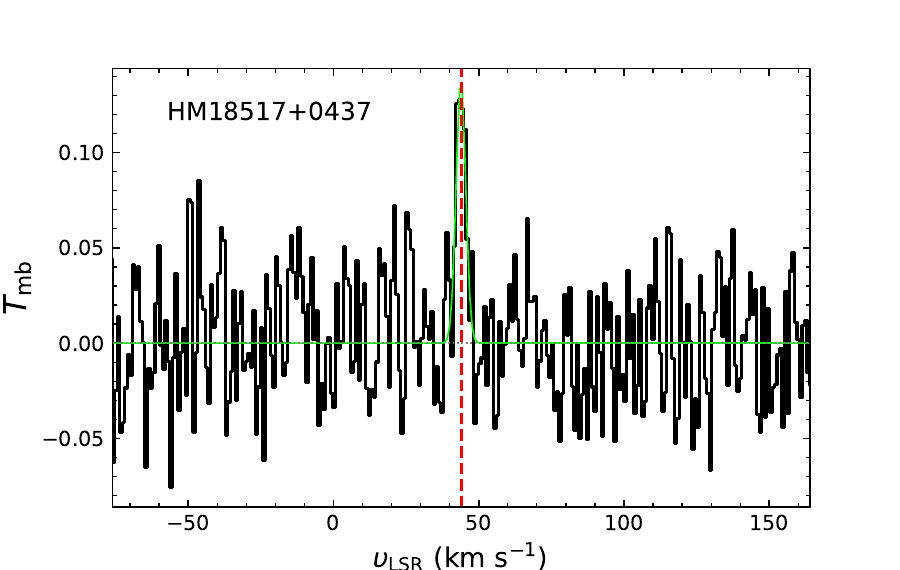}
    \caption{Continuation of Fig.\,\ref{appedix:sio_spectra1}} 
    \label{appendix:sio_spectra4}
\end{figure*}

\begin{figure*}[h!]
    \centering
    \includegraphics[width=0.33\textwidth]{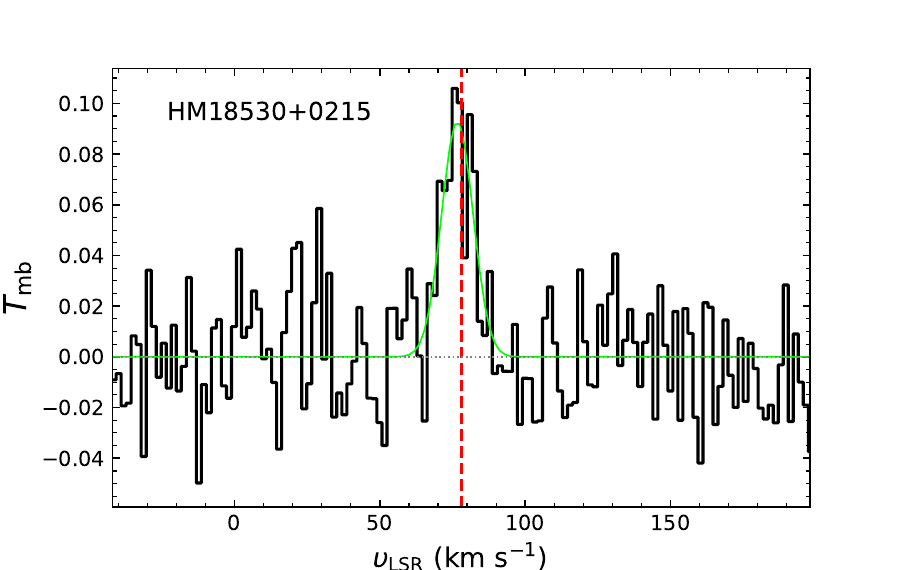}
    \includegraphics[width=0.33\textwidth]{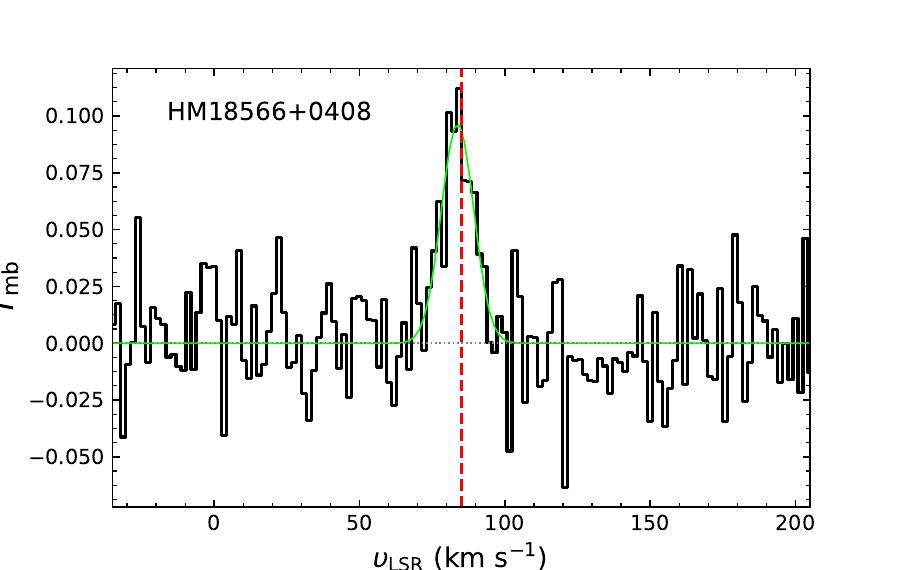}
    \includegraphics[width=0.33\textwidth]{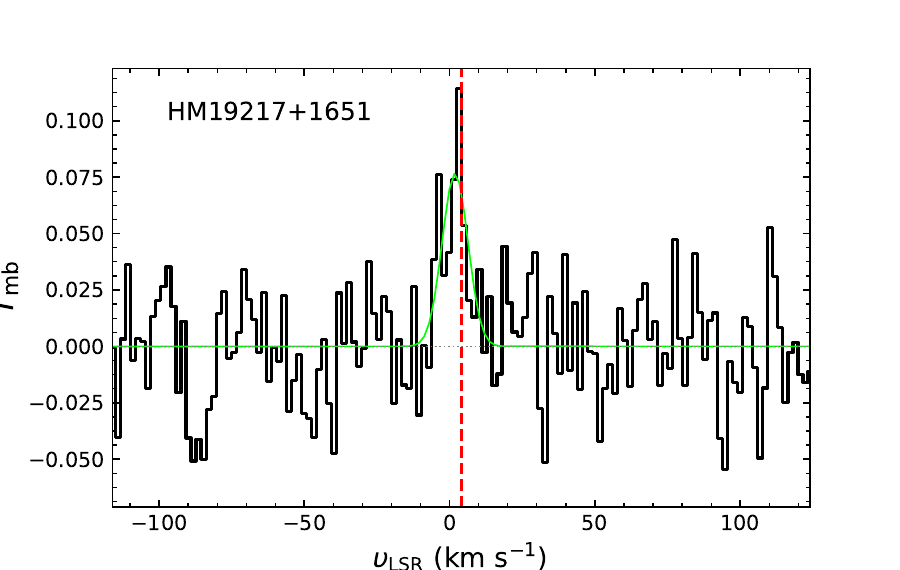}
    \includegraphics[width=0.33\textwidth]{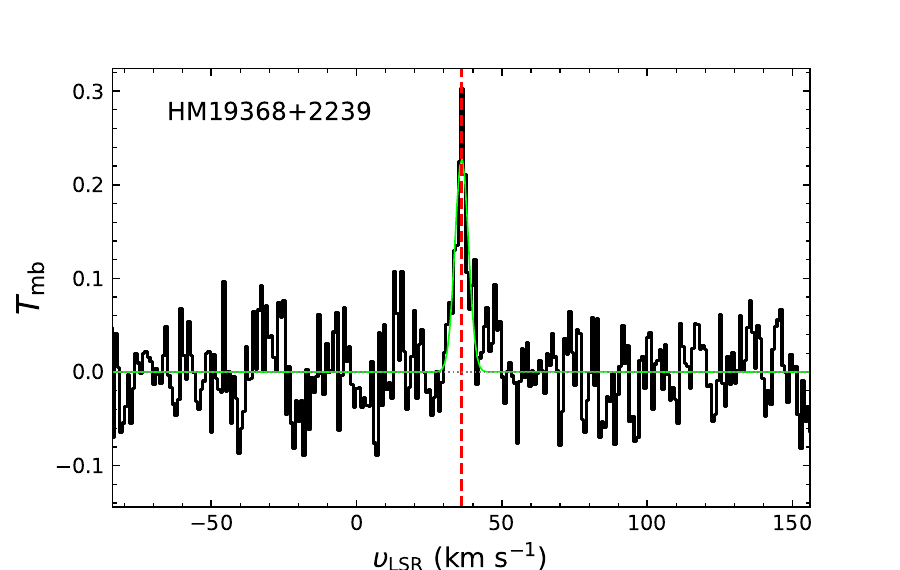}
    \includegraphics[width=0.33\textwidth]{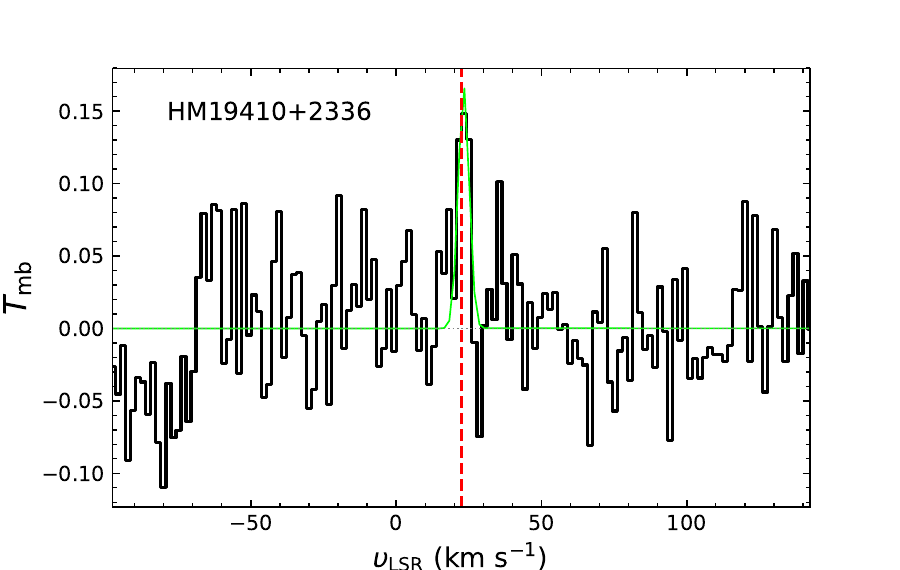}
    \includegraphics[width=0.33\textwidth]{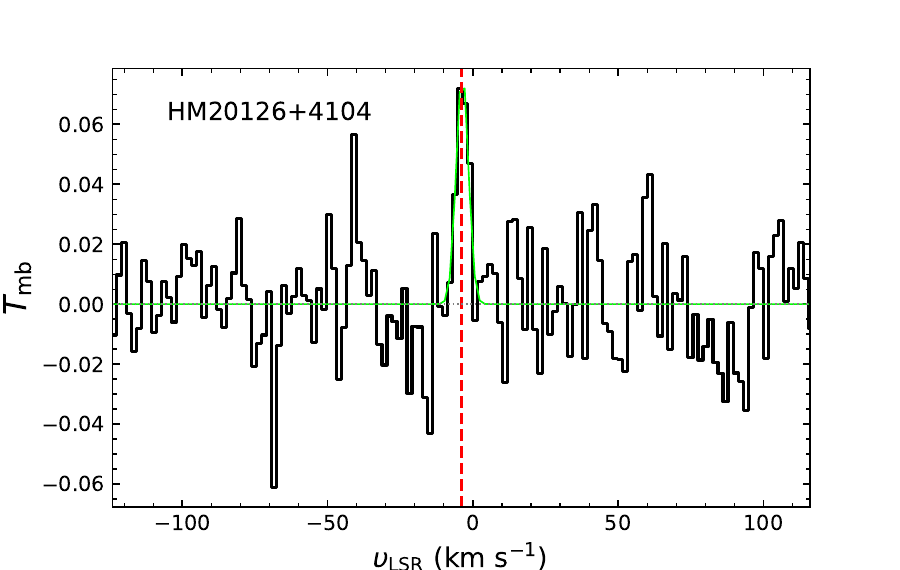}
    \includegraphics[width=0.33\textwidth]{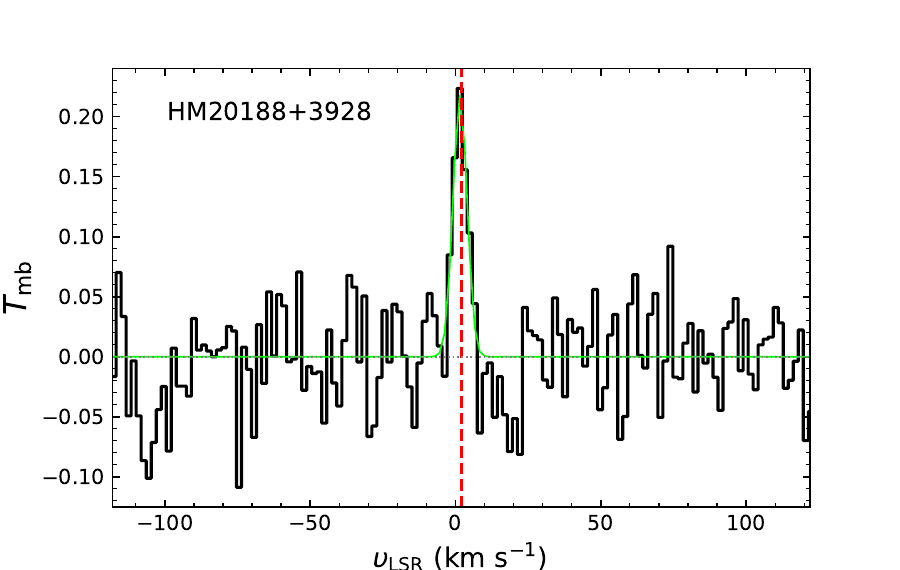}
    \includegraphics[width=0.33\textwidth]{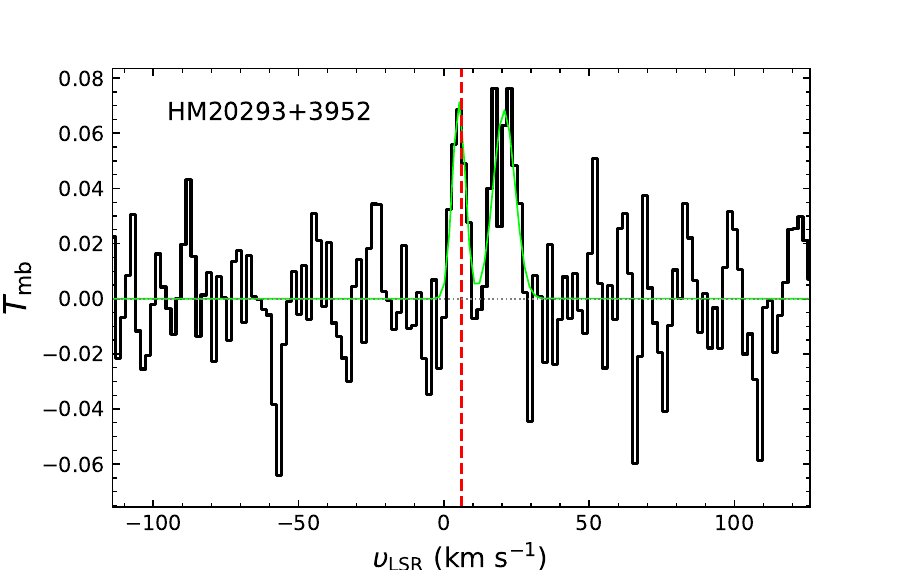}
    \includegraphics[width=0.33\textwidth]{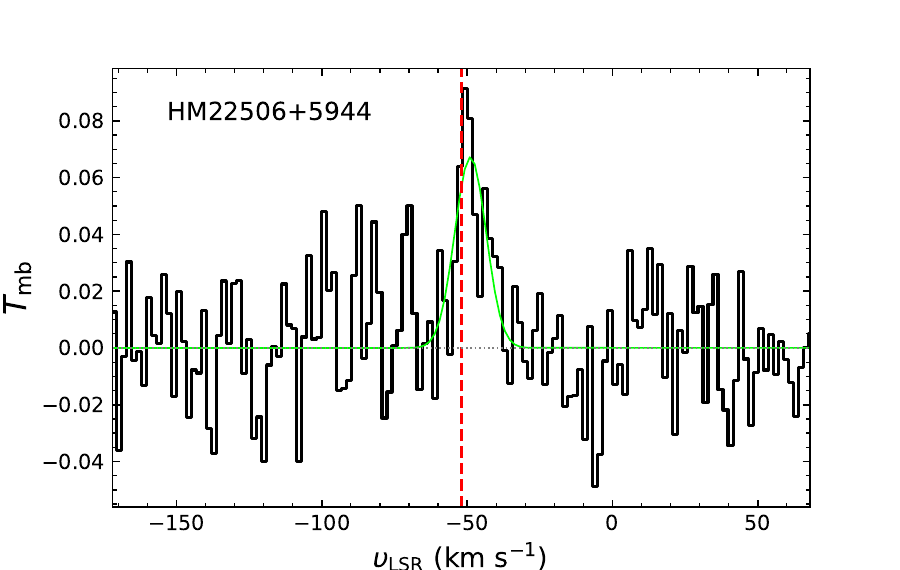}
    \includegraphics[width=0.33\textwidth]{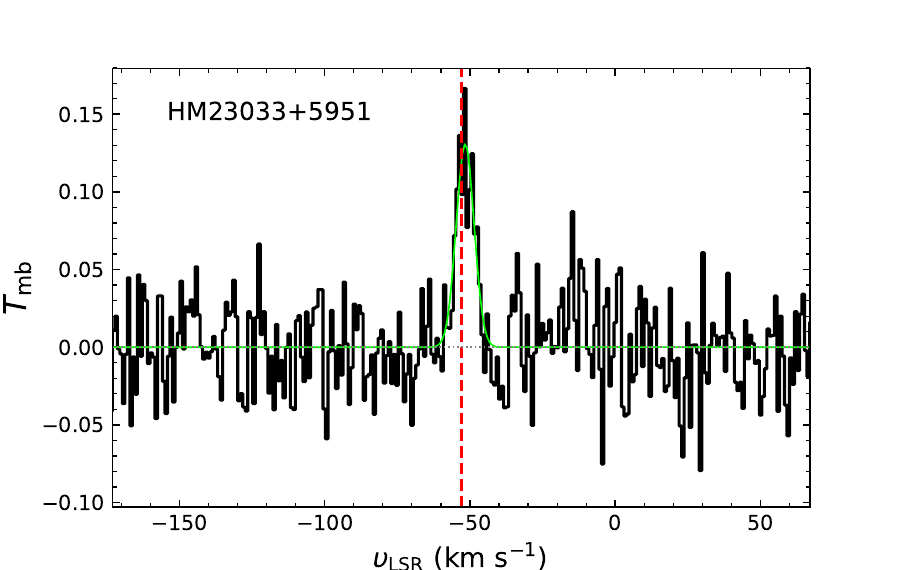}
    \includegraphics[width=0.33\textwidth]{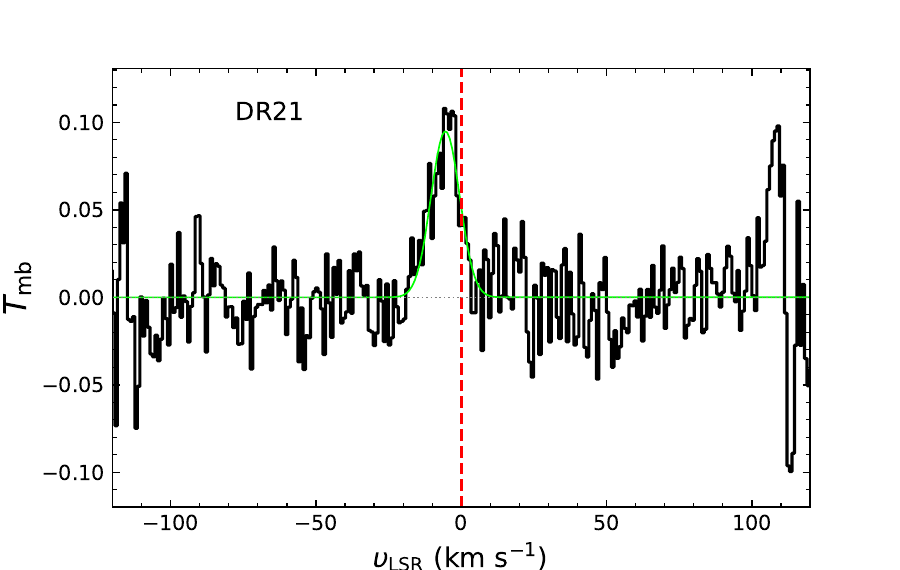}
    \includegraphics[width=0.33\textwidth]{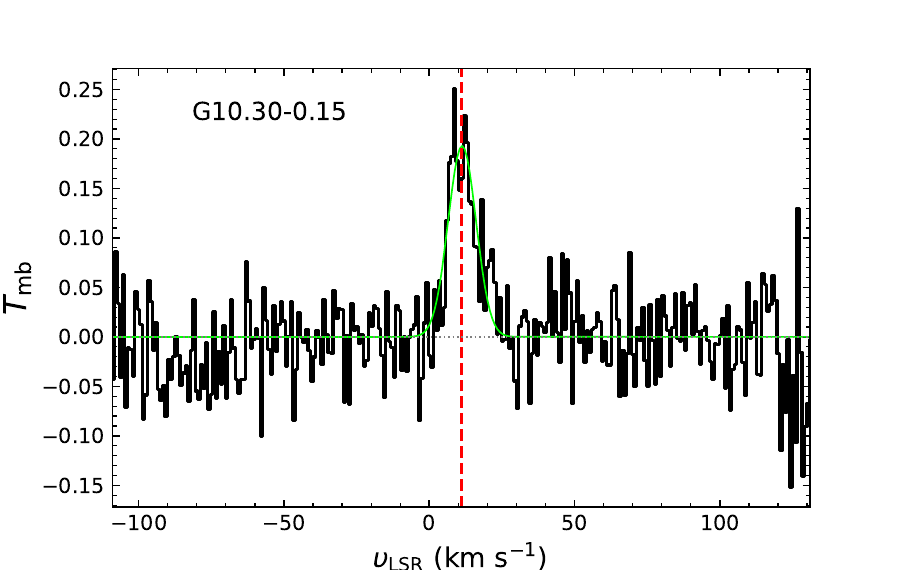}
    \includegraphics[width=0.33\textwidth]{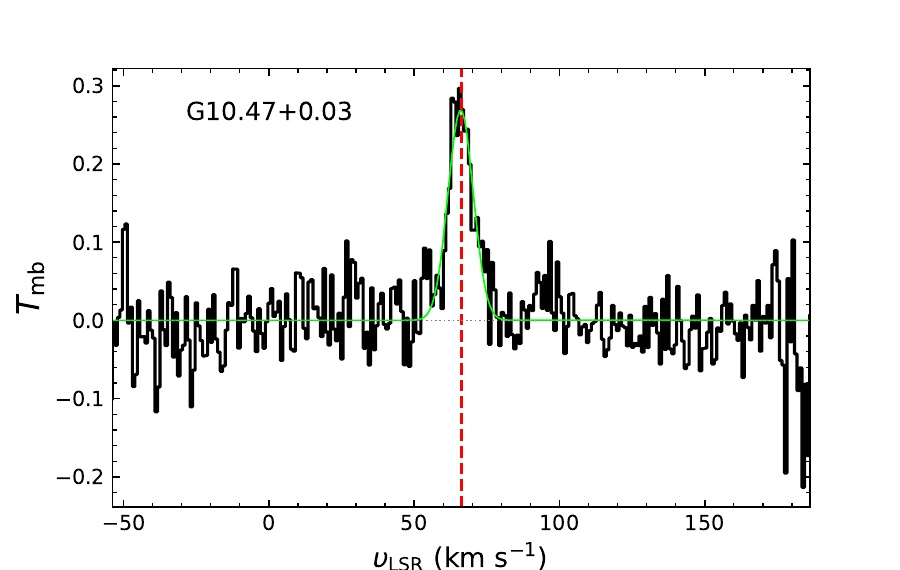}
    \includegraphics[width=0.33\textwidth]{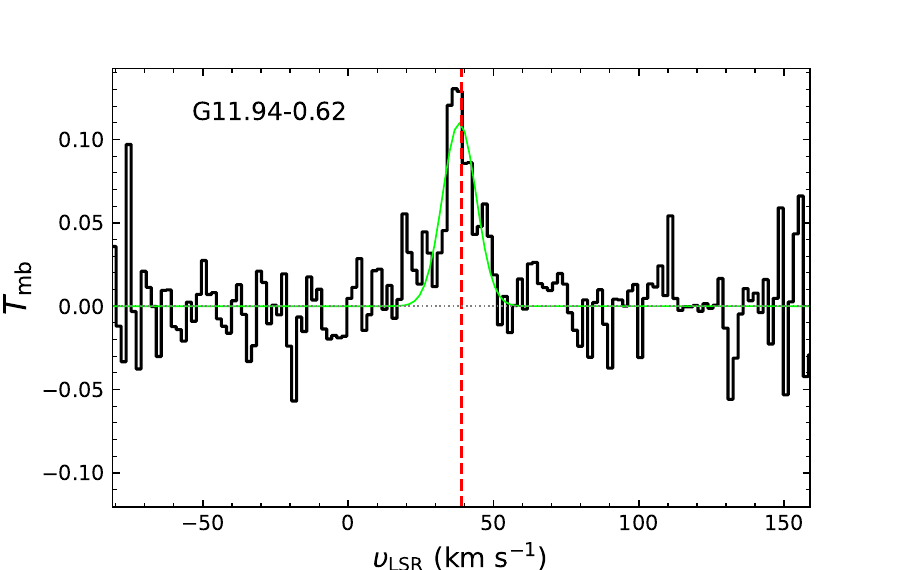}
    \includegraphics[width=0.33\textwidth]{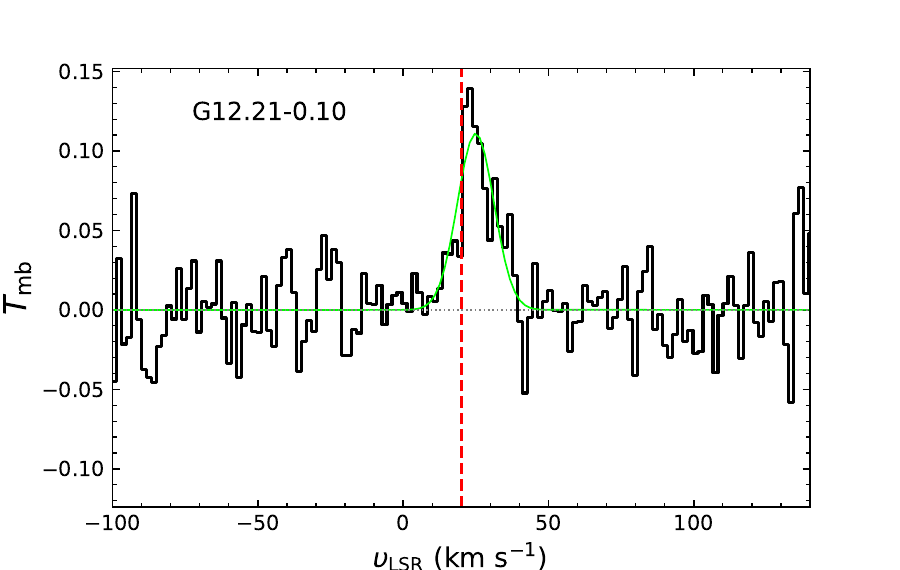}
    \includegraphics[width=0.33\textwidth]{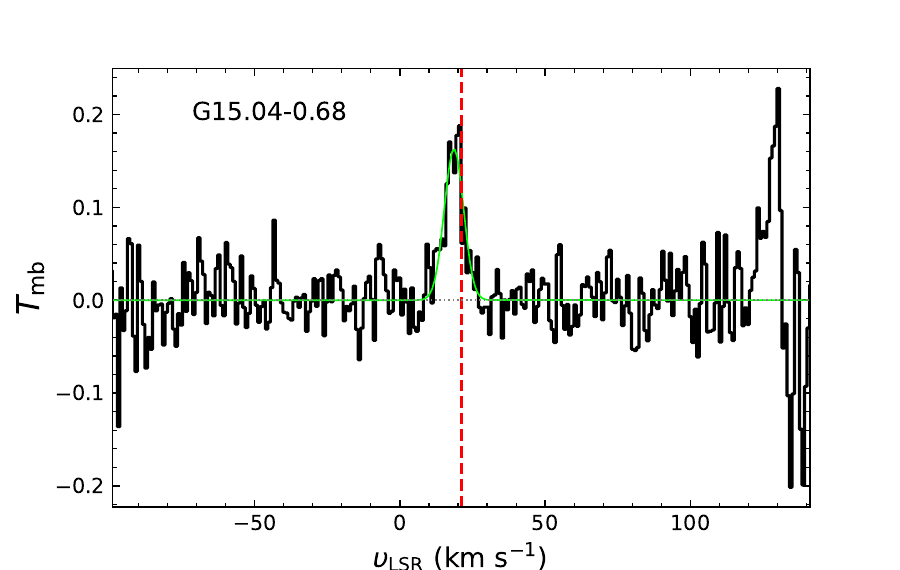}
    \includegraphics[width=0.33\textwidth]{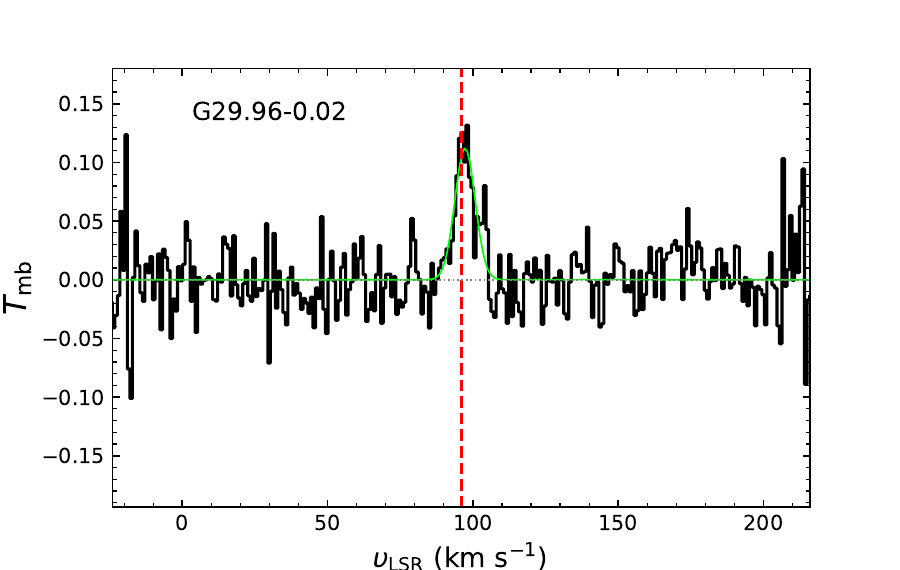}
    \includegraphics[width=0.33\textwidth]{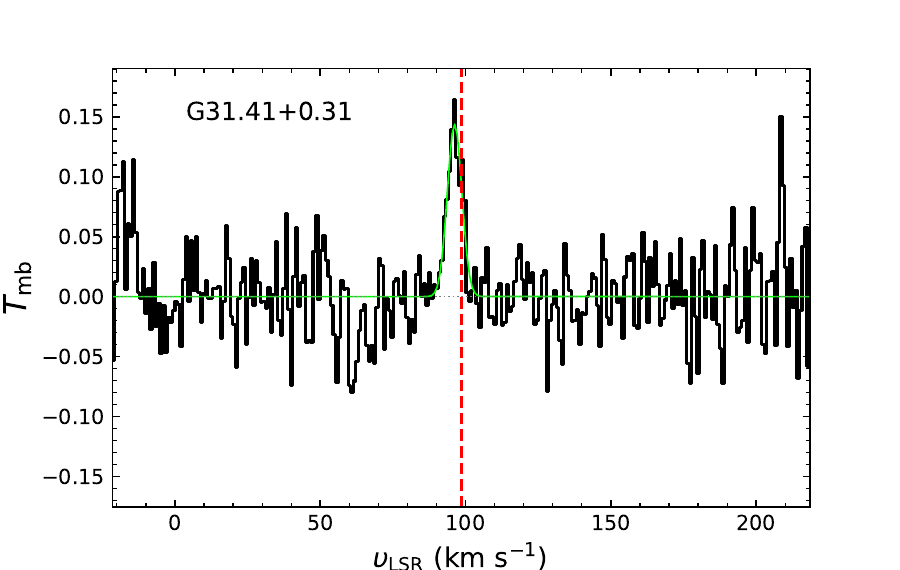}
    \caption{Continuation of Fig.\,\ref{appedix:sio_spectra1}}
    \label{appendix:sio_spectra5}
\end{figure*}

\begin{figure*}[h!]
    \centering
    \includegraphics[width=0.33\textwidth]{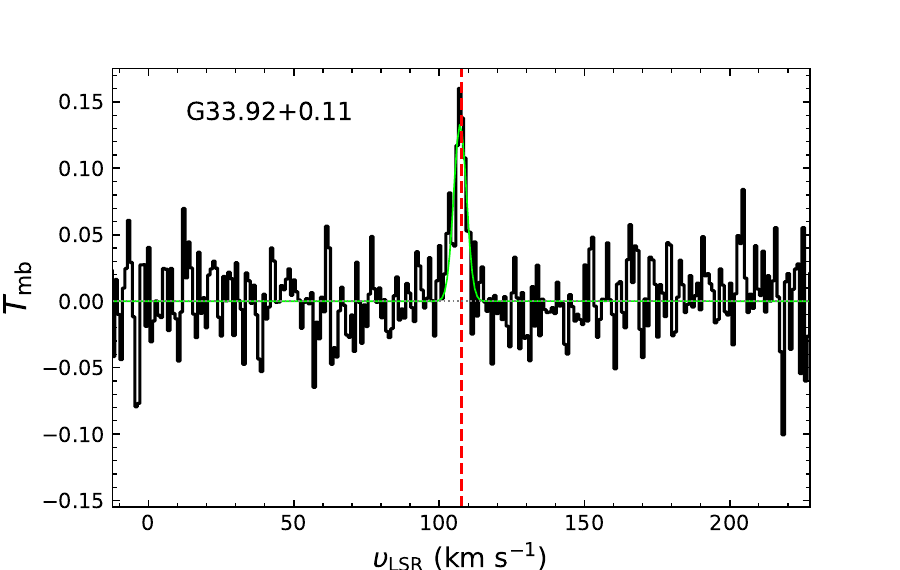}
    \includegraphics[width=0.33\textwidth]{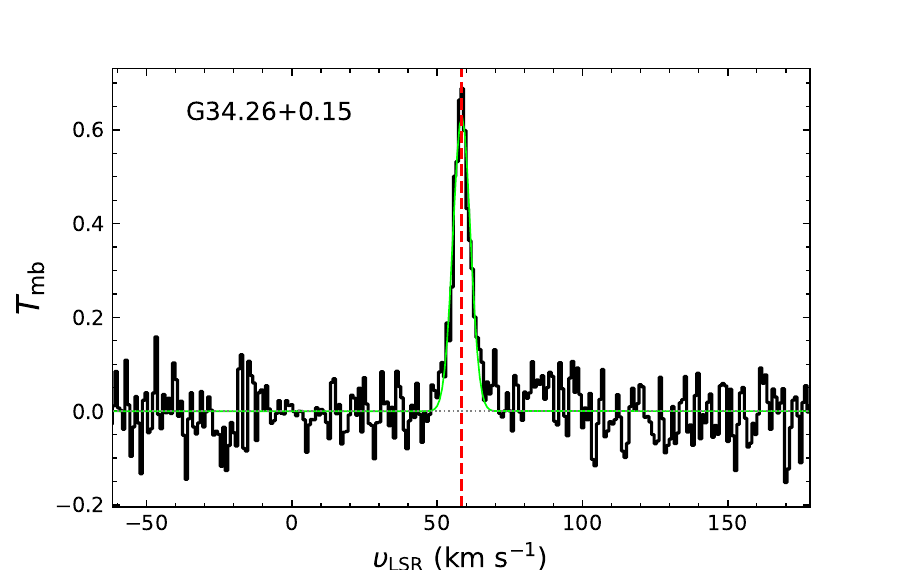}
    \includegraphics[width=0.33\textwidth]{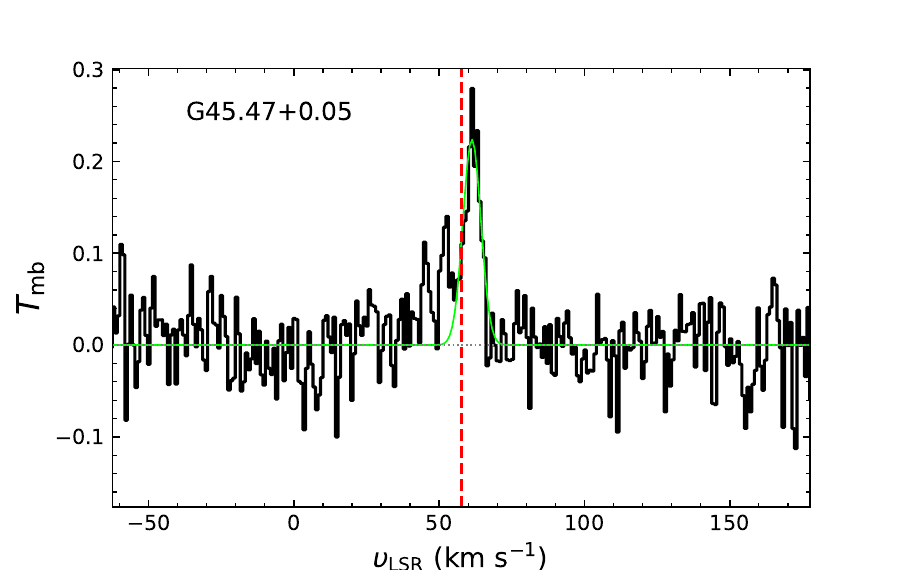}
    \includegraphics[width=0.33\textwidth]{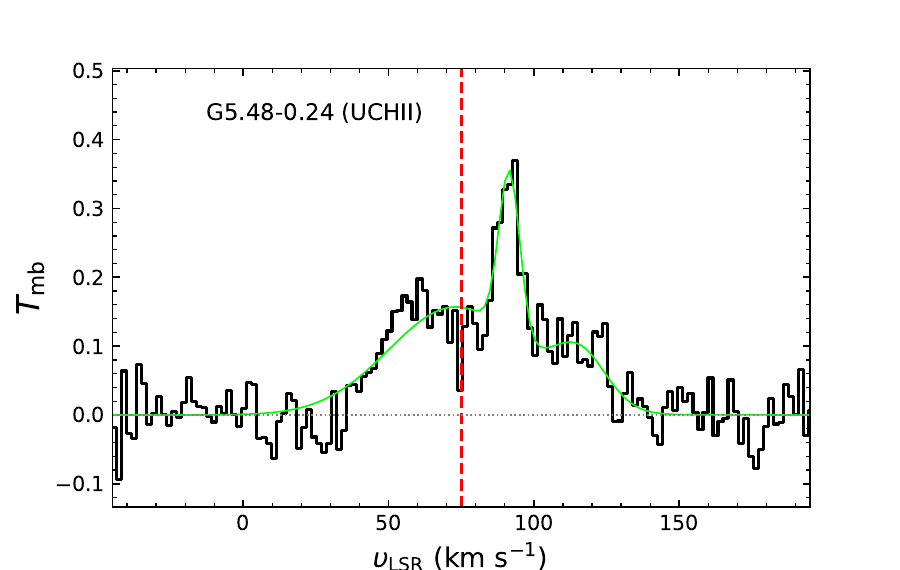}
    \includegraphics[width=0.33\textwidth]{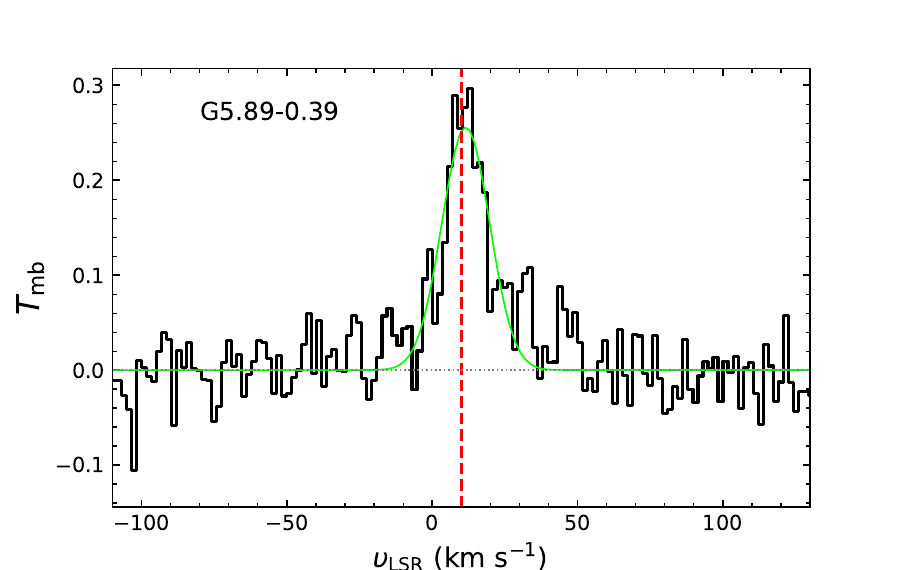}
    \includegraphics[width=0.33\textwidth]{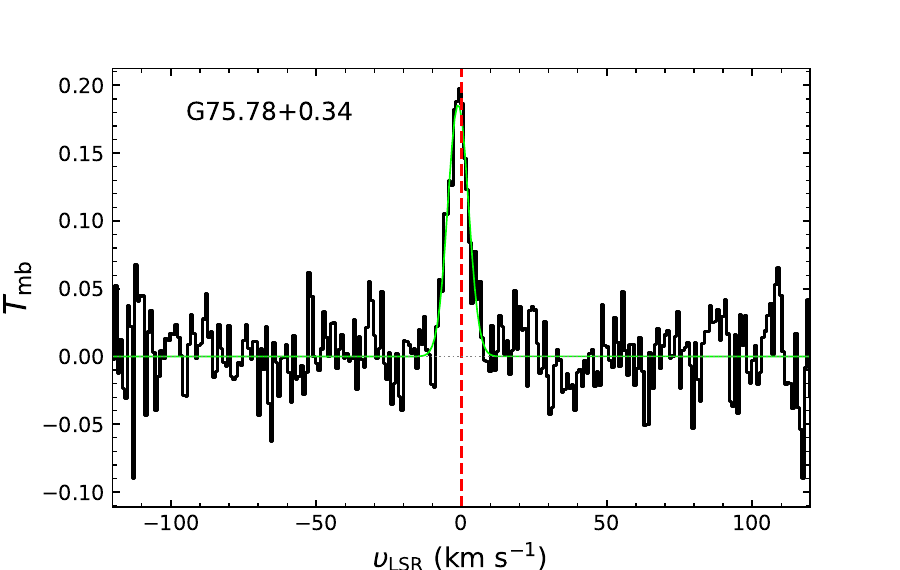}
    \includegraphics[width=0.33\textwidth]{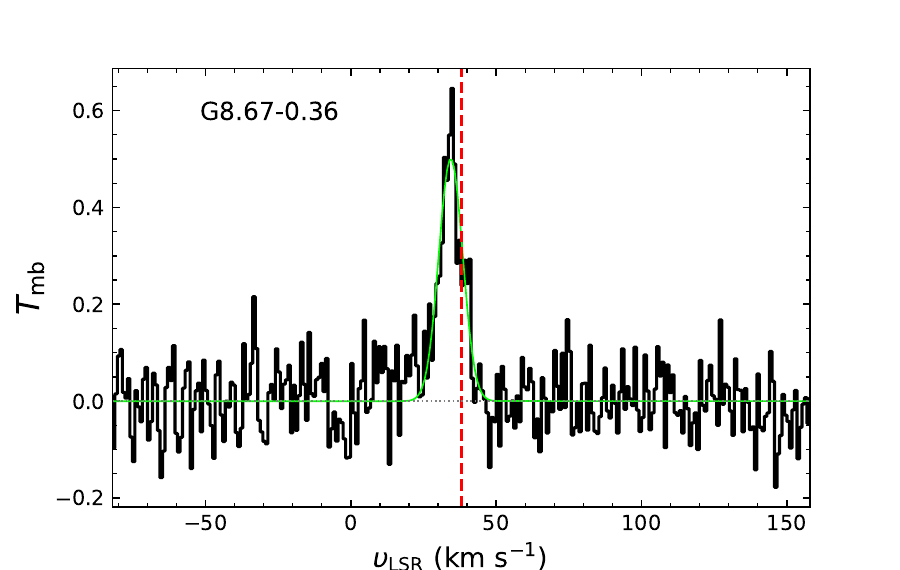}
    \includegraphics[width=0.33\textwidth]{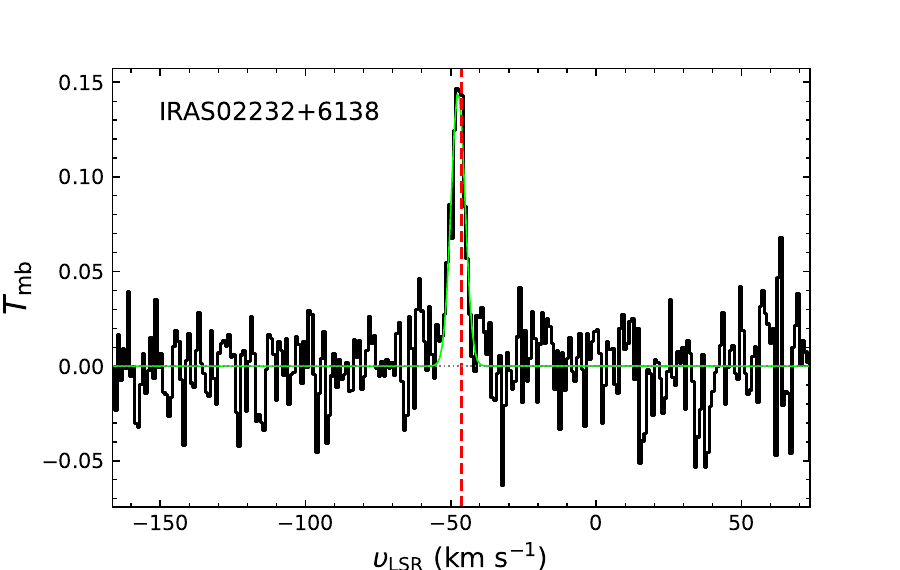}
    \includegraphics[width=0.33\textwidth]{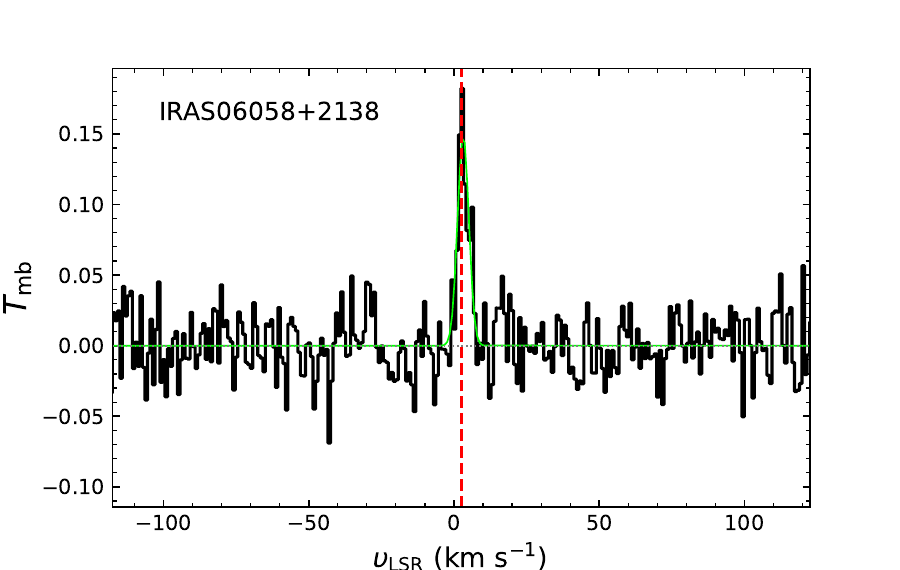}
    \includegraphics[width=0.33\textwidth]{./figures/IRAS18032-2032_SiO_tmb.pdf}
    \includegraphics[width=0.33\textwidth]{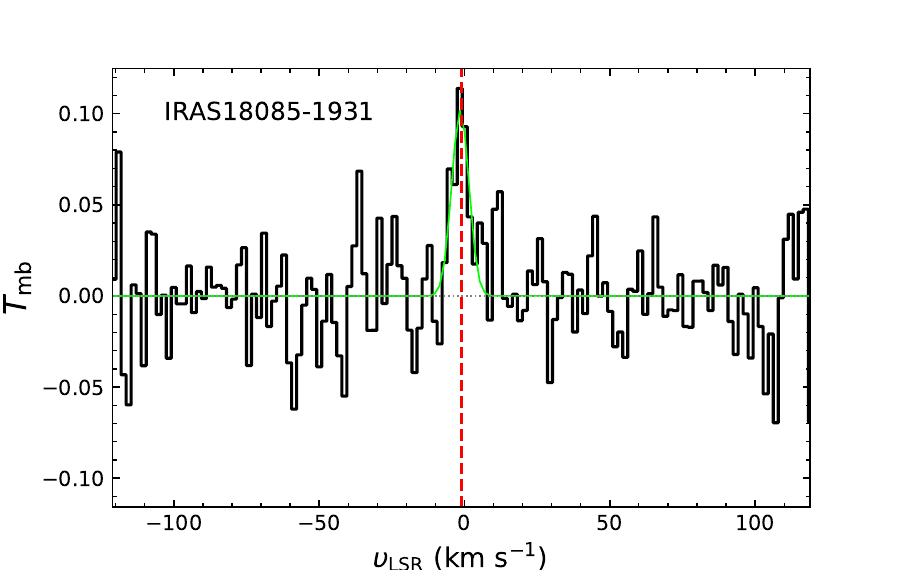}
    \includegraphics[width=0.33\textwidth]{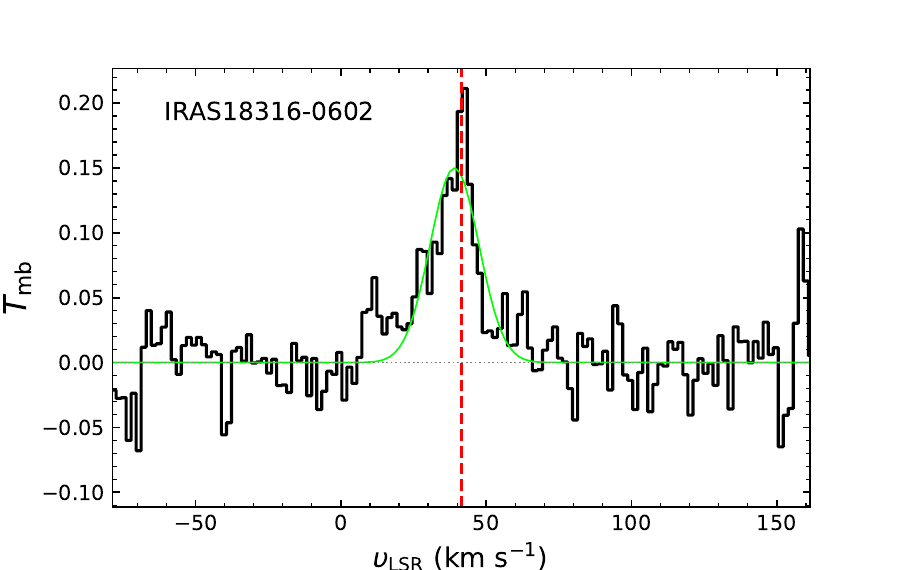}
    \includegraphics[width=0.33\textwidth]{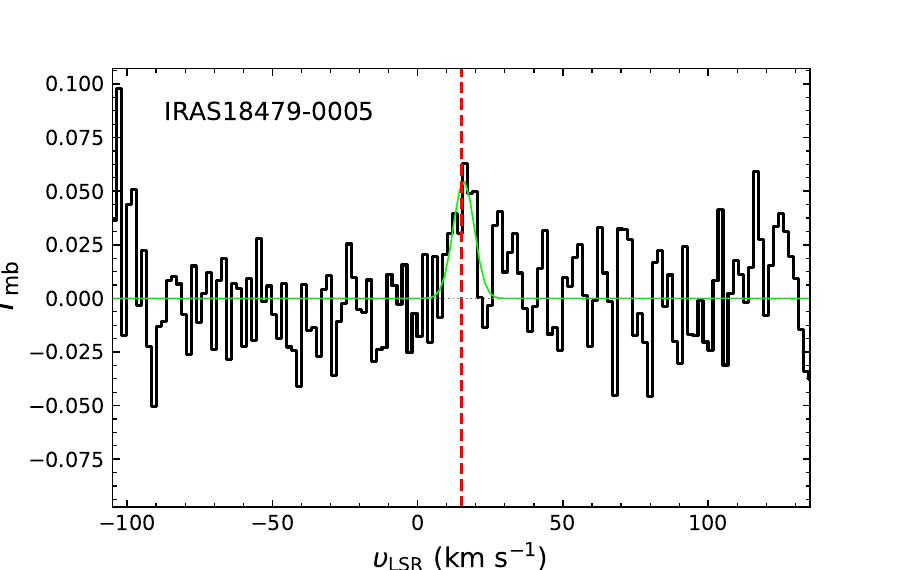}
    \includegraphics[width=0.33\textwidth]{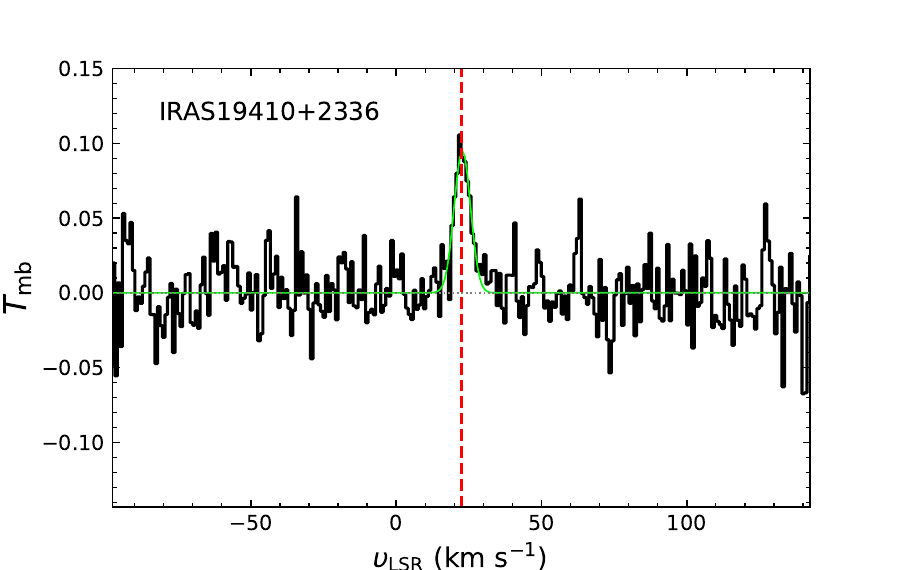}
    \includegraphics[width=0.33\textwidth]{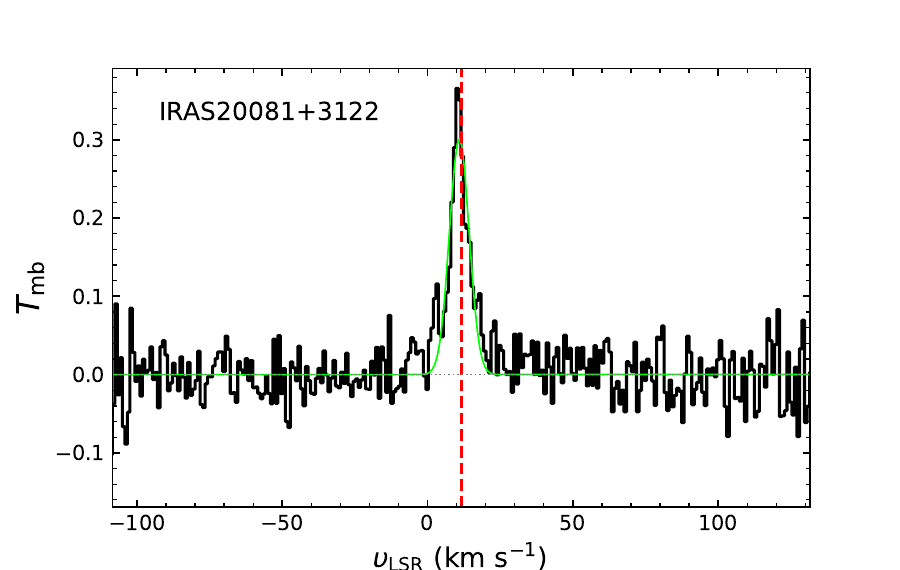}
    \includegraphics[width=0.33\textwidth]{./figures/IRAS22176+6303_SiO_tmb.pdf}
    \includegraphics[width=0.33\textwidth]{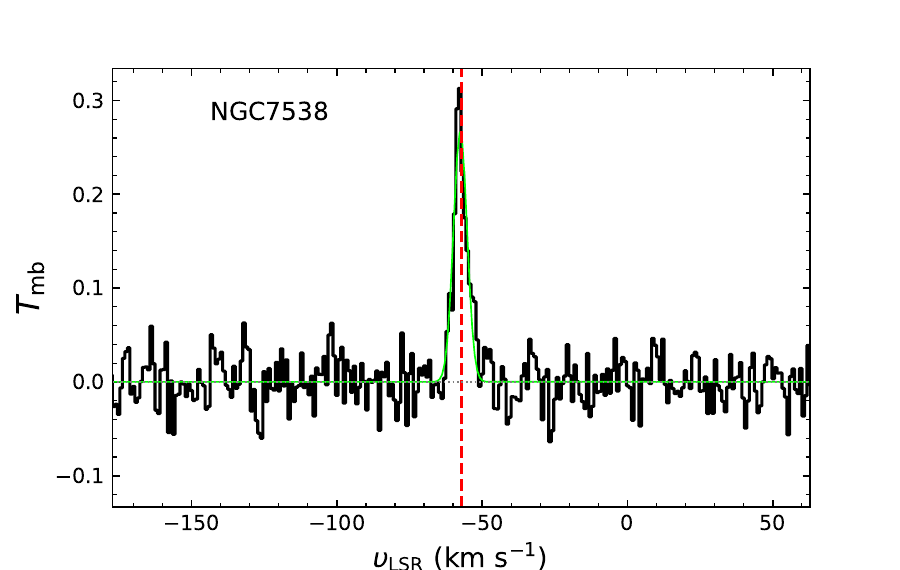}
    \includegraphics[width=0.33\textwidth]{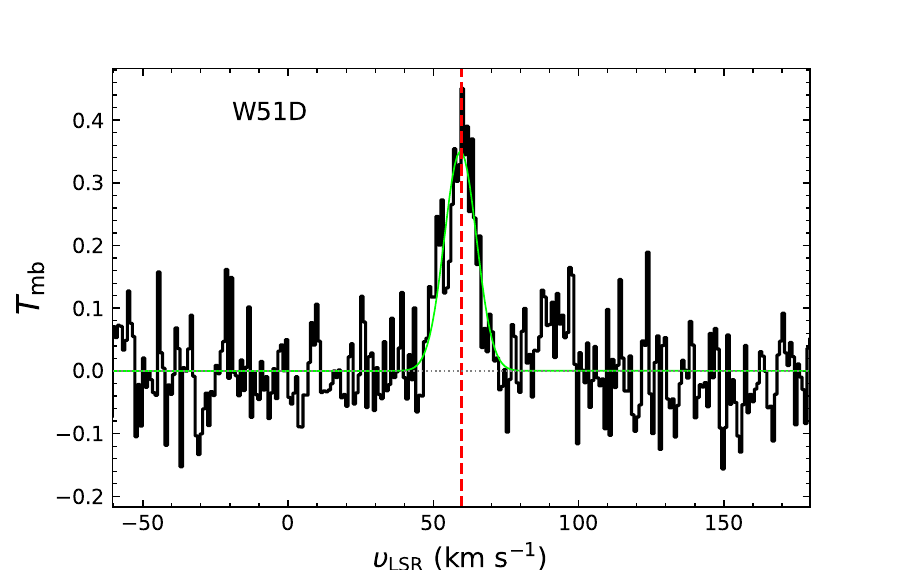}
    \caption{Continuation of Fig.\,\ref{appedix:sio_spectra1}}
    \label{appendix:sio_spectra6}
\end{figure*}
\clearpage
\newpage
\onecolumn

\clearpage
\newpage
\section{Histograms}
\begin{figure}[h!]
    \centering
    \includegraphics[width=0.48\textwidth]{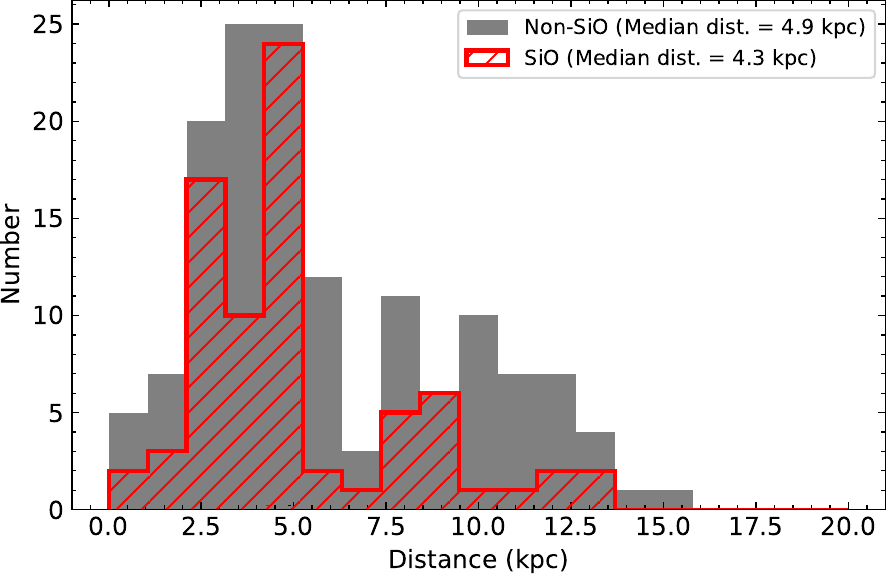}
    \includegraphics[width=0.48\textwidth]{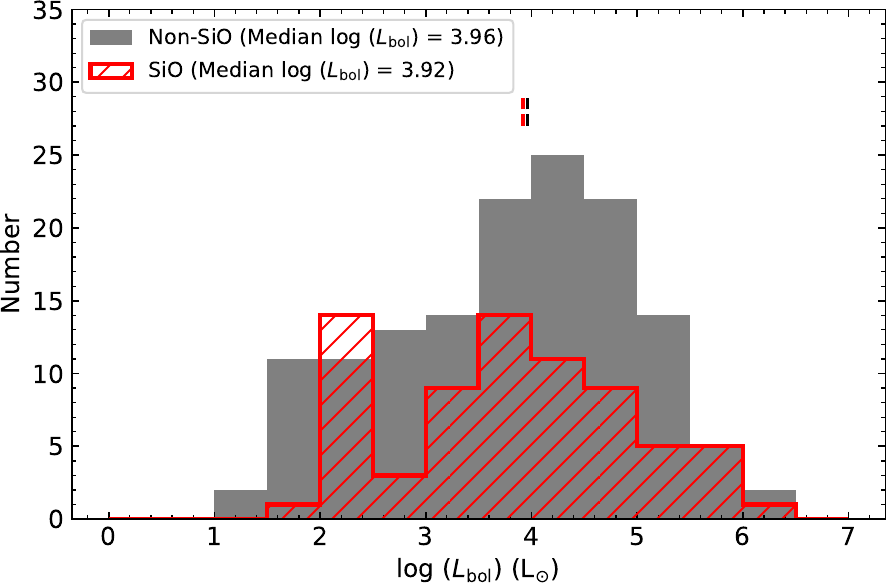}
    \caption{Histograms of distance (left) and bolometric luminosity (right)for sources with SiO detection (red) and non-detection (gray).}
    \label{fig:hist_dist_lbol_mass}
\end{figure}

\end{appendix}

\end{document}